\documentclass[comsoc ]{IEEEtran}
\IEEEoverridecommandlockouts

\usepackage[utf8]{inputenc}
\usepackage{xspace}
\usepackage{color}
\usepackage{soul}
\usepackage{newtxmath}
\usepackage{hyperref}
\usepackage{enumitem}
\usepackage{multirow}
\usepackage{authblk}
\usepackage{graphicx}
\usepackage{dirtytalk}
\usepackage{tablefootnote}
\usepackage{fancyhdr}

\newtoggle{draftmode}
\settoggle{draftmode}{false}

\newif\ifarXiv
\arXivtrue 

\newcommand{\ourname}{\textsc{TraceCORONA}\xspace}
\newcommand{\TraceCORONA}{\textsc{TraceCORONA}\xspace}
\newcommand{\ProntoCC}{\textsc{Pronto-C2}\xspace}
\newcommand{\dct}{DCT\xspace}
\definecolor{forestgreen}{rgb}{0.13, 0.55, 0.13}

\newcommand{\advUser}{{\ensuremath{\mathcal{A}^u}}\xspace}
\newcommand{\advServer}{{\ensuremath{\mathcal{A}^s}}\xspace}
\newcommand{\advWormhole}{{\ensuremath{\mathcal{A}^w}}\xspace}
\newcommand{\advEavesdropper}{{\ensuremath{\mathcal{A}^e}}\xspace}

\newcommand{\aSurveillance}{Surveillance\xspace}
\newcommand{\aMassSurveillance}{Mass Surveillance\xspace}
\newcommand{\aRelayAttack}{Relay/Replay Attack\xspace}
\newcommand{\aProfilingAttack}{Profiling Attacks\xspace}

\newcommand{\server}{\mathit{SP}}
\newcommand{\healthauthority}{\mathit{HA}}
\newcommand{\service}{\mathit{SP}}
\newcommand{\useri}{\mathit{U_i}}
\newcommand{\userj}{\mathit{U_j}}
\newcommand{\kijl}{k^l_{ij}\xspace}
\newcommand{\kjil}{k^l_{ji}\xspace}
\newcommand{\tijl}{t^l_{ij}\xspace}
\newcommand{\tjil}{t^l_{ji}\xspace}
\newcommand{\mtan}{\mathit{TAN}\xspace}
\newcommand{\tani}{\mathit{TAN_i}\xspace}

\newcommand{\userId}{\mathit{UserID}}
\newcommand{\tempId}{\mathit{TempID}}
\newcommand{\tek}{\mathit{TEK}\xspace}
\newcommand{\et}{ET\xspace}

\newcommand{\app}{TC-App\xspace}

\newcommand{\changeT}[1]{\textcolor{black}{#1}}

\newcommand{\fixedMC}[1]{\textcolor{black}{#1}}
\newcommand{\fixedAlC}[1]{\textcolor{black}{#1}}
\newcommand{\fixedAC}[1]{\textcolor{black}{#1}}

\newcommand{\gap}{GAEN\xspace}

\newcommand{\rAccuracy}{Accuracy\xspace}
\newcommand{\rAccuracyIdx}{R-Ef1\xspace}

\newcommand{\rSuperspreader}{Superspreader\xspace}
\newcommand{\rSuperspreaderIdx}{R-Ef2\xspace}
\newcommand{\rCAII}{CAII\xspace}

\newcommand{\rAccountability}{Accountability\xspace}
\newcommand{\rAccountabilityIdx}{R-Ef3\xspace}

\newcommand{\rIdentifying}{Identifying users\xspace}
\newcommand{\rIdentifyingIdx}{R-P1\xspace}

\newcommand{\rTracking}{Tracing users\xspace}
\newcommand{\rTrackingIdx}{R-P2\xspace}

\newcommand{\rSocialGraph}{Extracting social graph\xspace}
\newcommand{\rSocialGraphIdx}{R-P3\xspace}

\newcommand{\rFakeClaim}{Fake exposure claim\xspace}
\newcommand{\rFakeClaimIdx}{R-S1\xspace}

\newcommand{\rFakeInjection}{Fake exposure Injection\xspace}
\newcommand{\rFakeInjectionIdx}{R-S2\xspace}

\newcommand{\rTransparancyIdx}{R-Et1\xspace}

\newcommand{\rIndependence}{Independence\xspace}
\newcommand{\rIndependenceIdx}{R-Et2\xspace}

\iftoggle{draftmode} {
	\newcommand\dnote[1]{\textcolor{magenta}{#1}}
	\newcommand\bnote[1]{\textcolor{blue}{#1}}
	\newcommand\dknote[1]{\textcolor{violet}{#1}}
	\newcommand\pjnote[1]{\textcolor{orange}{#1}}
	\newcommand\tfnote[1]{\textcolor{cyan}{#1}}
	\newcommand\snote[1]{\TODO{\textcolor{olive}{#1}}}
	\setlength{\marginparwidth}{16mm}
	\setlength{\marginparsep}{.5mm}
	\newcommand{\TODO}[1]{\marginpar{\textcolor{red}{TODO\\\footnotesize #1}}}
} {
	\newcommand\dnote[1]{}
	\newcommand\bnote[1]{}
	\newcommand\dknote[1]{}
	\newcommand\pjnote[1]{}
	\newcommand\tfnote[1]{}
	\newcommand\snote[1]{}
	\newcommand{\TODO}[1]{}
	\hypersetup{hidelinks}
}

\begin{document}

\title{Digital Contact Tracing Solutions: Promises, Pitfalls and Challenges}

\author[1]{Thien Duc Nguyen}
\author[1]{Markus Miettinen}
\author[2]{Alexandra Dmitrienko}
\author[1]{Ahmad-Reza Sadeghi}
\author[3] {Ivan Visconti}
\affil[1]{Technical University of Darmstadt, Germany - \{ducthien.nguyen, markus.miettinen, ahmad.sadeghi\}@trust.tu-darmstadt.de}
\affil[2]{JMU Würzburg, Germany - alexandra.dmitrienko@uni-wuerzburg.de}
\affil[3]{University of Salerno, Italy - visconti@unisa.it}

\setcounter{Maxaffil}{0}
\renewcommand\Affilfont{\itshape\small}

%

\maketitle

\thispagestyle{fancy}
\chead{\small{A core part of this paper is to be published in IEEE Transactions on Emerging Topics in Computing, DOI: 10.1109/TETC.2022.3216473}}

\newcommand{\protocolName}{\Pi_{FO}\xspace}

\newcommand{\upcoming}[1]{\textcolor{gray}{(TODO: #1)}}

\newcommand{\txt}[1]{\mathit{#1}}
\newcommand{\txtMsg}[1]{\mathtt{#1}}

\newcommand{\define}{:=}
\newcommand{\inRand}{\in_{R}}
\newcommand{\seed}{\text{s}}
\newcommand{\abs}[1]{\lvert#1\rvert}

\newcommand{\pk}{\txt{pk}}
\newcommand{\sk}{\txt{sk}}
\newcommand{\pubk}{\txt{pubk}}
\newcommand{\GenSig}{\txt{GenSig}}
\newcommand{\GenPK}{\txt{GenPK}}
\newcommand{\Enc}{\txt{Enc}}
\newcommand{\Dec}{\txt{Dec}}
\newcommand{\Sign}{\txt{Sign}}
\newcommand{\Verify}{\txt{Verify}}
\newcommand{\Hash}{H}
\newcommand{\HKDF}{\mathit{HKDF}}

\newcommand{\poolsize}{n}

\newcommand{\BC}{BC}
\newcommand{\ledger}{\mathsf{BC}}
\newcommand{\enclave}{\mathcal{E}}
\newcommand{\tee}{T}	
\newcommand{\teeSet}{\mathcal{T}}
\newcommand{\manager}{M}
\newcommand{\contract}{C}
\newcommand{\contractSet}{\mathcal{C}}
\newcommand{\user}{U}

\newcommand{\TEE}{\txt{TEE}}
\newcommand{\install}{\txt{install}}
\newcommand{\resume}{\txt{resume}}
\newcommand{\prog}{\txt{prog}}
\newcommand{\VerifyQuote}{\txt{VerifyQuote}}

\newcommand{\latestBock}{\txt{now}}
\newcommand{\nbConfBlocks}{\gamma}
\newcommand{\blockTime}{\tau}
\newcommand{\confTime}{\Delta}

\newcommand{\tees}{\txt{registered}}	
\newcommand{\contracts}{\txt{contracts}}

\newcommand{\conCreator}{\txt{creator}}
\newcommand{\conCodeHash}{\txt{codeHash}}
\newcommand{\conPool}{\txt{pool}}
\newcommand{\conExecChalMsg}{\txt{c1Msg}}
\newcommand{\conWatchChalMsg}{\txt{c2Msg}}
\newcommand{\conExecChalBlock}{\txt{c1Time}}
\newcommand{\conWatchChalBlock}{\txt{c2Time}}
\newcommand{\conExecChalRes}{\txt{c1Res}}
\newcommand{\conWatchChalRes}{\txt{c2Res}}
\newcommand{\conPayoutLevel}{\txt{payouts}}
\newcommand{\conBalance}{\txt{balance}}

\newcommand{\received}{\txt{Rec}}
\newcommand{\checkpoint}{p}
\newcommand{\conId}{\txt{id}}
\newcommand{\state}{\txt{state}}
\newcommand{\code}{\txt{code}}
\newcommand{\move}{\txt{move}}
\newcommand{\blockHash}{\txt{bh}}
\newcommand{\flag}{\txt{flag}}
\newcommand{\openResponses}{\teeSet_{\txt{wait}}}
\newcommand{\key}{\txt{key}}
\newcommand{\pool}{\teeSet}
\newcommand{\coins}{\txt{amount}}

\newcommand{\timeout}{\delta}
\newcommand{\timeOffchainExecution}{\timeout^1_{\txt{off}}}
\newcommand{\timeOffchainPropagation}{\timeout^2_{\txt{off}}}
\newcommand{\timeOnchainExecution}{\timeout^1_{\txt{on}}}
\newcommand{\timeOnchainPropagation}{\timeout^2_{\txt{on}}}
\newcommand{\timeOnchainCreation}{\timeOnchainExecution}
\newcommand{\timeOffchainCreation}{\timeOffchainExecution}
\newcommand{\timeOnchainCreationPropagation}{\timeOnchainPropagation}

\newcommand{\protCreationChallenge}{\txt{CreationChallenge}}
\newcommand{\protWatchdogChallenge}{\txt{WatchdogChallenge}}
\newcommand{\protWatchdogCreationChallenge}{\txt{WatchdogCreationChallenge}}
\newcommand{\protExecutiveChallenge}{\txt{ExecutiveChallenge}}
\newcommand{\Validate}{\txt{Validate}}
\newcommand{\processChain}{\txt{processChain}}
\newcommand{\initContract}{\txt{initContract}}

\newcommand{\conNextState}{\txt{nextState}}
\newcommand{\conGetState}{\txt{getState}}
\newcommand{\conUpdateState}{\txt{update}}

\newcommand{\pre}{\txt{pre}}
\newcommand{\conf}{\txt{conf}}
\newcommand{\res}{\txt{res}}

\newcommand{\msgDeposit}{\txtMsg{deposit}}
\newcommand{\msgWithdraw}{\txtMsg{withdraw}}
\newcommand{\msgRegister}{\txtMsg{register}}
\newcommand{\msgDeregister}{\txtMsg{deregister}}
\newcommand{\msgCreate}{\txtMsg{create}}
\newcommand{\msgInit}{\txtMsg{init}}
\newcommand{\msgExecute}{\txtMsg{execute}}
\newcommand{\msgBad}{\txtMsg{bad}}
\newcommand{\msgUpdate}{\txtMsg{update}}
\newcommand{\msgConfirm}{\txtMsg{confirm}}
\newcommand{\msgOk}{\txtMsg{ok}}
\newcommand{\msgWait}{\txtMsg{wait}}
\newcommand{\msgFinalize}{\txtMsg{finalize}}
\newcommand{\msgFailed}{\txtMsg{fail}}

\begin{abstract}
The COVID-19 pandemic has caused many countries to deploy novel digital contact tracing (\dct) systems to boost the efficiency of manual tracing of infection chains.      
In this paper, we systematically analyze \dct solutions and categorize them based on their design approaches and architectures.
We analyze them with regard to effectiveness, security, privacy and ethical aspects and compare prominent solutions with regard to these requirements. 
In particular, we discuss shortcomings of the Google and Apple Exposure Notification API (GAEN) that is currently widely adopted
all over the world. We find that the security and privacy of \gap has considerable deficiencies as it can be compromised by severe large-scale attacks.

We also discuss other proposed approaches for contact tracing, including our proposal \ourname, 
that are based on Diffie-Hellman (DH) key exchange and aim at tackling shortcomings of existing solutions. 
Our extensive analysis shows that \ourname fulfills the above security requirements better than deployed state-of-the-art approaches. We have implemented \ourname and its beta test version has been used by more than 2000 users without any major functional problems\footnote{https://tracecorona.net/download-tracecorona/}, demonstrating that there are no technical reasons requiring to make compromises with regard to the requirements of \dct approaches.
\end{abstract}



\section{Introduction}
\label{sec:intro}

The pandemic caused by the SARS-CoV-2 corona virus has still the world in its grip since it was officially announced by the World Health Organization (WTO) on March 11, 2020. At the time of writing, we have been witnessing the surge of several infection waves all around the world.  Reliable and efficient contact tracing for containing the spread of infections has therefore become more important than ever. In many countries, digital contact tracing apps on smartphones have already been rolled out to support manual contact tracing with the hope of significantly improving its effectiveness in breaking infection chains and preventing the virus from spreading further. Particularly 
automatic notifications can be faster and can reach
at-risk individuals in situations related to random encounters between strangers as they occur, e.g., in public transportation, shops, and other indoor activities.
There also exist contrary opinions\footnote{Contact Tracing in the Real World, \url{https://tinyurl.com/2p9765ra}} on the generic usefulness of digital contact tracing. 
In this paper, however, we do not seek to resolve this controversy, but focus instead on analyzing how theoretical results of epidemiologists (e.g.,~\cite{ferretti2020quantifying}) 
are taken into account in current proposals for identifying at-risk contacts in the presence of technological errors, data pollution attacks and privacy and ethics regulations.
Initially we analyze deployed solutions, as many countries are currently actively employing them and millions of users are 
affected by such systems. Regardless of the potential usefulness of digital contact tracing or a lack thereof, contact tracing apps have become a reality in many countries. At the time of writing, 49 countries around the world (including, e.g., most European countries, Australia, China, Singapore) and 27 states in the USA have deployed contact tracing apps \footnote{MIT Covid Tracing Tracker, \url{https://tinyurl.com/3ey44r5c}}. 
Many of these systems in use today were designed, implemented and rolled out in great haste with the goal of containing the spread of the pandemic as quickly as possible. It is therefore ever more important to take a step back and try to obtain a critical view of the benefits and disadvantages of individual approaches.

In this context, \emph{effectiveness}, \emph{security}, \emph{privacy} and \emph{ethics} are key aspects that need to be considered thoroughly: (i)~the system should be \emph{effective}, i.e., able to provide acceptable detection accuracy (high true positive and low false positive rate), (ii)~it should be \emph{secure} so that malicious adversaries cannot manipulate the system to trigger false alarms, (iii)~it should protect \emph{privacy} to increase users' trust in the \dct system, 
and (iv) it should consider ethical aspects as it should be transparent and based on voluntary use.
Ensuring all above properties is necessary to achieve high adoption rates to then significantly contain the spread of the virus. Users need to be confident that the system is reliable in identifying at-risk contacts, and minimizes the use of users' data and provides strong security and privacy measures to protect them from malicious
institutional operators as well as external attackers who may seek to track or to damage users.
Otherwise, users will not be willing to use contact tracing apps, negatively impacting their adoption rate that would be crucial for their effectiveness in practice (ideally higher than 60\%)~\cite{wymant2021NHSapp}. 

While the first countries (predominantly in Asia) that deployed tracing apps adopted centralized approaches, and extensively collected sensitive user information (e.g., names, addresses, mobile phone numbers, location), a widespread and heated debate on user privacy broke out in Europe and the USA\footnote{In the course of this debate about 300 security and privacy researchers from 26 countries signed an open letter criticizing the 
specific privacy risks of some centralized contact tracing approaches, advocating privacy-preserving solutions whenever
better privacy can be obtained without penalizing effectiveness (\url{https://drive.google.com/file/d/1OQg2dxPu-x-RZzETlpV3lFa259Nrpk1J/view}). This signed letter
has been often abused claiming that centralized systems
are bad and decentralized systems do what is needed to detect
at-risk contacts, and moreover they do it protecting privacy.}. 
In this turmoil of evolving contact tracing approaches, somewhat surprisingly, Google and Apple established an unprecedented collaboration and provided their special application programming interface for decentralized contact tracing called \emph{Exposure Notification API (GAEN)} \cite{GAEN:crypto} which they rapidly integrated into their mobile operating systems. Google and Apple give in each country access to this interface only to one organization that is authorized by the local government. 
GAEN runs an almost complete contact tracing solution as a part of the underlying mobile operating systems, so that the role of national organizations is reduced to developing a user interface to GAEN through a smartphone app and providing the backend server infrastructure required for acquiring and distributing \fixedMC{information about at-risk contacts. Further, although Apple and Google initially promised not to get directly involved in contact tracing by developing their own backend server and app,} later they did so by providing the \gap Express solution that is used in several US states, e.g., Maryland and Utah \footnote{MD COVID Alert, \url{https://tinyurl.com/yeymtrm2}}.
Unfortunately, as we will discuss in more detail in Sect.~\ref{sec:analysis}, it is known that existing rolled out Digital Contact Tracing (\dct) systems exhibit a number of important security and privacy risks~\cite{gvili2020security,baumgartner2020mind,Iovino2020timetravel,avitabile2020tenku}.

The controversial debate on tracing apps goes significantly beyond purely technical concepts and includes also questions related to effectiveness, socio-political impact, surveillance capitalism, and technological sovereignty (see, e.g., \cite{ White2021why,lanzing2020ethical}).
In this paper, we contribute to this debate by providing a systematization of the development, deployment, effectiveness, security, and privacy of contact tracing apps aiming to cover these relevant aspects. Further, in order to tackle the shortcomings of existing approaches, we introduce a novel user-controlled privacy-preserving contact tracing system called \ourname. It leverages a robust privacy architecture based on Diffie-Hellman key exchange to provide a level of security and anonymity unparalleled by any of the other systems proposed so far. It also improves the effectiveness and accuracy of the overall system and its resilience to misuse through the ability to \emph{verify} all critical encounters.  

In particular, we provide following contributions:
\begin{itemize}
    \item We introduce a categorization of the requirements on Digital Contact Tracing (\dct) systems in four dimensions, namely:  effectiveness, privacy, and security as well as ethical considerations (Sect. \ref{sec:requirements}).  
    
    \item We review existing contact tracing approaches by analyzing prominent schemes with respect to these aforementioned aspects and construct a generic framework for categorizing them (Sect. \ref{sec:existing-schemes}). We point out that the advantages and disadvantages of a \dct system should not be prejudged by certain design choices. For example, contrary to commonly voiced opinions, one cannot claim that decentralized approaches satisfy "privacy by design". \fixedMC{We argue that more emphasis} should be given to data quality since decentralized solutions are limited in their capabilities to mitigate errors due to technological limitations and in preventing data pollution attacks by third parties.

    \item We extensively analyze to what degree prominent and widely deployed 
    \dct approaches fulfill these requirements (Sect. \ref{sec:analysis}) providing a comprehensive and insightful view on each type of these approaches. 
    In contrast to the vast endorsement received from many governments and scientists, we show that \gap\ performs 
    poorly in satisfying almost all requirements. 
    When taking into account various costs associated with privacy leaks and failures due to data pollution, the justification for an approach like \gap becomes highly questionable.

    \item Based on our analyses and findings, we discuss other schemes and technologies that could fulfill those requirements (Sect. \ref{sec:proposal}). In particular, we elaborate on approaches based on Diffie-Hellman (DH) key exchange that have not yet been deployed in practice. Our analysis shows that DH-based systems provide better security and privacy guarantees than \gap while maintaining comparable effectiveness (Sect. \ref{sec:proposal}). We propose a novel distributed contact tracing system, \ourname, providing strong security and privacy guarantees. In contrast to almost all existing approaches that are based on exchanging pseudonymous proximity identifiers, our approach leverages advanced cryptographic algorithms to establish and verify encounter tokens that are unique to each encounter between two users. Further, we propose various use cases and deployments of \ourname including a hybrid approach (cf. Sect. \ref{sec:tc-hybrid}). We implemented, deployed, and published \ourname for beta user test. More than 2000 users have used the TRACECORONA app without any functional problems.
\end{itemize}

\fixedAlC{In summary, we provide a comprehensive set of requirements to evaluate \dct systems. We show that current approaches do not fulfill such requirements at large, e.g., have number of security, privacy and effectiveness issues. Hence, we propose \ourname, a novel approach that address the deficiencies of existing \dct system.}

\section{Digital Contact Tracing}
\label{sec:background}
In this section, we present system models, architectures and technologies of Digital Contact Tracing (\dct) Systems.
\begin{table}[ht]
    \centering
    \caption{Notations.}
    \label{tab:notation}
    \begin{tabular}{l|l}
    \hline
    
    User ($U$)  & A Person that uses a \dct App \\ \hline
    User App ($App$) & A \dct app installed on users' devices\\ \hline
    \begin{tabular}[c]{@{}l@{}}Tracing Service \\Provider ($\service$) \end{tabular} & \begin{tabular}[c]{@{}l@{}}Providing a system (e.g., servers and apps) \\ for identifying at-risk contacts \end{tabular}\\ \hline
    Health Authority ($\healthauthority$) &  Authenticating the user infection  status\\ \hline
    Infected user & A user that has tested positive for COVID-19 \\ \hline
    Affected user & A user that has encountered an infected user\\ \hline
    Indirect contacts & A user that has encountered an affected user\\

    \hline 
    \end{tabular}
\end{table}

\subsection{System Model}
\begin{figure}[ht]
    \centering
    \includegraphics[width=\columnwidth]{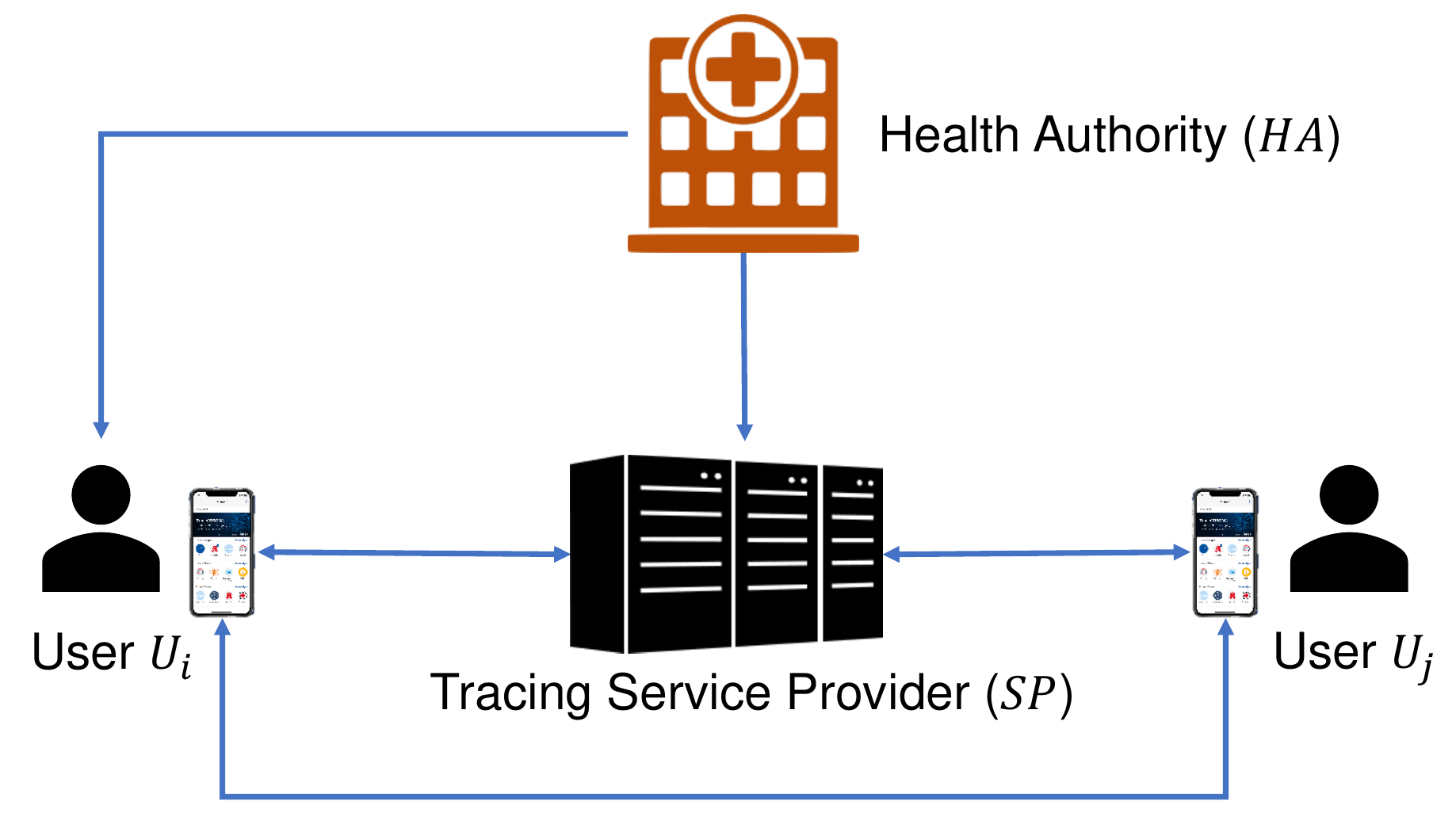}
    \caption{System model of Digital Contract Tracing (DCT).}
    \label{fig:system-model}
\end{figure}

Figure~\ref{fig:system-model} shows the typical system model of contact tracing schemes. There are three types of entities: \fixedAlC{Users $U$ (e.g., $\useri$ and $\userj$)} of the tracing system (app), a contact tracing Service Provider ($\service$), as well as a health authority ($\healthauthority$). In the following, we discuss these roles in more detail.

\subsubsection{Users}
\fixedAlC{A user $\useri$ uses a dedicated \emph{contact tracing app} installed on its device (typically a smartphone) to collect information required to determine contacts with other users of the system}. Different technologies can be used for this purpose, e.g., directly through exchange of specific information over a proximity communication protocol like Bluetooth LE, or, indirectly with the help of a trace of location information obtained from a positioning system like GPS, by determining simultaneous co-presence of the users at the same location at the same time. We will discuss various technologies in Sect. \ref{sec:technology}.
Users' contact tracing apps collect and store this information about contacts of users locally on users' mobile devices. In case a user $\useri$ is tested positive with a disease (like COVID-19), the user is expected to use the contact tracing app to warn other users of the system by uploading the collected information about his/her contacts to the contact tracing service provider $\service$.

\subsubsection{Tracing Service Provider}
 The Tracing Service Provider $\service$ is responsible for collecting and distributing information necessary for identifying contacts with infected users and/or notifying other users of such contacts. In  \emph{centralized} systems, the $\service$ determines contacts between infected users and other users and issues notifications to them, whereas in \emph{decentralized} systems, the determination of possible contacts is performed by the users' contact tracing apps.

\subsubsection{Health Authority}
The Health Authority $\healthauthority$ is responsible for identifying infected users (e.g., through administered medical tests) and authenticating their infection status towards $\service$. \fixedAlC{This is necessary to prevent malicious users \advUser from pretending to be infected and thereby triggering false alarms with users they have had contacts with. To do this, $\healthauthority$ will issue a user-specific unique authenticator, e.g., a transaction authentication number (TAN) (a form of single use one-time password (OTP)) to an infected user $\useri$, who can subsequently present this authenticator when uploading their information to $\service$. By verifying the authenticator with $\healthauthority$, the $\service$ can verify the infection status of the user $\useri$.}

\subsection{Centralized vs. Decentralized Architectures} 
\label{sec:cen-or-decentralized}
\begin{figure}[htb]
	\centering
	\includegraphics[width=0.8\columnwidth]{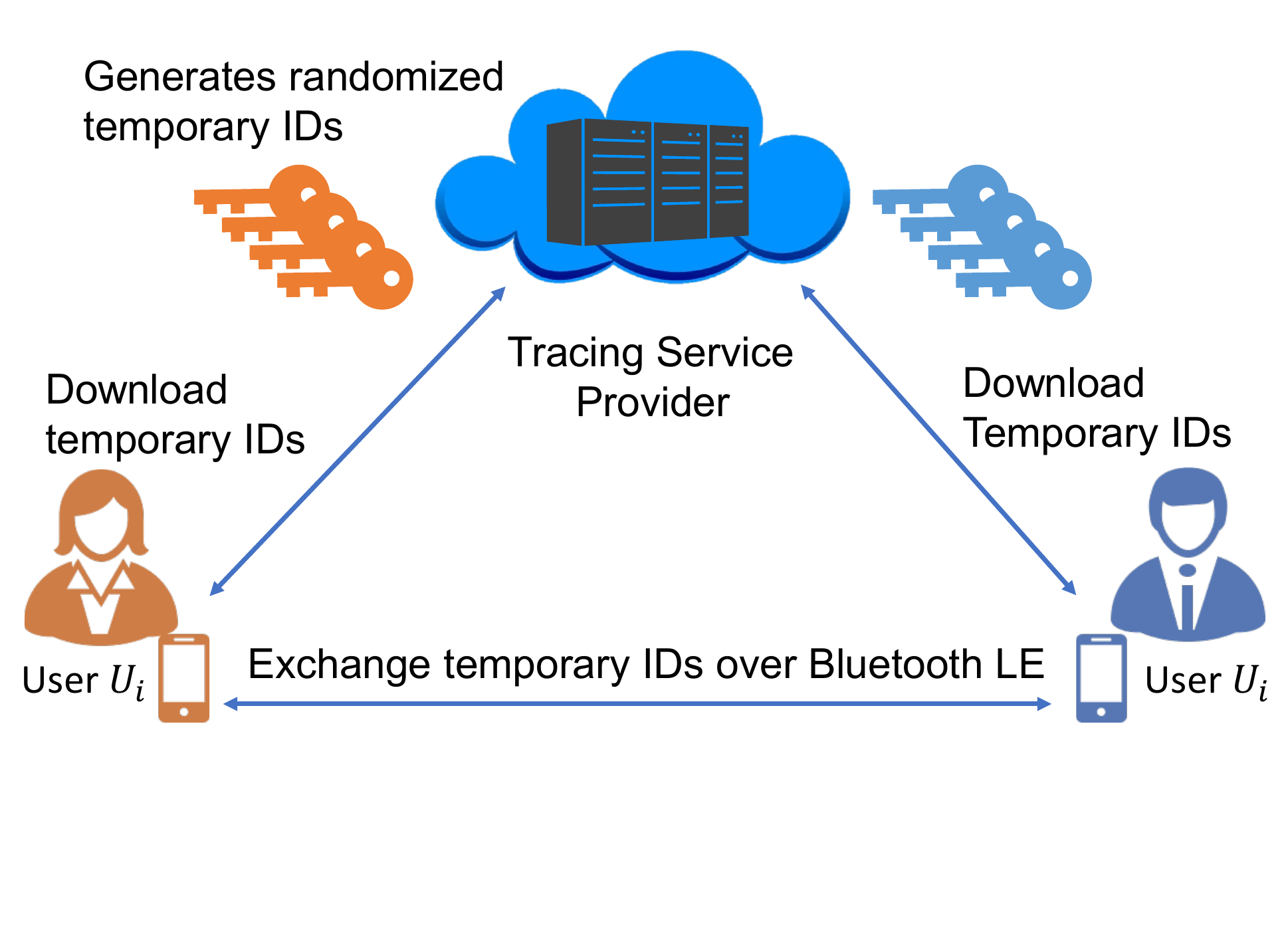}
	\caption{Centralized Architecture}
	\label{fig:centralized}
\end{figure}

\begin{figure}[htb]
	\centering
	\includegraphics[width=0.8\columnwidth]{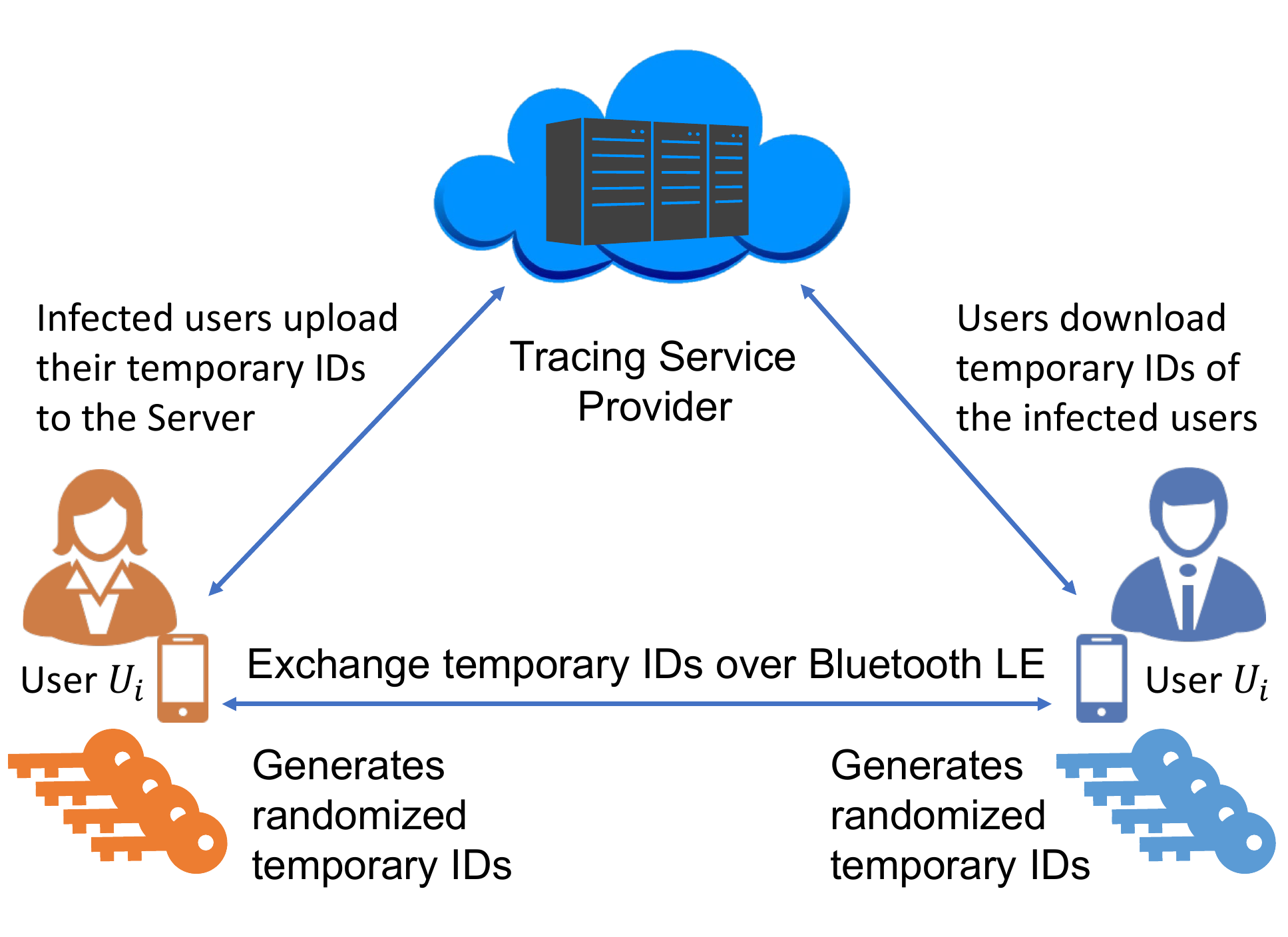}
	\caption{Decentralized Architecture}
	\label{fig:decentralized}
\end{figure}

In general, contact tracing approaches can be divided into two main design architectures, \emph{centralized} and \emph{decentralized}, based on whether the identification of encounters between users is performed by the tracing service provider $\service$ or by the tracing apps of users $U$. Figures~\ref{fig:centralized} and \ref{fig:decentralized} show an overview of both architectures.
Both approaches are based on individual users' tracing apps recording temporary identifiers ($\tempId$s) of other devices they encounter. In the case a user $\useri$ is infected, he uses his tracing app to upload identifiers to $SP$. In centralized systems, the recorded identifiers of \emph{other} apps will be uploaded, whereas in decentralized systems, the $\tempId$s used by the tracing app \emph{itself} in the recent past will be uploaded.

The main difference between these schemes is the fact that in the centralized system the service provider $\service$ generates all $\tempId$s centrally and is therefore able to link the infected user with the (pseudonymous) identities of other users, whereas in the decentralized approach, the $\tempId$s are generated individually by each tracing app. The determination of contacts can therefore only be performed by the actual tracing apps involved in an encounter. The tracing app conducts this by downloading the $\tempId$s of infected users, e.g.,  $\useri$ from $\service$ and comparing these to the $\tempId$s the tracing app has encountered in the past. This approach therefore effectively limits the exposure of sensitive information about encounters to $\service$. 

In contrast to common belief, however, this difference does not directly guarantee "privacy by design" for decentralized systems and susceptibility to "mass surveillance" in centralized systems. The actual evaluation of these models highly depends on the underlying architectural decisions and on the various threat models considered.
We will analyze the effectiveness, privacy and security of these architectural approaches in more detail in Sect.~\ref{sec:analysis}.

Before considering these approaches in more detail, we will elaborate on the security and privacy requirements in Sect.~\ref{sec:requirements} that are important driving factors behind the decisions to adopt particular contact tracing approaches as these are also highly influenced by political and legal decisions taken by governments in individual countries.

\subsection{Technologies to Determine Encounters}
\label{sec:technology}

In general, there are two types of technologies to determine encounters: (1) location-based technologies such as GPS and QR-codes used for venue check-ins and (2) peer-to-peer proximity detection-based technologies like Bluetooth, Ultra-wideband (UWB), and ultrasound. Currently, Bluetooth is the most dominant technology deployed in contact tracing. 

\noindent\textbf{GPS-based Location.}
GPS tracking can be used to determine the location of individuals and hence encounters between persons present at the same location at the same time (e.g., \cite{IsraelHashomer}). However, in practice, using GPS for this purpose is challenging, as it is relatively inaccurate, especially in indoor areas. Indoor encounters are particularly important to be captured accurately due to the higher risk of contagion in closed spaces. More importantly, GPS traces of users can reveal a lot of sensitive information about users and their habits\footnote{"They Stormed the Capitol. Their Apps Tracked Them.", \url{https://tinyurl.com/5ddun4dm}}. In a centralized GPS-based approach, GPS traces of all users are typically sent to the $\service$. This information, however, can be misused for surveillance and profiling. In a decentralized approach, infected users have to upload their GPS traces, i.e., make them publicly available so that other users can check their potential encounters with the infected person by comparing their own location trace with the released locations of the infected person. 
The use of these solutions requires enabling the location service on the mobile device.  This implies, however, that also other apps that have been given the permission to use the user's location, including\fixedMC{, e.g., Google's location service on Android devices,} can collect detailed location traces potentially along with other data such as IP and MAC addresses \cite{leith2020gaendata}.

Privacy protection in such schemes can be addressed to some degree by allowing infected users to redact locations that they deem sensitive from their individual traces before sharing them with others. This, however, can have serious negative effects. Indeed, many potential encounters may be lost if users redact sensitive locations like workplace or specific entertainment venues like bars or clubs from their traces, thus \fixedMC{reducing the} utility of the system for contact tracing.

\noindent\textbf{QR Code-based Location.} Instead of using geolocation information to identify the presence of a user at a location, QR codes placed at specific venues can be used to allow users to "check-in" at the venue and thus record information about the presence at the venue at a specific point in time. This can be either done by recording identifiable contact information of users and venues including names, phone numbers, and addresses (of the guests and restaurants) in a QR Code 
so that it can be exchanged easily by scanning the QR code using a mobile app~\cite{lucaApp2021}, similar to check-in approaches in hotels. In some other solutions, users scan a QR code associated with a venue. The QR code contains a venue-specific URL linking to a cloud-based service $\service$ associated with the venue, which the mobile or web app uses to enter the contact information of the user. \changeT{To the best of our knowledge,} deployed QR-code based apps follow a centralized architecture, where all data (encounter information) are stored by $\service$ \cite{lucaApp2021}. \fixedMC{Some approaches, such as \cite{lucaApp2021}, enable users (including individual users and the venues) to encrypt the data before sending them to $\service$ (the encryption keys are kept locally). In particular, individual users encrypt the contact information of venues (e.g., addresses, phone numbers, and the time of visit) and vice-versa, i.e., the venues also encrypt the contact information of users). If a user $\useri$ is tested positive for COVID-19, the user will send the encryption keys to health authority $\healthauthority$ who can decrypt the encounter information stored by $\service$ and know which venues the user has visited and at what time. Afterwards, $\healthauthority$ will ask those venues to provide keys to decrypt contact information of all users (guests) also stored by $\service$ who visited the venues around the given time. Finally, $\healthauthority$ notifies the corresponding users.} However, similar to GPS-based approaches, QR Code-based approaches also reveal the movement traces and encounter information (who they have met) of infected users \cite{stadler2021preliminary}. Another drawback of QR-code-based systems is that for it to be effective, many venues such as restaurants need to adopt the system. 

\noindent\textbf{Bluetooth Low Energy (BLE).}
 BLE can be used for sensing the proximity between individual users' devices, e.g., \cite{GAEN:crypto, BlueTrace}. Indeed, many recent approaches for contact tracing on smartphones use Bluetooth proximity detection. The participating smartphones beacon out information like temporary identifiers ($\tempId$s) that can be sensed by other devices. In addition, also related metadata like the signal strength of the beacon may be recorded. Using the signal strength information, some approaches seek to provide estimates about the distance of the encounter. However, it has been shown that signal strength can provide only a very rough estimate about the actual distance of devices, as it is influenced by other factors like device orientation and surrounding structures~\cite{ leith2020measurement}. Nevertheless, since BLE is widely available on most recent smartphone versions, it seems the most viable alternative for implementing proximity detection on smartphones that are widely used by the population in many countries.
 
 \fixedMC{Compared to GPS and QR-code based approaches, BLE would seem to reveal the least amount of information about the users because $\healthauthority$ and $\server$ do not collect physical locations as well as actual encounter times. Thus, only anonymized random strings are shared among the apps using BLE. However, BLE-based approaches still have several security, privacy, effectiveness, and ethical problems. For example, they are susceptible to fake exposure injection attacks, e.g., relay attacks, or user profiling, e.g., movement tracking and user identification. We will elaborate all of these problems in detail in Sect.~\ref{sec:analysis}.}

\noindent\textbf{Ultra-Wideband (UWB).}
In contrast to Bluetooth and other technologies that cannot measure distances among users precisely, UWB radio technology can be used to measure the distance at an accuracy level of a few centimeters. Therefore, some approaches propose using UWB \cite{istomin2021janus}. 
However, using UWB faces some challenges such as the fact that very few smartphones support UWB and that it is not energy efficient \cite{istomin2021janus}. This makes UWB-based approaches less practical. Istomin et al. \cite{istomin2021janus} propose a hybrid approach that combines Bluetooth and UWB, so that Bluetooth is used to constantly broadcast BLE signals and to identify other devices in vicinity. Once an encounter is identified (i.e., a device detects another device via BLE), UWB is used to measure the exact distance between two devices in order to estimate the risk of exposure.     
\noindent\textbf{Ultrasound.} Since ultrasonic communication range is very short (only effective for distances shorter than one meter) and likely blocked by walls or other objects, it can also be used for determining encounters \cite{MIT-SonicPACT}. However, Meklenburg et al.~\cite{MIT-SonicPACT} also raise many open questions about the practicality of ultrasound-based approaches such as scalability \fixedMC{(as it appears nearly impossible to handle multiple connections because it is hard to transfer and separate several ultrasonic signals simultaneously), low accuracy (due to ambient noise), or high power consumption. Further, this approach requires microphones and speakers of smartphones to be enabled. This would make devices vulnerable to several known privacy and security attacks such as tracking user's activities, leaking user's conversations, or injecting fake voice commands \cite{Trivedi2020digital}.} 

\section{Quadrilemma Requirements}
\label{sec:requirements}
As mentioned above, digital contact tracing (\dct) schemes need to collect information about infected individuals. Although many countries have deployed contact tracing apps, the effectiveness of \dct is so far still unclear. Moreover, \dct poses a number of privacy and security challenges on the underlying scheme design, since it collects and processes sensitive information which is related to users' health and users' contacts to some extent. In this section, we systematically consider the requirements for \dct based on four pillars: effectiveness, privacy, security, and ethical aspects. These requirements are broken down and listed in Tab. \ref{tab:requirments}. Next, we will discuss each of them in detail.
 
\subsection{Effectiveness}
\label{sec:req-effect}

\subsubsection{Accuracy (\rAccuracyIdx) - Distance and Duration}
\label{sec:req-accuracy}
\fixedMC{For accurately estimating the risk of contagion it is necessary to estimate the duration of each contact (in minutes)
along with a good estimate of the distance between the users involved in the encounter. The duration of contacts ideally could be detected
by continuously scanning for the presence of BLE devices in proximity to verify the continued presence of other devices. 
This aggressive approach will, however, lead to significant energy consumption draining the smartphone battery quickly. 
In practice, one needs therefore to pause the scanning for several seconds before the next scan to preserve energy. 
Computing a good estimate of the distance between devices is even more challenging since there are multiple factors (e.g., positioning of the antenna in the smartphone, obstacles
in between smartphones, and their orientations)} that introduce significant errors to distance estimates.
Indeed, experiments performed by Leith and Farrel~\cite{leith2020measurement} showed that GAEN is
quite imprecise in estimating the distance of devices of potential at-risk exposures.

\subsubsection{Superspreaders (\rSuperspreaderIdx)}
\label{sec:req-superspreader}
The mere capability of detecting at-risk exposures was initially considered sufficient by many endorsers of decentralized systems like, e.g., the team around the influential DP-3T~\cite{DP3T:WhitePaper} contact tracing approach, which also had a considerable influence on the GAEN design adopted by Google and Apple. However, along the way, more epidemiological
insights about the behavior of SARS-CoV-2 have been discovered. Among them is the fact that a very relevant aspect for understanding the spread of the virus is the important role of so-called \emph{superspreaders}. Indeed, Reichert et al.~\cite{reichert2020lighthouses} showed that while there is a large percentage of infected individuals that do not transmit the virus at all, there is a small fraction of infected individuals that instead are very contagious and cause numerous further infections. 
A \dct system aiming at effectively
defeating SARS-CoV-2 should therefore also take into account the importance of superspreaders and provide mechanisms allowing to detect them \fixedMC{and their potential contacts.} 

\noindent\textbf{Contagious asymptomatic infected individuals (\rCAII{s}).}
Particularly problematic are so-called asymptomatic infected individuals, i.e., persons that are infected and contagious, but asymptomatic and thus may unwillingly spread the disease. Such individuals have \fixedMC{a very low chance of being} tested positive since they do not show any symptoms of being sick and therefore \fixedMC{will not likely seek to be tested. Even if they want to be tested,} in many countries, they will not be prioritized in testing. \fixedMC{Hence,} they can have an active role in spreading the virus.
However, as such individuals are unlikely to be tested and receive a positive diagnosis from $\healthauthority$ (which is a prerequisite for uploading information about contacts to the service provider $\service$), it is unlikely that such persons will ever be able to use the \dct system to warn other users about possible at-risk contacts with them.

\subsubsection{Accountability (\rAccountabilityIdx)}
\label{sec:req-accountability}
Implementing, deploying, and operating a \dct system can be very costly and requires a majority of the population to participate in its operation. Therefore, the system should provide adequate and valid information about its effectiveness in a privacy-preserving way. For example, the system should be able to provide basic statistics about the number of active users, infected users, users notified about potential at-risk exposures, as well as false positive rates, etc. At a minimum, the system should be able to demonstrate clear benefits in comparison to a purely random selection of users to be quarantined in specific at-risk groups (e.g., where
the infection rate is higher)~\cite{leith2020measurement}. Although some \gap-based apps do provide reports on some measures related to the system's effectiveness, such measures can be biased, unreliable or misleading \cite{vaudenay2020analysis, White2021why} 
as we will discuss in detail in Sect.~\ref{sec:gaen-effectiveneness}.

 \subsection{Privacy} 
 \label{sec:sec-req-privacy}
 
 The main privacy concerns relate to the abuse of a \dct in order to \emph{identify} users, \emph{track} users, or \emph{extract the social graph} of users. 
 \fixedMC{Information that is emitted to the user's surroundings by contact tracing apps and shared with other involved parties should not introduce such privacy risks as elaborated next.}
 
 \subsubsection{Identifying users (\rIdentifyingIdx)}
 \label{sec:req-identify}
 
 \dct systems aim at identifying encounters, not users. Therefore, the systems should not leak any information that can be used to establish the true identity of any individual user.
 
 \subsubsection{Tracking users (\rTrackingIdx)}
 \label{sec:req-tracking}
 \dct apps work by continuously beaconing pseudonymous identifiers into their surroundings. These identifiers should not be linkable, i.e., it should not be possible to trace the movements of any user over time, as this may potentially enable to deduce facts about the user's behaviour and lead to an identification of the user.
 
 \subsubsection{Extracting the social graph (\rSocialGraphIdx)}
 \label{sec:req-graph}
In general, contacts (especially long encounters), are often related to social relationships (i.e., users that decide to be close to each other). 
When handling contact information, a \dct system should make sure that one cannot abuse information collected by it to generate a relevant part of the
social graph of any user, since this may enable to draw conclusions about social relationships between users and thus potentially identify them.
 
\noindent\textbf{Note:} Obviously, there exists in some cases inherent information leakage  due  to  specific  circumstances, e.g., in situations in which the  adversary is in the proximity only to one specific person. If the adversary later receives an at-risk notification, it will be trivial for the adversary to conclude that this  one person is indeed the infected person.  Therefore, when  considering  the above  three privacy  requirements,  we  will  always  focus  on  \emph{large-scale  attacks} and \changeT{will  in  particular focus on identifying attacks affecting potentially many users}.
 
\subsection{Security}
\label{sec:sec-req-sec}

\fixedMC{The effectiveness of a DCT system is severely impacted} if a system is not resilient to large-scale data pollution attacks. Such attacks can generate, for instance, false at-risk notifications (false positives) therefore jeopardizing the correctness of the contact tracing system.
\fixedMC{Indeed, massive false at-risk notifications could result in spreading panic among the general population.
Moreover, this could also cause unnecessary strain on the health system through unnecessary testing and negative impact on the society due to unnecessary self-quarantining.}

\subsubsection{Fake exposure claims (\rFakeClaimIdx)} 
\label{sec:req-fake}
The system should prevent a \fixedMC{malicious or dishonest user \advUser that aims to circumvent the \dct system to claim that he or she has encountered an infected user}. 
There can be different motivations for this attack: (i) \fixedMC{\advUser aims to harm the reliability of the system by manipulating encounter checking results, (ii) \advUser uses the fake exposure status as an excuse to stay at home instead of going to work or participating in an event,} and (iii) \advUser intentionally shares wrong encounter information to epidemiologists, thus sabotaging their analysis of the epidemiological situation.

\subsubsection{Fake exposure injection - Relay/replay attacks (\rFakeInjectionIdx)}
\label{sec:req-relay}
This attack aims to inject fake contacts on a large scale resulting in many false exposure notifications. \fixedMC{Here, a fake contact indicates the state that the \dct system incorrectly determines that two users were in ``close contact'' at a specific time although they were not.
It affects the main goal of \dct system as to identify contacts that potentially cause high exposure risks. Relay attacks are a typical example of fake exposure injection attacks. In a relay attack, the adversary captures the temporary IDs of a user $\useri$ and broadcasts them in other locations (e.g., other cities). As a result, the system incorrectly identifies the users in the other locations who captured those temporary IDs to have encountered $\useri$.}

\subsection{Ethics}
\subsubsection{Transparency and voluntary participation (\rTransparancyIdx)}
\label{sec:req-transparency}
The whole process (design, development, deployment, and operation) of a contact tracing system must be transparent to users and the systems must be removed immediately when the pandemic is over \fixedMC{to avoid misuse.} 
Further, users should be free to decide whether they want to participate in the system or not, and be free to withdraw their participation anytime they wish. \fixedMC{Otherwise, users will not trust, and thus will not be willing to use \dct apps. This will affect the crucial need of a high adoption rate of \dct.} 

\subsubsection{Independence (\rIndependenceIdx)}
\label{sec:req-sovereignty}
The contact tracing process (design, development, deployment and operation) in a particular region \fixedMC{should be independent of any parties with potential vested interests}. Procedural controls of the contact tracing system should underlie a transparent public scrutiny and be solely under the control of democratically-elected governments. In particular, giant technology corporations (e.g., Mobile OS vendors) should not be allowed to use their technological or market dominance to control or drive \dct systems since they might be biased in it for the sake of their own subjective benefits, \fixedMC{e.g., using \dct data for business purposes could undermine the de-facto ability of legitimate governments to oversee the use of data collected for contact tracing purposes.}

\begin{table}
\caption{List of requirements for digital contact tracing.}
\label{tab:requirments}
\begin{tabular}{|l|l|l|}
\hline
      & Requirement                                                           & Description                                                                                                                         \\ \hline
      & \multicolumn{2}{l|}{\textbf{Effectiveness}}                                                                                                                                                                         \\ \hline
\rAccuracyIdx  & \rAccuracy                                                             & \begin{tabular}[c]{@{}l@{}}Specifying distance and\\ duration of encounters\end{tabular}                                            \\ \hline
\rSuperspreaderIdx  &  \rSuperspreader & \begin{tabular}[c]{@{}l@{}}Identifying superspreaders\\ and their contacts\end{tabular}                                             \\ \hline

\rAccountabilityIdx  & \rAccountability                                                       & \begin{tabular}[c]{@{}l@{}}Providing statistics to evaluate\\ the actual effectiveness\end{tabular}                                 \\ \hline

 & \multicolumn{2}{l|}{\textbf{Privacy}}                                                                                                                                                                               \\ \hline
\rIdentifyingIdx  & \rIdentifying                                                    & \begin{tabular}[c]{@{}l@{}}Users should always\\ remain anonymous\end{tabular}                                                          \\ \hline
\rTrackingIdx  & \rTracking                                                       & \begin{tabular}[c]{@{}l@{}}Users should not\\ be tracked\end{tabular}                                                               \\ \hline
\rSocialGraphIdx  & \begin{tabular}[c]{@{}l@{}}Extracting\\social graph\end{tabular}                                                & \begin{tabular}[c]{@{}l@{}}Making sure that no social\\ graph can be extracted\end{tabular}                                         \\ \hline
      & \multicolumn{2}{l|}{\textbf{Security}}                                                                                                                                                                              \\ \hline
\rFakeClaimIdx  & \begin{tabular}[c]{@{}l@{}}Fake\\ exposure claim\end{tabular}                                                     & \begin{tabular}[c]{@{}l@{}}Preventing malicious users to\\ lie about their exposure status\end{tabular}                             \\ \hline
\rFakeInjectionIdx  & \begin{tabular}[c]{@{}l@{}}Fake\\ exposure injection \end{tabular}                                             & Preventing relay/replay attacks                                                                                                     \\ \hline
      & \multicolumn{2}{l|}{\textbf{Ethics}}                                                                                                                                                                                 \\ \hline
\rTransparancyIdx & \begin{tabular}[c]{@{}l@{}}Transparency and\\ voluntary use\end{tabular} & \begin{tabular}[c]{@{}l@{}}The system must be transparent \\ and based on voluntary use\end{tabular}                                \\ \hline
\rIndependenceIdx & \rIndependence                                                        & \begin{tabular}[c]{@{}l@{}}\fixedMC{Ones should not be allowed to use} \\ \fixedMC{their technological or market dominance} \\ \fixedMC{to control \dct systems in their favour}\end{tabular} \\
\hline
      
\end{tabular}
\end{table}

\section{Systematization of Contact Tracing Schemes}
\label{sec:existing-schemes}
In this section, we systematize state-of-the-art contact tracing schemes and their adoptions.
\subsection{Architectures}
\subsubsection{Centralized Architecture}
\label{sec:central}
Centralized contact tracing approaches have been already deployed in many countries, e.g., Singapore (TraceToGether\footnote{\url{https://www.tracetogether.gov.sg/}}) and France (TousAntiCovid\footnote{\url{https://www.gouvernement.fr/info-coronavirus/tousanticovid}}). \fixedMC{Some of these schemes can be considered privacy-invasive as they collect sensitive personal data like GPS/locations, and phone numbers, e.g., AarogyaSetu\footnote{ \url{https://www.mygov.in/aarogya-setu-app/}}, the Indian App}. 
In this section, we primarily focus on schemes that use BLE technology (as apposed to GPS) to determine users' encounters, because it is the the most prevalent and allows for building less privacy-invasive tracing systems.

\begin{figure}[ht]
	\centering
	\includegraphics[width=\columnwidth]{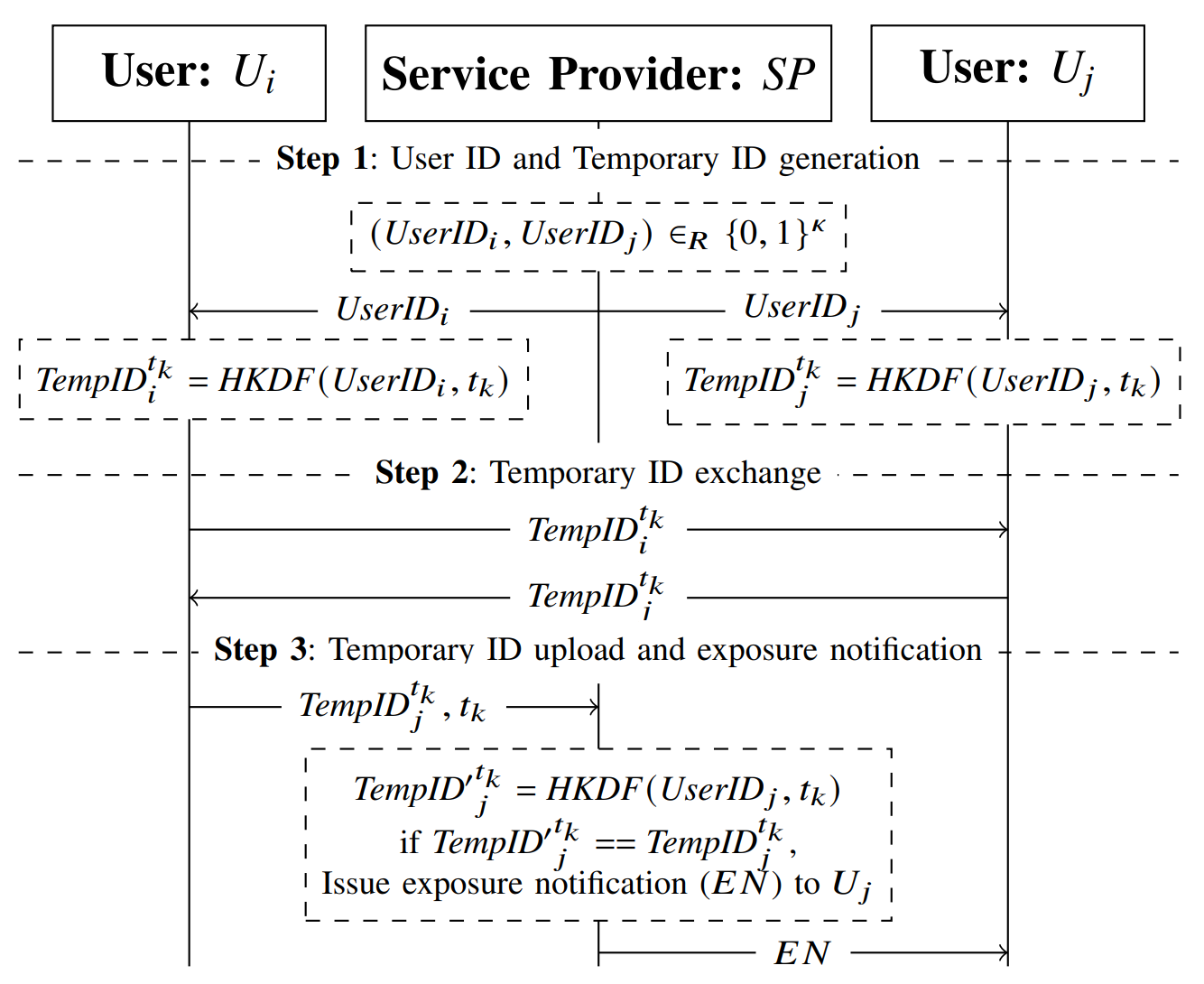}
	\caption{Generic framework of centralized approaches.}
	
	\label{fig:generic-centralized}
\end{figure}

\noindent\textbf{Generic Framework.} 
We will systematize prominent contact tracing solutions based on centralized architecture, e.g., BlueTrace~\cite{BlueTrace}, PEPP-PT~\cite{pepppt} and TousAntiCovid that are conceptually very similar. We first present a generic framework for 
centralized schemes and then discuss how it is implemented in practice. Figure~\ref{fig:generic-centralized} shows a typical generic centralized protocol. The system consists of a Tracing Service Provider $\service$ and users, e.g., User $\useri$  and User $\userj$. In \textbf{Step 1}, users register their contact tracing app to the system either in an anonymous~\cite{pepppt} or identifiable way (e.g., using their mobile phone number)~\cite{BlueTrace}. The Tracing Service Provider $\service$ generates and sends a unique pseudonymous $\userId$ associated to each user. Each user generates batches of temporary identifiers $\tempId$s associated with each $\userId$ by encrypting the $\userId$ (using an HMAC-based Key Derivation Function - HKDF) along with validity timestamps $t_k$ as follows: $\tempId_i^{t_k} = \HKDF(\userId_i, t_k)$ and $\tempId_j^{t_k} = \HKDF(\userId_j, t_k)$. In \textbf{Step 2}, the user's tracing app beacons $\tempId$s into its proximity and simultaneously listens for other user apps' $\tempId$s and stores them locally. \textbf{Step 3} shows the matching and notification steps in case $\useri$ is tested positive for SARS-CoV-2: $\useri$ uploads the $\tempId$s of the users that $\useri$ has encountered along with corresponding timestamps (e.g., during the last 14 days) to the $\service$. $\service$ then identifies the users who have encountered $\useri$ by comparing the received temporary identifiers to ones derived on the server side. If there is a match, $\service$ notifies the identified users.   

\begin{figure}[ht]
	\centering
	\includegraphics[width=\columnwidth]{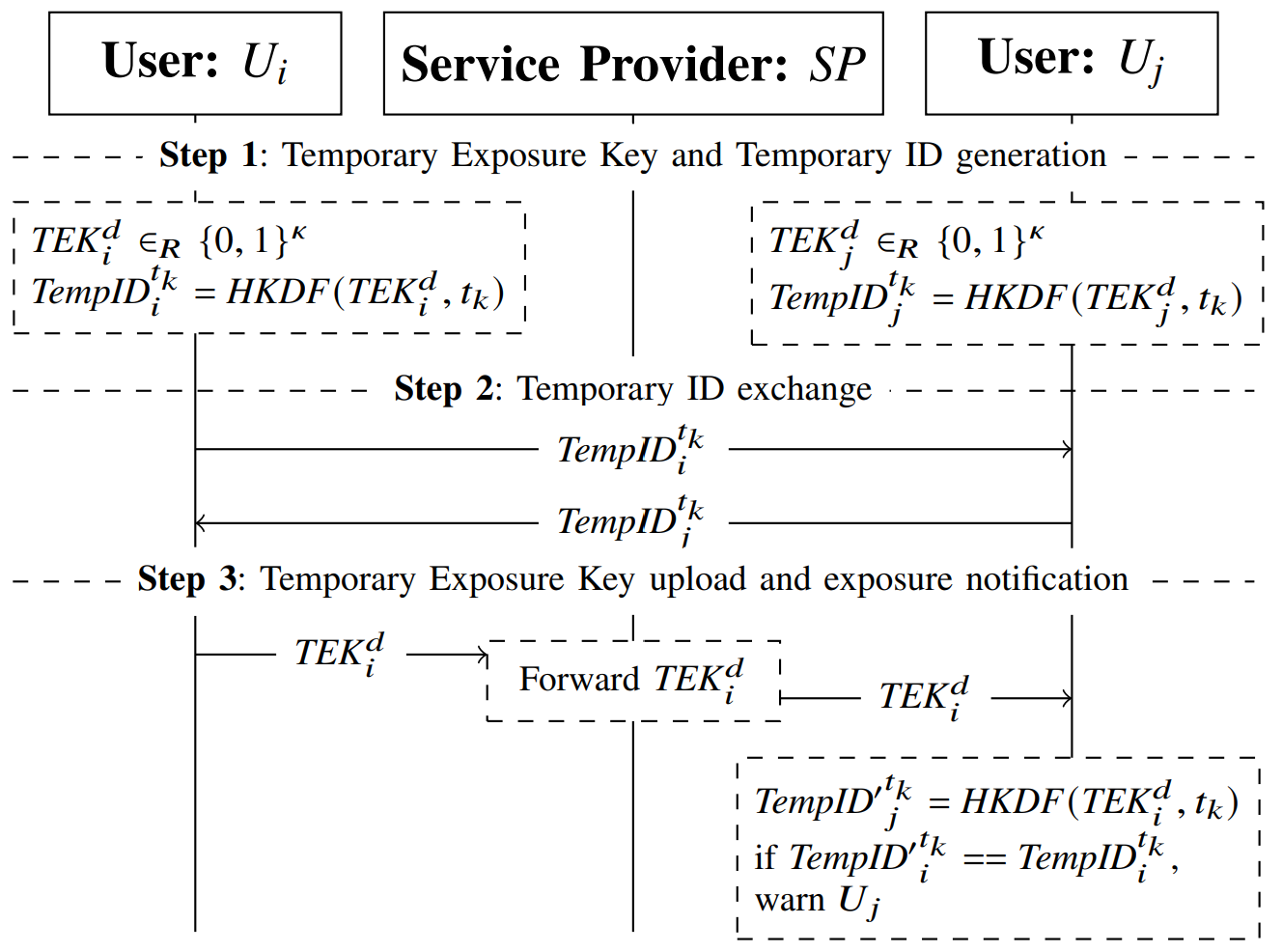}
	\caption{Generic framework of decentralized approaches.}
	\label{fig:generic-decentralized}
\end{figure}

\paragraph{PEPP-PT and TousAntiCovid} \fixedAlC{TousAntiCovid is the official app in France and it is based on PEPP-PT}. These approaches follow the generic framework shown in Fig.~\ref{fig:generic-centralized}. In contrast to BlueTrace discussed below, users do not need to provide personal information like mobile phone numbers to register with $\service$. Therefore, $\service$ only notifies users (as identified by their $\userId$) via the contact tracing app.

\paragraph{BlueTrace \cite{BlueTrace}} This is a contact tracing framework adopted by, e.g., Singapore (TraceTogether) and Australia (CovidSafe). BlueTrace follows the generic framework with some modifications. Firstly, it requires users to register using their mobile phone numbers so that $\service$ can then contact the corresponding users directly by telephone and notify them about the contact with an infected person. 
Secondly, users' tracing apps do not generate $\tempId$s by themselves but receive pre-generated $\tempId$s from $\service$. To derive the $\tempId$s, BlueTrace uses the $\HKDF$ function with four inputs: a user ID $\userId$, a timestamp $t_k$, a random initialization vector $ \mathit{IV} $, and an authenticity tag $\mathit{AuthTag}$ for integrity verification using a secret master key $K$ known only to $\service$, as shown in (\ref{eq:tempid-gen}):
\begin{equation}\label{eq:tempid-gen}
\mathit{\tempId} = \HKDF(\mathit{\userId}\Vert \mathit{t_k}  \Vert \mathit{IV} \Vert \mathit{AuthTag})
\end{equation}

In addition to BlueTrace, PEPP-PT, and TousAntiCovid, a number of other centralized tracing approaches have been proposed.
\ifarXiv
We discuss these approaches in detail in Appendix \ref{sec:list-centralized-app}.
\fi

\noindent\textbf{Hybrid approaches or centralized approaches}. Castelluccia et al. \cite{castelluccia2020desire} propose a \emph{hybrid solution} called Desire that combines centralized and decentralized techniques. The main differences to the generic framework are in \textbf{Step 1}, where instead of temporary identifiers, Diffie-Hellman (DH) keys are generated by the apps, and in \textbf{Step 3}, where infected users upload encounter tokens derived from the DH keys in \emph{an anonymous way}, e.g., using an anonymization network like Tor or another Mix network. However, according to our classification introduced in Sect.~\ref{sec:cen-or-decentralized}, Desire should still be considered a centralized approach, since the Tracing Service Provider $\service$ still (1) receives all encounter tokens, i.e., $\tempId$s from all users, (2) performs encounter token matching, and (3) is responsible for sending exposure notifications to corresponding users based on the users' App IDs. We will discuss the effectiveness of Desire further in Sect.~\ref{sec:analysis}.

\subsubsection{Decentralized Architecture}
\label{sec:decentral}

A number of decentralized contact tracing approaches have been proposed, e.g., DP3T~\cite{DP3T:WhitePaper}, MIT-PACT~\cite{PACT-MIT}, and the \emph{Exposure Notification API}~\cite{GAEN:crypto} of Apple and Google (\gap), which is closely related to the so-called Design 1 of the DP3T project (DP3T-1). Conceptually, these approaches are highly similar and follow the basic structure shown in Fig.~\ref{fig:generic-decentralized}.

Similar to centralized approaches, the decentralized approaches are based on temporary identifiers $\tempId$ (called \emph{Rolling Proximity Identifiers} in \gap, \emph{Ephemeral Identifiers} in DP3T or \emph{Chirps} in MIT-PACT) that tracing apps beacon into their surroundings over BLE. In \textbf{Step 1}, the $\tempId$s are generated locally by using an $\HKDF$ function that encrypts a random \emph{Temporary Exposure Key ($\tek)$} along with a timestamp $t_k$, e.g., $\tempId_i^{t_k} = \HKDF(\tek_i^d, t_k)$ \fixedAlC{where d is the epoch, e.g., the day that the key is valid in \gap.} 
Existing approaches usually define individual $\tempId$s to be valid for 10 to 15 minutes.  In \textbf{Step 2}, similar to centralized approaches, user Apps broadcast their $\tempId$s and record $\tempId$s from other apps in their vicinity. The fundamental difference to the centralized architecture is in \textbf{Step 3}, as the checking for possible encounters and notifying the user in case an at-risk contact is identified are performed locally by user Apps. 
As shown in Fig.~\ref{fig:generic-decentralized}, an infected user $\useri$ uploads her $\tek$s, e.g., $\tek_i^d$ to $\service$ which forwards them to all other users. A non-infected user $\userj$ downloads the $\tek$s and extracts $\tempId'$s, e.g., $\tempId{'}_j^{t_k} = \HKDF(\tek_i^d, t_k)$. $\userj$ then checks whether there is a match with the $\tempId$s that he has collected from other users who he has encountered. If there is a match (i.e., $\tempId{'}_i^{t_k}== \tempId_i^{t_k}$), it means an at-risk exposure is identified and the user is notified.

\paragraph{\gap and DP3T design 1 (DP3T-1)} These schemes follow the generic framework. In these schemes, the $\tek$ is also called Daily Tracing Key (DTK) referring to the fact that $\tek$ will be changed daily. $\tempId$s are changed every 10 minutes so that 24x6=144 $\tempId$s will be generated per day.

\paragraph{DP3T design 2 (DP3T-2) \cite{DP3T:WhitePaper}, and MIT-PACT \cite{PACT-MIT}} In these designs, the $\tek$ changes at the same frequency as $\tempId$s, so that the $\tempId$ of an infected user $\userj$ are unlikable, even if the associated $\tek$s are published by the $\service$.

In addition to DP3T~\cite{DP3T:WhitePaper}, MIT-PACT~\cite{PACT-MIT},
and \gap ~\cite{GAEN:crypto}, there are several other decentralized tracing approaches, e.g., Pronto-C2~\cite{avitabile2020ProntoC2}, ClerverParrot~\cite{canetti2020cleverParrot}, and Epione~\cite{trieu2020epione}. 
\ifarXiv
We discuss these approaches in detail in Appendix \ref{sec:list-decentralized-app}.
\fi

\noindent\textbf{Advanced Privacy Techniques}.
There are several (mainly cryptography-based) privacy techniques that have been considered to enhance privacy of \dct, e.g.,  Blind signatures \cite{reichert2021ovid, avitabile2020ProntoC2}, Blockchain \cite{avitabile2020ProntoC2}, Private Set Intersection (PSI)/filter \cite{trieu2020epione}, and secret sharing \cite{DP3T:WhitePaper}. Unfortunately, none of these contact tracing approaches have been used in practice yet.
\ifarXiv
We discuss these approaches in detail in Appendix \ref{sec:privacy-tech}.
\fi

\section{Contact Tracing Schemes and Quadrilemma}
\label{sec:analysis}

In this section, we systematically analyze the most prominent centralized (BlueTrace and its derivatives) (Sect.~\ref{sec:central}) and decentralized DCT approaches (DP3T, \gap and PACT) (Sect.~\ref{sec:decentral}) with regard to the aspects laid out in the quadrilemma introduced in Sect.~\ref{sec:requirements}. We summarize our analysis in Tab.~\ref{tab:comparison-esp}.
 
\noindent\textbf{Note on technical limitations of BLE \rAccuracy.} As discussed in Sect.~\ref{sec:req-accuracy}, measuring the distance between smartphones using BLE is not very reliable due to its the inherent technical limitations. Hence, we note that all approaches based on BLE-proximity sensing share the same challenge of not being able to reliably estimate the distance between devices involved in a contact. Therefore, none of such approaches will be able to entirely fulfill the \rAccuracy requirement \rAccuracyIdx. We will therefore exclude this aspect from our subsequent analysis.
 
 \subsection{BlueTrace and Others}
\subsubsection{Effectiveness}
In this subsection, we evaluate the effectiveness of BlueTrace and its variants based on the requirements introduced in Sect. \ref{sec:req-effect}.

\noindent\textbf{\rSuperspreader.} In BlueTrace, the $\service$ receives full information about all encounters of all infected users, and can thus use this information to identify potential superspreaders of \rCAII{s}.
If $\service$ detects that several infected users have encountered the same user, then the the likelihood of that user being a superspreader is relatively high. Hence, $\service$ can recommend such users to get tested and, perhaps more importantly, can immediately notify other users who have encountered the CAII user. Hence, the requirement \rSuperspreaderIdx is fulfilled. 

\noindent\textbf{\rAccountability.} Centralized approaches and hybrid approaches like Desire \cite{castelluccia2020desire} also provide useful information for epidemiologists and policy makers as follows:

\begin{itemize}
    \item Encounter information, w.r.t. lockdown/social distancing measures: The number of encounters and metadata (e.g., duration of each encounter) of individual users can provide information about the effectiveness of different lockdown or social distancing measures, i.e., how the number of encounters increases/decreases, or, how encounter durations change.

    \item Exposure information: How frequently and for how long users encounter infected users.

    \item Hotspot prediction: The $\service$ can predict hotspots based on encounter and exposure information if it keeps track of where a user is (only rough information, e.g., the postcode or the city). The $\service$ can do this by checking the IP addresses of client apps from ISPs; the server does not need to store the IP addresses. It is practical since, e.g., most existing web servers/content delivery networks (CDNs) have the IP addresses of their visitors. Alternatively, the system can ask users to provide their postcodes/cities when they install the App, which is unlikely to be a major privacy concern for most users.

\end{itemize}
\fixedAlC{Therefore, centralized approaches fulfill Requirement \rAccountabilityIdx.}

\subsubsection{Privacy} In the following, we evaluate how BlueTrace fulfills privacy requirements introduced in Sect. \ref{sec:sec-req-privacy}.
 
 \noindent\textbf{\rIdentifying and \rSocialGraph.} Existing centralized contact tracing approaches (Sect.~\ref{sec:central}) are vulnerable to this attack because $\service$ has knowledge of temporary identifiers $\mathit{TempID}$s of users involved in contacts. Hence, \rIdentifyingIdx is not achieved. $\service$ can use this information to extract the encounter graph of infected users (i.e., which users have been in contact with infected users) since the matching is performed by $\service$ \cite{vaudenay2020dilemma, White2021why}. This could allow $\service$ to build the social graph of the infected users~\cite{dp3tPeppptSecurity}. As such, \rSocialGraphIdx is not achieved either. BlueTrace is vulnerable to linkability attacks \cite{knight2021linkability} in which the $\tempId$s of a device can be linked based on their corresponding BLE MAC addresses (cf. Sect. \ref{sec:gaen-privacy}). 
 
\noindent\textbf{\rTracking.} In a centralized contact tracing system such as BlueTrace, a malicious tracing service provider \advServer knows all $\mathit{TempID}$s of each user while an eavesdropping adversary \advEavesdropper knows where and when a temporary identifier $\mathit{TempID}$ was observed. Therefore, the collusion of these two adversaries can lead to significant violations of user privacy as the movements of all users and their contacts can be tracked \cite{vaudenay2020dilemma, avitabile2020ProntoC2}. It also means the adversary can derive the social graph (e.g., knows where and when users have met) of all users so that the \rTrackingIdx requirement is not fulfilled. 
 
 \subsubsection{Security} In the following, we will summarize how existing attacks affect the security of BlueTrace its variants.
  
 
 \noindent\textbf{\rFakeInjection.}
 BlueTrace and others are vulnerable to \textit{large-scale} relay attacks because the adversary \advWormhole can easily capture and relay $\mathit{TempID}$s that the apps frequently beacon via Bluetooth. It does therefore not satisfy requirement \rFakeInjectionIdx ~\cite{vaudenay2020dilemma, avitabile2020ProntoB2}.
 \subsubsection{Ethics}
 BlueTrace is open-source and provides a solution for stand-alone apps (i.e, it does not depend on any built-in contact tracing APIs running in mobile operating systems such as Android and iOS). This makes the app transparent to users, and the health authorities have full control of the app's functionalities and deployment, satisfying the \rTransparancyIdx and \rIndependenceIdx requirements.

\subsection{\gap}
In this section, we analyze decentralized approaches based on client-derived $\tempId${s} presented in Sect. \ref{sec:decentral}. 

 \subsubsection{Effectiveness}
 \label{sec:gaen-effectiveneness}
 Although \gap-based apps have been deployed in many countries since 2020, their contributions with regard to confining the spread of the pandemic have been heavily disputed. 
Indeed, several existing studies raise serious questions concerning the effectiveness of \gap-based apps ~\cite{White2021why, white2021privacy, vaudenay2020dark} as well as the trustworthiness of the self-reported effectiveness of some \gap-based apps. For example, a report of Wymant et al.~\cite{wymant2021NHSapp} states that 6,1\% of users who got notified via the NHS Covid-19 App were eventually tested positive. However, since the study took place in the UK during strict lock-down measures, most contacts at that time likely took take place within households or essential working places, so that a large fraction of the affected persons were likely tested due to the regular manual contact tracing process, regardless of possible app notifications. 
 There is therefore no clear evidence whether \gap-based apps have provided any significant additional benefits with regard to the regular manual contact tracing processes.

 \noindent\textbf{\rSuperspreader{s}.} In contrast to BlueTrace, in \gap-based systems, $\service$ does not have encounter information. Therefore, it is hard to see how \gap could determine \rSuperspreader{s} and \rCAII users. Potentially one could use the user's App to roughly determine whether the user is a \rSuperspreader or \rCAII user based on the numbers of $\tek$s with infected users. \fixedAlC{However, this solution can be abused since the number of matching TEKs is not reliable, i.e., self-injection of fake at-risk contacts are trivial in \gap, e.g., by applying the fake exposure claim attack (cf. Sect. \ref{sec:gaen-security}), or, a relay attack such as the time-travel machine attack \cite{Iovino2020timetravel}. Hence, \gap does not achieve \rAccountabilityIdx.}

 \subsubsection{Privacy}
 \label{sec:gaen-privacy}
\noindent\textbf{\aProfilingAttack.}
Several works show that \gap is vulnerable to \aProfilingAttack ~\cite{baumgartner2020mind, vaudenay2020analysis, knight2021linkability, hoepman2021critique}. An eavesdropping adversary \advEavesdropper can possibly track the movements of infected users in the previous days and in some cases it can also figure out who they have been in contact with, still taking into account recently infected users only. Based on these data, the adversary can build partial profiles and social graphs of the 
infected users \cite{vaudenay2020analysis, DP3T:WhitePaper}. 
To achieve this, \advEavesdropper deploys a Bluetooth sensor network to collect (sniff) $\mathit{\tempId}$s emitted by the \gap. \advEavesdropper then links the collected $\mathit{\tempId}$s to each infected user, resulting in a movement graph of the users. From this data, \advEavesdropper can infer sensitive information about infected users and their activities, e.g., where infected users live (locations where they stay during evenings and at night), work and leisure activities. 
\fixedAlC{Hence, requirements \rTrackingIdx and \rSocialGraphIdx are not satisfied.} 

Furthermore, the movements of infected users can be traced by exploiting a weakness in the scheme requiring infected users to upload their $\tek${s} 
from which all rolling proximity identifiers ($\mathit{\tempId}$s) of a whole day can be derived. \advEavesdropper can thus link the $\mathit{\tempId}$s that are derived from the same $\tek$, so that \advEavesdropper can get to know the movement profiles of the infected users within its Bluetooth sensing network. Moreover, since movement patterns of a user may have marked similarities over several different days, \advEavesdropper may also be able to link movement profiles of the same user over more than one day (\rTrackingIdx is not achieved). 

To mitigate this attack vector, Troncoso et al.~\cite{DP3T:WhitePaper} propose a more advanced "design 2" version of their DP3T protocol and  Rivest et al.~\cite{PACT-MIT} propose a solution named PACT that allows infected users to only upload short-lived ephemeral IDs (that are changed every 15 minutes or even faster). This means \advEavesdropper can only track users for less than 15 minutes, which is not enough to build informative movement profiles of the user.

Yet, Knight et al. \cite{knight2021linkability} show that \gap and other BLE-based approaches on Android are vulnerable to tracing due to linkability of $\tempId$s. 

In this attack, an adversary can link $\tempId$s based on their corresponding random BLE MAC addresses when $\tempId$s are advertised (beaconed) using resolvable Random Private Addresses (RPAs). This is because all random BLE MAC addresses of a device are derived from the Identity Resolving Key (IRK) of the Bluetooth adapter that is unchanged unless the device is factory-reset. This attack affects all Android apps since Android only supports resolvable RPA protocol while iOS also supports non-resolvable RPA protocol.

\noindent\textbf{\rTracking.}
In decentralized contact tracing approaches like \gap, \fixedAlC{adversary \advServer (a dishonest $\service$) can get to know the Temporary Exposure Keys ($\tek$s) of infected users (for a maximum of 14 days). Hence, movements of infected users can be tracked up to 14 days, i.e., \rTrackingIdx is not satisfied.} 

 \noindent\textbf{Privacy problems caused by logging.}
 Reardon et al., \cite{reardon2021why} show that \gap keeps the log of Bluetooth communications, so that any built-in apps have access to it. This can be misused to identify users, tracking users movements as well as social graphs of the users. Hence, all three privacy requirements are not fulfilled. 
 
 \subsubsection{Security}
 \label{sec:gaen-security}
 Existing works show that \gap is vulnerable to several attacks. Next, we will elaborate how such attacks affect the security of \gap. 
 
\noindent\textbf{\rFakeClaim{s}.}
Existing distributed tracing schemes (e.g., \cite{GAEN:crypto, DP3T:WhitePaper, PACT-MIT}) that are based on simply exchanging temporary identifiers $\mathit{TempID}$s (called rolling proximity identifiers or RPIs in \gap), are vulnerable to fake exposure claim attacks because a malicious user \advUser can falsely claim that a $\mathit{\tempId}$ of an infected user downloaded from the tracing server is actually one that it has collected via the Bluetooth channel, thus incorrectly indicating a contact with the infected user, in violation of requirement \rFakeClaimIdx. 

\noindent\textbf{\aRelayAttack{s}.}
Similar to BlueTrace, \gap, DP3T-1, DP3T-2 and PACT are also vulnerable to \textit{large-scale} \aRelayAttack{s}, in which the adversary \advWormhole captures and relays rolling proximity identifiers $\mathit{\tempId}$s beaconed by the apps. Even worse, the \gap API~\cite{GAEN:crypto} allows for a two-hour time window for \advWormhole to conduct relay attacks. Recently, various flavors of relay attacks against \gap were introduced ~\cite{baumgartner2020mind, vaudenay2020analysis, github2020immunirelay, avitabile2020tenku, vaudenay2020dilemma, Gennaro2020, dehaye2020swisscovid} For instance, Baumgärtner et al.~\cite{baumgartner2020mind} realized this attack on the German Corona-Warm-App \cite{CoronaWarnAppGermany} which is based on \gap. They showed that \advWormhole can even use off-the-shelf smartphones to capture and relay $\mathit{\tempId}$s between three different cities in Germany. 
Morover, Dehaye and Reardon~\cite{dehaye2020swisscovid} introduce a new effective \textit{large-scale} relay attack against the \emph{SwissCovid} app also based on the \gap API. They demonstrate that a malicious Software Development Kit (SDK) can transform benign devices into malicious relay stations for relaying $\mathit{\tempId}$s without the knowledge of the device owners. Hence,  requirement R-S2 is clearly not satisfied by \gap.

\textit{Time travel attacks.} Iovino et al.~\cite{Iovino2020timetravel} demonstrate that an adversary \advWormhole does not need to capture $\mathit{\tempId}$s from potentially infected users, but can replay a $\mathit{\tempId}$ derived from an ''outdated" $\mathit{TEK}$ key advertised on the contact tracing server directly. In order to make the outdated $\mathit{\tempId}$ valid, they introduce a time machine attack in which the adversary \advWormhole in proximity of a victim remotely manipulates the clock of the victim's phone by setting it back with the help of a simple commodity device (that costs only US \$10). When the victim's clock is later eventually restored to the correct time, the tracing app will associate the replayed $\mathit{\tempId}$ to a valid $\mathit{TEK}$ and raise an exposure notification alert.

\textit{The KISS attack.} In the same work \cite{Iovino2020timetravel}, Iovino et al. also pointed out a bug in \gap-based contract tracing apps that falsely accept $\mathit{\tempId}$s captured during the same day when the $\mathit{TEK}$ was published. Therefore, the adversary only needs to download a $\mathit{TEK}$ of the current day, derive $\mathit{\tempId}$s from the $\mathit{TEK}$ and replay them.

\textit{Tracing Forgery: The Terrorist Attack.} In this class of attacks (see~\cite{avitabile2020tenku}), the adversary colludes with infected users (who want to monetize their infection status) to obtain their Temporary Exposure Key ($\mathit{TEK}$). The adversary then replays $\mathit{\tempId}$s derived from the $\mathit{TEK}$ at the adversary-chosen (targeted) places. Finally, the colluding infected user will upload the $\mathit{TEK}$ resulting in false exposure alarms in the devices that have captured the replayed $\mathit{\tempId}$s. 
 
 \subsubsection{Ethics}
 \gap based apps have received criticism about ethical aspects such as lack of transparency and coercion \cite{lanzing2020ethical,White2021why}. These works indicate that the requirements R-Et1 and R-Et2 are not achieved.

\section{Proposed Approach - \ourname}
\label{sec:proposal}

In this section, we first provide a generic framework for Diffie-Hellman (DH)-based schemes. We then present our novel scheme, \ourname, a fully fledged example of a DH-based approach and highlight its benefits compared to the prominent approaches analyzed in Sect. \ref{sec:tc-security}.

\begin{figure}[ht]
	\begin{center}
		\includegraphics[width=\columnwidth]{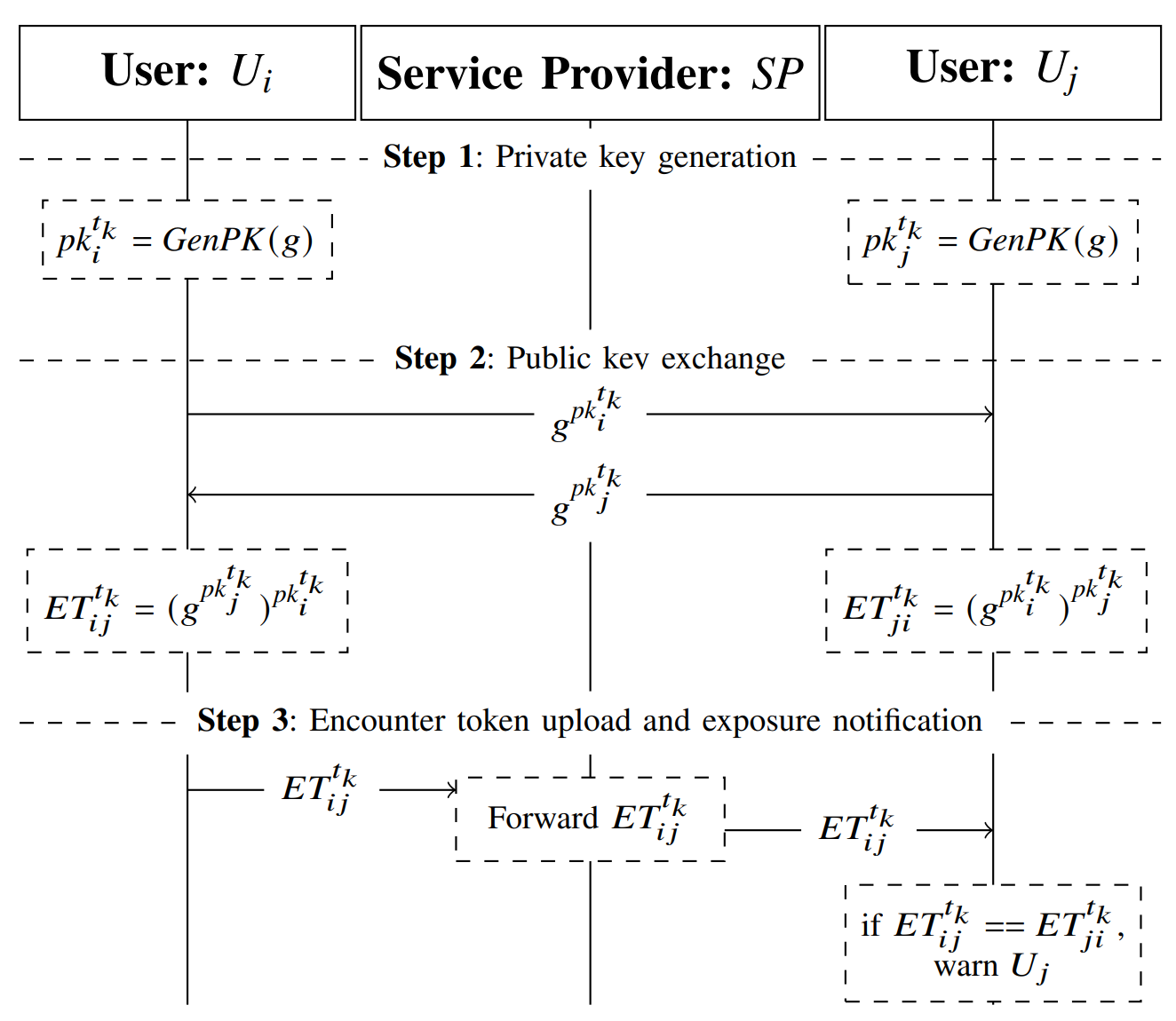}
	\end{center}
	\caption{Generic framework of DH-based Approaches.}
	\label{fig:generic-dh}
\end{figure}

\subsection{Generic framework of DH-based approaches.}
The core idea of decentralized approaches based on asymmetric key cryptography like Diffie-Hellman is that two users establish a \textit{unique and secret} Encounter Token ($\et$) using a key exchange protocol when they are in proximity by exchanging short-lived random public keys via BLE. In this paper, we use Diffie-Hellman as a key exchange protocol.
Figure \ref{fig:generic-dh} shows an overview of the use of DH-based encounter tokens in a contact tracing scheme. 
In \textbf{Step 1}, users $\useri$ and $\userj$ generate their own private keys $\pk_i^{t_k}$ and $\pk_j^{t_k}$ respectively for each time interval $t_k$ that is changing  every $T$ (e.g., 15) minutes. These private keys are used to derive corresponding public keys $\pubk_i^{t_k} = g^{\pk_i^{t_k}}$ and $\pubk_j^{t_k} = g^{\pk_j^{t_k}}$. In \textbf{Step 2}, the public keys are exchanged via BLE when two devices are in vicinity. For encounters surpassing a specified minimal duration, e.g., 5 minutes, an $\et$ will be calculated, e.g., $\useri$ calculates $\et_{ij}^{t_k}$ from $\useri$'s private key $\pk_i^{t_k}$ and $\userj$'s public key $\pubk_j^{t_k}$ as follows: $\et_{ij}^{t_k}= (g^{\pk_j^{t_k}})^{\pk_i^{t_k}}$. Since $\useri$ and  $\userj$ never share their private keys, only they can know their secret encounter token $\et_{ij}^{t_k}$. It is worth noting that the DH key generation and encounter token calculation processes do not need to happen on-line. For saving battery, it can be deferred to the next time when the smartphone is being charged. In \textbf{Step 3}, when a user (e.g., $\useri$) is tested positive for COVID-19, $\useri$ sends its encounter token $\et_{ij}^{t_k}$ to the $\service$ which will forward $\et_{ij}^{t_k}$ to other users. Once $\userj$ receives $\et_{ij}^{t_k}$, it will compare $\et_{ij}^{t_k}$ to the $\et$s it has calculated. If $\et_{ij}^{t_k}$ is equal to $\et_{ji}^{t_k}$, $\userj$ is notified that it has encountered an infected user.

Although we use the well-known DH-based approach for illustrative purposes, any other two-party key-exchange protocols where parties send only one short message to each other are applicable. Thus, existing proposals like CleverParrot~\cite{canetti2020cleverParrot}, \ProntoCC~\cite{avitabile2020ProntoC2}, and Epione \cite{trieu2020epione} 
use Elliptic-curve DH (ECDH). Further, these approaches provide several modifications and optimizations to improve the effectiveness, security and privacy of the system (cf. Sect. \ref{sec:existing-DH}).
\subsection{Limitations of DH-based approaches}
\label{sec:design-consideration}
Our proposed approach \ourname seeks to address three technical limitations of DH-based approaches as follows:
\begin{itemize}
    \item \textbf{Size restriction of BLE beacon message.} Since public keys are in general too big for BLE beacon messages, existing solutions apply workarounds, e.g., \ProntoCC needs to handle a bulletin board, or CleverParrot has to reduce the key size and requires operating systems to enable special BLE advertising messages. 

    \item \textbf{Sharing encounter tokens $\et${s}.} Uploading $\et${s} directly may raise privacy risk. Hence, we aim to keep $\et${s} always secret. 
    
    \item \textbf{No time window restriction.} Existing approaches do not limit limit time window that would open opportunity for two-way relay attacks.

\end{itemize}
In the following, we will present \ourname and discuss how we address those limitations in detail.

\subsection{\ourname Design}
\label{sec:tc-design}

\begin{figure}[ht]
	\centering
	\includegraphics[width=\columnwidth]{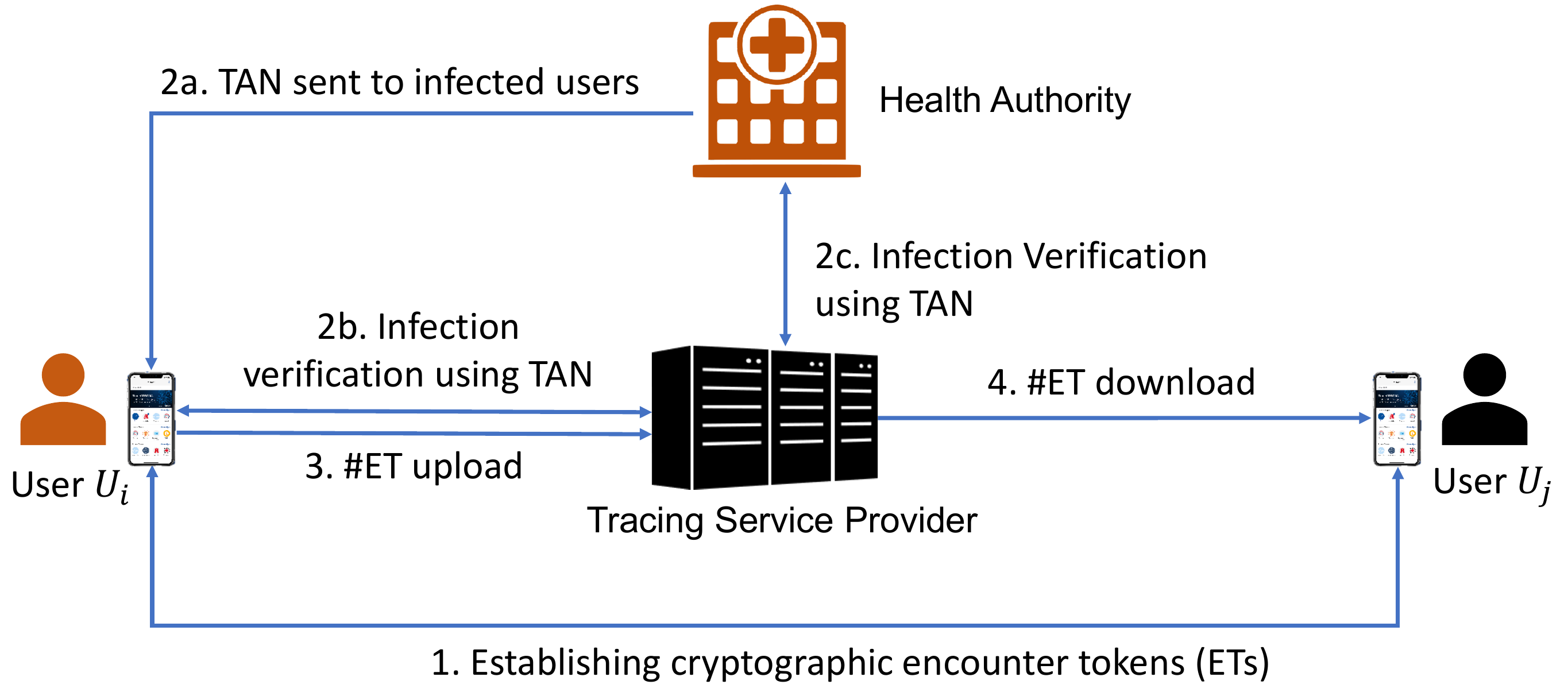}
	\caption{\ourname system overview.}
	\label{fig:system-overview}
\end{figure}

\subsubsection{System Overview}
\label{sec:system-component}
Our design follows the system model (cf. Fig. \ref{fig:system-model}) and the generic framework for DH-based schemes shown in Fig.~\ref{fig:generic-dh}.
An overview of the basic usage scenario of \ourname is shown in Fig.~\ref{fig:system-overview}. For a discussion on complementary application scenarios like wearable devices and private contact tracing please refer to  
\ifarXiv 
 Appendix \ref{sec:discussion}. 
\fi
The functionality of \ourname can be divided into four phases: (1)~Encounter token establishment, (2)~infection verification, (3)~token information upload, and (4)~token information download and contact verification. Next, we will describe each of these phases in detail.

\subsubsection{Encounter Token Establishment}
\label{sec:et-establish}
\begin{figure}[ht]
	\centering
	\includegraphics[width= \columnwidth]{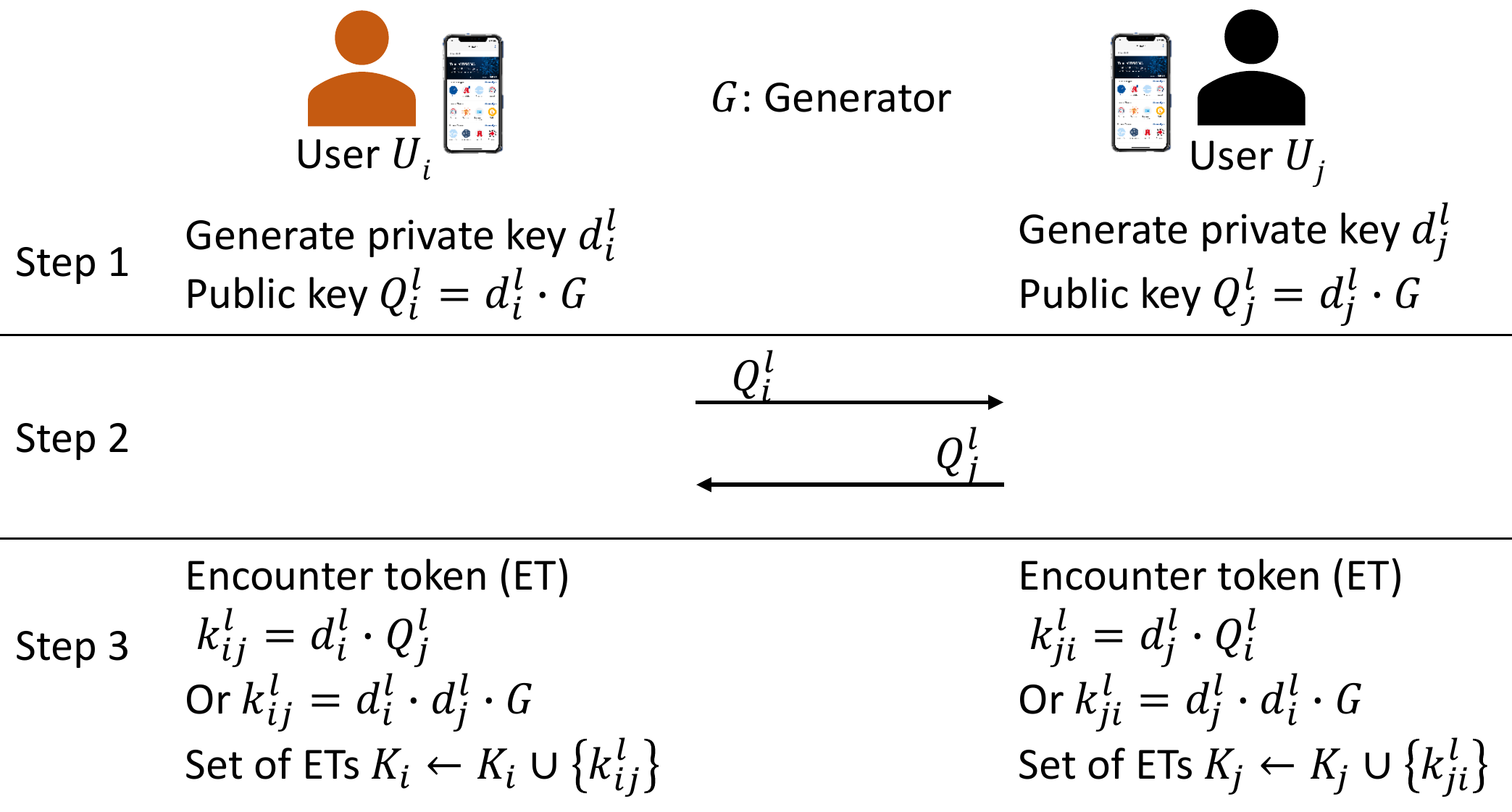}
	\caption{Elliptic-curve Diffie–Hellman (ECDH)-based encounter token establishment.}
	\label{fig:tc-ecdh}
\end{figure}

\ourname App uses BLE as a proximity communication protocol to advertise a random ephemeral identifier to other devices in the environment and to scan for the identifiers of other apps. Once an ephemeral identifier of another app has been observed for a minimum duration (e.g., 5 minutes), a connection over BLE to the other app is opened and an Encounter Token (ET) is established using the Elliptic Curve Diffie-Hellman (ECDH) key exchange protocol. Figure \ref{fig:tc-ecdh} shows the token establishment protocol in detail for two users $\useri$ and $\userj$. \fixedAlC{Following typical ECDH notation, let $Q$ denote the public key, $d$ the private key and $G$ the generator.} Let $T$ denote the period of a rolling key time frame and $l$ be the index of the time frame $f^l= [l*T, (l+1)*T]$. Let $K_i$ and $K_j$ be the sets of ETs of users $U_i$ and $\userj$, respectively. Let $k^l_{ij}$ be an ET established between two user Apps $\useri$ and $\userj$ at time point $\tijl$, i.e., a timestamp falling in  time frame $f^l$. 
The process of establishing an ET is then as follows:
\begin{enumerate}
	\item \textbf{Step 1:} For every time frame $f^l$, users $\useri$ and $\userj$ generate a ECDH keypair including private keys $d^l_i$ and $d^l_j$, and public keys $Q^l_i = d^l_i*G $ and $Q^l_j = d^l_j*G $, respectively, where $G$ is the generator defining the used cyclic subgroup of the elliptic curve.
	
	\item \textbf{Step 2:} $\useri$ and $\userj$ exchange their public keys $Q^l_i$ and $Q^l_j$ via Bluetooth LE.
	
	\item \textbf{Step 3:} Each user calculates the encounter token based on its private key and the received public key. In particular, $\useri$ calculates $\kijl = d^l_i * Q^l_j$ while $\userj$ calculates $\kjil = d^l_j * Q^l_i$. Obviously, $\kijl = \kijl = d^l_i * d^l_j * G$. Each user then adds the encounter token into its encounter token set: $K_i \gets K_i \cup \{ \kijl\}$ for $\useri$ and  $K_j \gets K_j \cup \{\kjil\} $ for $\userj$. After $\kijl$ is established, $\useri$ and $\userj$ continue exchanging their ephemeral identifiers 
	periodically to monitor the duration $D^l_{ij}$ of the encounter and the strength of the Bluetooth signals $S^l_{ij}$ (which roughly correlate with how far or near two users are from each other). In summary, the data recording the start of the encounter $\tijl$, the duration of the encounter $D_{\kijl}$ and the strength of the Bluetooth signal $S^l_{ij}$, are stored as metadata associated with token $\kijl$.         
\end{enumerate}

It is worth noting that in order to preserve battery lifetime, \textbf{Step 1} and \textbf{Step 3} can be done offline, i.e., when the smartphones are being charged (e.g., during the night).

\begin{figure*}[ht]
	\centering
	\includegraphics[width= 1.8\columnwidth]{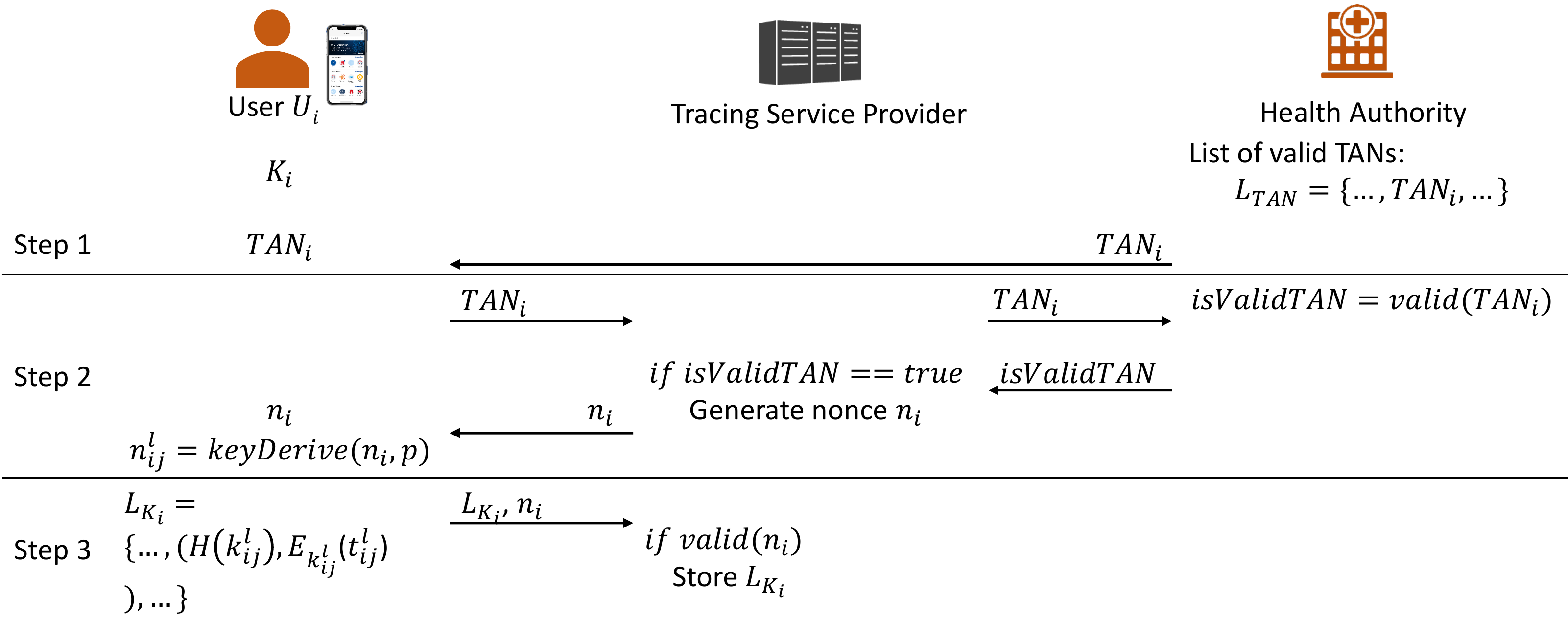}
	\caption{Infection verification and encounter token upload.}
	\label{fig:ver-upload}
\end{figure*}

\begin{figure*}[ht]
	\centering
	\includegraphics[width= 1.4\columnwidth]{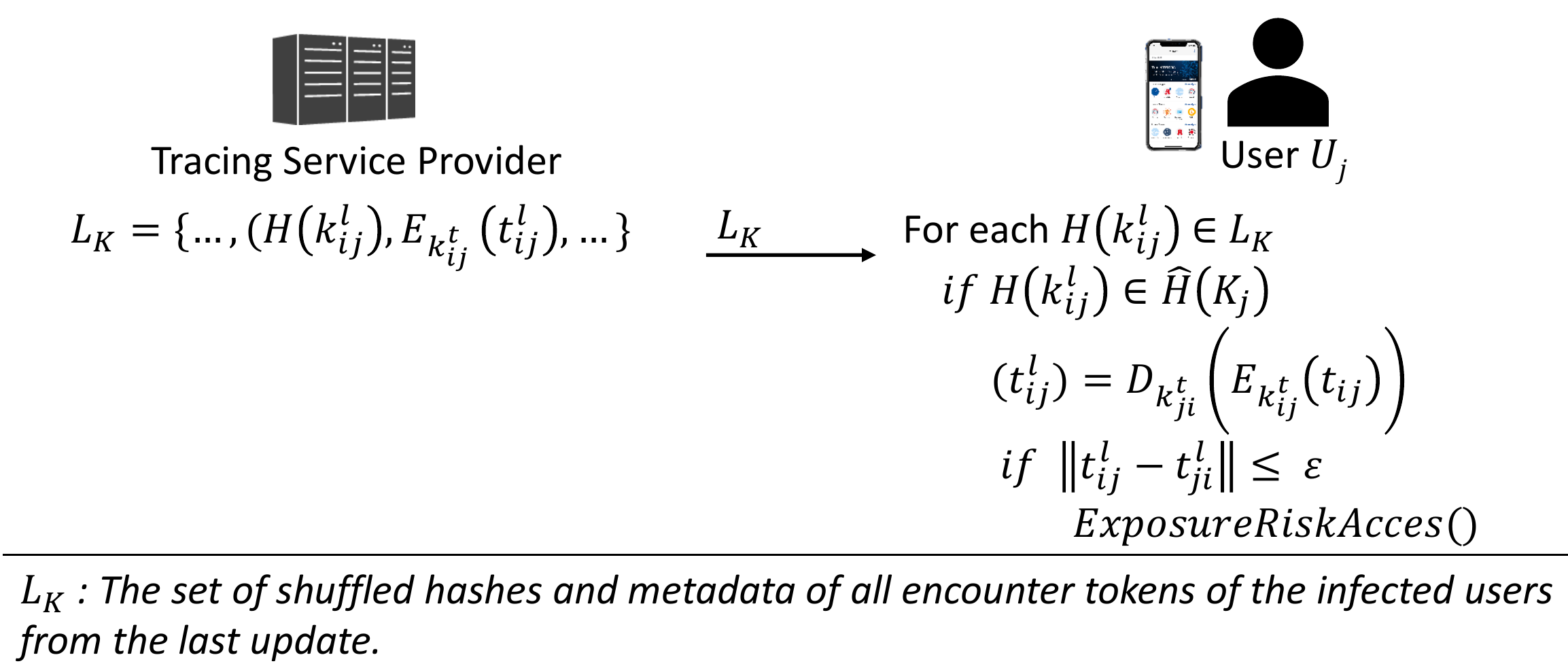}
	\caption{Encounter Token Download and Exposure notification.}
	\label{fig:et-download}
\end{figure*}

\subsubsection{Infection Verification and Encounter Token Upload}
\label{sec:verification-upload}
Since the main goal of the system is to notify users who have encountered infected users (tested positive for COVID-19), the system needs to make sure that only infected users can use the system to release their encounter tokens $K$. In our system, the Health Authority $\healthauthority$ issues for each infected user a unique authentication code, a so-called Transaction Authentication Number ($\mtan$). If an infected user wants to share their encounter tokens, it can use this $\mtan$ to prove its infection status by uploading the $\mtan$ along with their encounter token information. 

Fig. \ref{fig:ver-upload} illustrates the infection verification and encounter token uploading phases. 
\textbf{In Step 1} and \textbf{Step 2} $\healthauthority$ sends a $\tani$ to infected user $\useri$. This can be done by using any appropriate out-of-band channel: in person, via SMS, via regular mail or via e-mail. $\tani$ can also be sent along with the test results. The infected user can input their $\mtan$ directly by typing the number in or use their smartphone's camera to scan a QR code containing the $\mtan$.
\textbf{Step 3} shows how $\useri$ can upload its encounter token information. Timestamp $ \tijl$ is encrypted using AES encryption using the encounter token $\kijl$ as the key (or a key derivation function can be used to derive a key from $\kijl$). Let $m^l_{ij} = E_{\kijl}(\tijl)$ denote the encryption of $\tijl$. $\useri$ sends $\mtan_i$ and a list $L_{K_i}$ consisting of the $m^l_{ij}$ along with corresponding hashes $h^l_{ij} = H(\kijl)$ of the encounter tokens $\kijl$ to server $\server$. We have thus $L_{K_i} = \{\ldots, (m^l_{ij}, h^l_{ij}), \ldots \}$. $\server$ forwards $\tani$ to $\healthauthority$ to verify whether $\tani$ is valid or not. If $\tani$ is valid, it will extract and store each element $(m^l_{ij}, h^l_{ij})$ of $L_{K_i}$ separately. 

It is worth noting that \ourname provides both usability and privacy benefits by enabling infected users to remove specific unnecessary or sensitive encounter tokens that, e.g., (1) had only a short duration, thus being not essential for contracting the disease, or, 
(2) happened at a time or place that users do not want to disclose even anonymously, e.g., at a sensitive event or meeting.   

\subsubsection{Encounter Token Download}
\label{sec:et-download}

All \ourname Apps download regularly, e.g., every night, encounter token information from server $\server$ to identify potential exposure risks. Figure \ref{fig:et-download} shows the encounter download protocol. Let $L_k = \{\ldots,(H(\kijl),E_{\kijl}(\tijl)), \ldots \} $ be the list of the hashes and metadata of all encounter tokens of all infected users since the last update. To avoid linking entries related to a particular infected user together based on their position in the list, all entries in $L_k$ are shuffled before sending them to users. Once a user $\userj$ receives $L_k$, it compares the received token hashes to its own token hashes to discover matching encounters. If a matching encounter hash, e.g., $H(\kijl)$ is identified, $\userj$ decrypts the matching encounter token metadata using the associated encounter token $ \kijl $ as the key: $\tijl = D_{\kjil} (E_{\kijl}(n_i ||\tijl))$. $\userj$ then checks the validity of encounter token w.r.t. to encounter time $ \tijl $ to make sure that $\kijl$ and $\kjil$ were established during the same time frame. This will limit the time-window available for a relay attack as we will discuss in Sect. \ref{sec:dh-security}. Assuming that the clocks of the two devices are deviating by at most $\epsilon$ seconds, if $|\tijl - \tjil| \leq \epsilon$, $\kijl$ and $\kjil$ are considered to have been derived at the same time, i.e., the matching of $\kijl$ and $\kjil$ is valid. The system then uses metadata information, e.g., the time of the encounter $\tijl$, the duration of the encounter $D_{\kijl}$ and the signal strength $S^l_{ij}$ to assess the risk of this exposure.


\begin{table*}[ht]
\centering
\caption{The advantages of DH-based approaches in comparison to state-of-the-art approaches. (*) on the user side. (**) Possibly only infected users. (***) prevent one-way and limit real-time two-way attacks.  +/- means achieve/not achieve corresponding requirements.}
\label{tab:comparison-esp}

\begin{tabular}{|l|c|c|c|c|}
\hline
\multirow{2}{*}{}                                                    & Centralized                                                                                        & \multicolumn{3}{c|}{Decentralized}                                                                                                                                                                                                                                               \\ \cline{2-5} 
                                                                     & \multicolumn{1}{l|}{\begin{tabular}[c]{@{}l@{}}BlueTrace/\\ PEPP-PT/\\ TousAntiCovid\end{tabular}} & \multicolumn{1}{l|}{\begin{tabular}[c]{@{}l@{}}GAEN/\\ DP3T-1\end{tabular}} & \multicolumn{1}{l|}{\begin{tabular}[c]{@{}l@{}}DP3T-2/\\ MIT-PACT/\\ UW-PACT\end{tabular}} & \multicolumn{1}{l|}{\begin{tabular}[c]{@{}l@{}}TraceCORONA/\\ Pronto-C2/\\ CleverParrot\end{tabular}} \\ \hline
User identifier                                                      & \multicolumn{1}{l|}{\begin{tabular}[c]{@{}l@{}}Phone number\\ /App ID\end{tabular}}                & \multicolumn{1}{l|}{Random keys}                                            & \multicolumn{1}{l|}{Random keys}                                                           & \multicolumn{1}{l|}{Random keys}                                                                      \\ \hline
\begin{tabular}[c]{@{}l@{}}Life-time of \\ initial keys\end{tabular} & \multicolumn{1}{l|}{Long-lived}                                                                    & \multicolumn{1}{l|}{Daily}                                                  & \multicolumn{1}{l|}{Short-Lived}                                                           & \multicolumn{1}{l|}{Short-lived}                                                                      \\ \hline
Superspreader                                                        & +                                                                                                  & -*                                                                          & -*                                                                                         & +*                                                                                                    \\ \hline
CAII                                                                 & +                                                                                                  & -*                                                                          & -*                                                                                         & +*                                                                                                    \\ \hline
Identifying users                                                    & \textbf{-}                                                                                         & -**                                                                         & +                                                                                          & +                                                                                                     \\ \hline
Tracking users                                                       & -                                                                                                  & \textbf{-**}                                                                  & +                                                                                          & +                                                                                                     \\ \hline
Extracting social graph                                              & \textbf{-}                                                                                         & \textbf{-**}                                                                & +                                                                                          & +                                                                                                     \\ \hline
Fake exposure claim                                                  & -                                                                                                  & \textbf{-}                                                                  & \textbf{-}                                                                                 & +                                                                                                     \\ \hline
Relay attack                                                         & -                                                                                                  & \textbf{-}                                                                  & \textbf{-}                                                                                 & +***                                                                                                  \\ \hline
Transparency                                                         & +                                                                                                  & \textbf{-}                                                                  & +                                                                                          & +                                                                                                     \\ \hline
Independency                                                         & +                                                                                                  & \textbf{-}                                                                  & +                                                                                          & +                                                                                                     \\ \hline
\end{tabular}
\end{table*}

 \subsection{Hybrid Approach}
 \label{sec:tc-hybrid}
 In the following, we will present a hybrid approach that provides a trade-off between the effectiveness and the privacy requirements of centralized and decentralized architectures, i.e., maximizes effectiveness of the app while preserving privacy of the users. \fixedAlC{As discussed in Sect. \ref{sec:req-accountability}, the accountability requirement (\rAccountabilityIdx) refers to the possibility to evaluate the effectiveness of a \dct scheme. Therefore, we focus on this requirement 
by specifying what kind of data are needed to satisfy it and how they can be submitted to the health authority $\mathit{\healthauthority}$ and the tracing service provider $\service$.} 
Our design leverages the advantages of the DH-based scheme.

\subsubsection{Useful data}
To fulfill the requirement \rAccountabilityIdx (Accountability), the App needs to send authentic, but anonymized data in a privacy-preserving way to $\service$. Table \ref{tab:useful-data} shows potentially useful types of data that can help to evaluate and optimize the \dct system. Such types of data can also be helpful to epidemiologists and decision makers to understand the virus spreading patterns and, e.g,  deploy effective policies to limit the pandemic. 
\begin{table}[ht]
    \centering
    \caption{Useful information for epidemiological analysis and evaluation and optimization of a \dct system.}
    \label{tab:useful-data}
    \begin{tabular}{l}
    \hline
    Number of active users \\
    Number of infected users \\
    Number of encounters of infected users \\
    Number of affected users \\
    Number of encounters of affected users \\
    True positive rate \\
    Importance of notification \tablefootnote{Exposure notification from a \dct is less important if users already knew their exposure status before being notified by a \dct app, e.g., the affected users who live in the same household to an infected users are expected to be informed immediately when the test result is available.} \\
    Distribution of Risk score \\
    The correlation between risk score and true positive rate\\
    
    \hline 
    \end{tabular}
\end{table}

\subsubsection{Further levels of contacts - early notifications}
\label{sec:second-level-contacts}
Only considering first-level  contacts may not be enough. Affected users who were in contact with an infected user and thus may have gotten infected, may already spread the virus to other users (here referred to as second-level contacts) before ever being notified. Since there is a significant delay from the time that the direct (first-level) contact took place until the the direct contacts get an exposure notification and potentially a positive test result for themselves which is required to upload encounter tokens to warn second-level contacts, valuable time is lost, leading to potential new infection chains. Also, if direct contacts get infected but remain asymptomatic and do thus not get tested, second-level contacts may never receive exposure notifications in spite of the apparent risk of contagion. 

\begin{figure}[htb]
    \centering
    \includegraphics[width=\columnwidth]{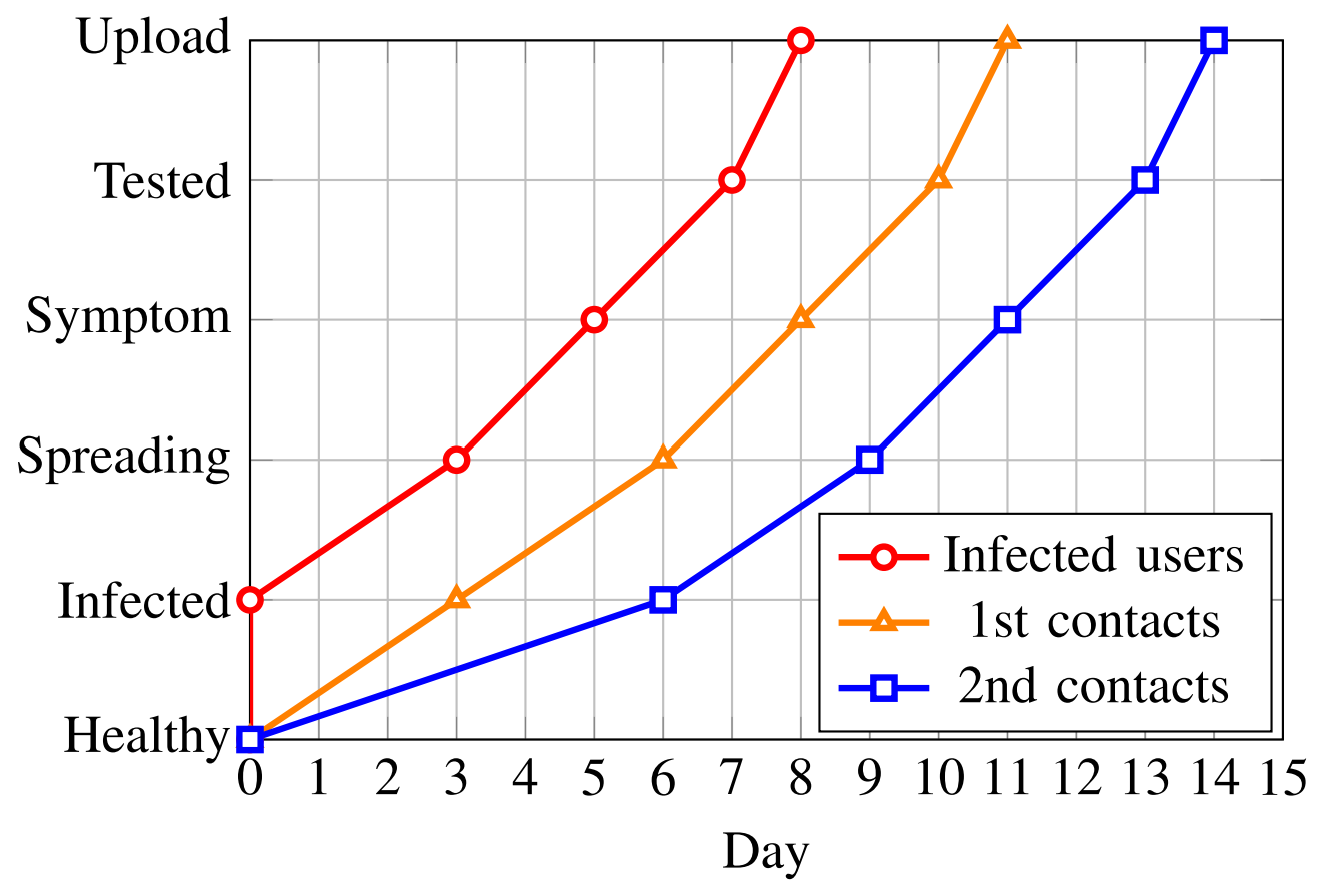}
    \caption{An example of the development of statuses of infected users, their first-level contacts (1st contacts) and second-level contacts (2nd contacts).}
    \label{fig:infection-chain}
\end{figure}

Figure \ref{fig:infection-chain} shows an example of a typical scenario in which a user $\useri$ gets infected on Day 0 and becomes contagious (spreading the virus) after around three days. However, it can take around four or five days until symptoms emerge. In the first year of the pandemic, it was typical in many countries, that it took around 2 days for a symptomatic user to be tested and another day (or even longer) to get the positive test result. 
Assuming that infected user $\useri$ uploads their encounter information and the affected users receive exposure notification immediately, it still takes 5 days from the time when $\useri$ starts spreading the virus until the affected users are notified. This means the affected users who got infected already can spread the virus for two to three days (from Day 6 to Day 8) to further second-level contacts. Therefore, to prevent potentially infected second-level contacts from spreading the virus, they should also be notified as early as possible. 

In a DH-based system this can be easily achieved by allowing direct contacts $\userj$ receiving exposure notifications to upload their encounter tokens even \emph{before} receiving a positive test result in order to warn second-level contacts. To authenticate the uploaded tokens, the direct contact $\userj$ only needs to provide the secret value of the encounter token $\et_{ij}$ of its contact with the infected person $\useri$ to $\service$, which can then verify it against the published hashed version of the token, and thus authenticate that a contact between the originator of the uploaded tokens ($\userj$) and an infected person ($\useri$) has in fact taken place. This prevents malicious users that are not direct contacts to claim fake exposures. However, encounter tokens concerning second-level contacts should be tagged by $\service$ as such to distinguish them from ones concerning direct contacts, as the risk of contagion for second-level contacts likely is lower. Therefore, also any notifications posted by the contact tracing app to second-level contacts should clearly communicate that the warning concerns second-level contacts and not direct exposure.

\subsubsection {Sharing Epidemiological Information with Health Authorities}
As discussed in the previous section, a direct contact $\userj$ can prove it exposure status with an infected user ($\useri$)  based on the possession of the secret value of the encounter token $\et_{ij}$. \ourname utilizes this to authenticate the correctness of exposure information that users may voluntarily want to share with health care research institutions, thereby preventing malicious users from corrupting the data by providing faked exposure information to the researchers. This helps in improving the accuracy and correctness of the epidemiological modelling used as as basis for political decision making in the crisis situation.

\subsubsection {Sharing Epidemiological Information via Healthcare Professionals}
Since healthcare professionals like doctors collect information about their patients that come for a COVID-19 test or for consultation for their symptoms, doctors can act as a source of reliable information for epidemiological analysis in a properly anonymized form. For example, the healthcare professional could provide for each patient following anonymous information to help in assessing the epidemiological situation as well as the effectiveness of the contact tracing system: whether the user was notified by the contact tracing app and what the possible risk score was, whether the user knew about a potential exposure status even before being notified by the app, possible symptoms, and the test result. These kind of data provided to the epidemiological analysis do as such not reveal any information about the true identity of individual patients, but they do provide crucial information necessary to evaluate the effectiveness of the contact tracing app.

\section{Security and Privacy Analysis of \ourname}
\label{sec:tc-security}

\subsection{Advantages of DH-based approaches}
\label{sec:dh-advantages}
In this section, we will analyze DH-based approaches in general and \ourname in particular in comparison to \gap and BlueTrace with regard to requirements laid out in Sect.~\ref{sec:requirements}.
\subsubsection{Effectiveness}
\label{sec:dh-effectiveness}
\textbf{\rSuperspreader.} Although the Tracing Service Provider $\service$ only receives anonymous encounter tokens that are not sufficient to detect \rSuperspreader{s} and \rCAII users, the contact tracing App itself can be used to warn its user in case the App identifies a large number of contacts with other infected users, since this can be an indication that actually the user itself is a \rSuperspreader or \rCAII who has been the source of contagion for those infections. As a result, the user could seek immediate testing, but also immediately upload their encounter tokens to warn others as discussed in Sect.~\ref{sec:second-level-contacts}. Further, the App can prove the user's status as a suspected \rSuperspreader or  \rCAII to $\service$ by uploading the secret encounter tokens it has in common with infected users. By verifying these against the published hashes of encounter tokens of infected users the $\service$ can verify that the user is indeed a person with many contacts with infected people and therefore a possible \rSuperspreader. The $\service$ can then tag the encounter tokens of the user accordingly, so that exposure notifications related these tokens can additionally be marked as being related to a 'possible superspreader' contact.
Hence, requirement \rSuperspreaderIdx related to the ability to identify \rSuperspreader{s} can be successfully addressed.

\subsubsection{Privacy}
\label{sec:dh-privacy}
In DH-based systems, the ECDH public keys change every 15 minutes. This means that an eavesdropper adversary \advEavesdropper cannot link public keys of a user, i.e., \advEavesdropper can only track the movement of a user for less than 15 minutes, which is not enough to build informative movement profiles of the user.  

\noindent\textbf{\aSurveillance.} Like other decentralized BLE systems, this attack fails against DH-based systems since the matching of contacts is done exclusively by the smartphone apps. A malicious service provider \advServer does not benefit from learning the \et{s} of infected users since the uploaded encounter tokens do not reveal any information about the counterparts of those encounters.

\noindent\textbf{\aMassSurveillance.}
In \TraceCORONA, even if a malicious service provider \advServer colludes with an eavesdropper \advEavesdropper, the adversaries only get to know the hashes of encounter tokens of infected users and possible locations where \advEavesdropper has collected them. However, as discussed in Sect.~\ref{sec:design-consideration}, since \advEavesdropper can obtain \et{s} only through direct interaction with the monitored users and \et{s} are created only if encounters last for a specific time (e.g., 5 minutes), \advEavesdropper is much more limited in its ability to obtain \et{s} associated with other users. In particular, \advEavesdropper will be unable to establish \emph{any} \et{s} with users that are just shortly passing by an eavesdropping station, so that the adversary's ability to track the movements of infected users is very limited. It is to be noted that this is a significant difference to both centralized and decentralized approaches presented in Sects.~\ref{sec:central} and \ref{sec:decentral}, since in these approaches the ability of the eavesdropping adversary \advEavesdropper is in this sense unlimited and it can effectively sense the presence \emph{all} users passing by its eavesdropping stations, even based on \emph{one single observation} of the user.

On the other hand, in the basic set-up, a malicious service provider \advServer acting as $\service$ could link encounter tokens \et of a specific infected user, as they would be submitted in one transaction when they are uploaded to the service provider $\service$. However, this threat can be effectively mitigated by applying appropriate anonymization techniques like blind signatures with an anonymous postbox service~\cite{reichert2021ovid} to the upload process of encounter tokens. This assures that even a malicious service provider \advServer cannot link individual encounter tokens of infected users, thereby limiting the trackability of individual system users to relatively short time frames of, e.g., 15 minutes. By applying aforementioned defense techniques, \ourname can therefore effectively address the requirements regarding providing protections against identifying (\rIdentifyingIdx) and tracking (\rTrackingIdx) users and extracting their social graphs (\rSocialGraphIdx).

\subsubsection{Security}
\label{sec:dh-security}
In the following, we will explain how DH-based systems can mitigate current attacks, hence, fulfill the security requirements.

\noindent\textbf{\rFakeClaim.} DH-based systems can easily mitigate fake exposure claims (requirement \rFakeClaimIdx). As mentioned in Sect.~\ref{sec:proposal}, infected users only share the hashes of encounter tokens meaning that the values of the encounter tokens themselves are always kept secret, so that only users actually participating in the encounter obtain the corresponding encounter token. Therefore, by proving possession of the (secret) encounter token, a user can prove that a contact with the counterpart has in fact taken place. 
The only way a dishonest user \advUser can make fake exposure claims is to obtain access to the phones of users having matching encounter tokens and extracting them. However, this attack requires compromising individual devices one-by-one and can therefore not be easily scaled. 

\noindent\textbf{Relay/Replay Attack.} Principally, all proximity-based approaches are vulnerable to this attack. This is a crucial to note that the main goal of this attack is to inject false exposure notifications on a large scale. 
However, DH-based systems provide two effective mitigation techniques that reduce the window of opportunity for attackers: on one hand, two-way communication is required for establishing contact tokens, prohibiting massive duplication of contact information by just copying beacon information, and, on the other hand, using limited time windows for validating contacts.

\textit{Two-way communication.}  
In contrast to existing approaches \cite{GAEN:crypto, DP3T:WhitePaper, pepppt, PACT-MIT, BlueTrace} that are vulnerable to \emph{one-way} relay attacks (cf. Sect. \ref{sec:analysis}), DH-based schemes utilize a \emph{handshake} protocol requiring \emph{two-way} communication to establish an encounter token. This means \advWormhole cannot simply capture the beaconed information in one place and replay it in many other places like it would be possible in other schemes. \advWormhole has to capture and relay messages at both places at the same time. This not only limits the time window of the attack but more importantly, it imposes a restriction on the scale of the attack since a mobile device cannot communicate with too many other devices at the same time, as Bluetooth LE provides only a limited number of channels and bandwidth. Based on our estimation, an average smartphone can only handle 8 Bluetooth LE connections simultaneously in a reliable manner. Therefore, in theory \advWormhole can relay the handshake of one device to at most 8 other remote devices, while this number is not limited in other approaches.

\textit{Limited time window.} 
In DH-based schemes, two users $\useri$ and $\userj$ in proximity of each other establish a unique secret encounter token $\et_{ij}$. An nfected user $\useri$ can use $\et_{ij}$ to encrypt any meta-data that only $\userj$ can decrypt. Leveraging this property, in a DH-based scheme, e.g., \TraceCORONA, the exact timestamp of an encounter can be encrypted and added to encounter token metadata so that user Apps checking encounter tokens can also check the exact encounter time. \fixedAlC{Therefore, only matching encounters that took place within a time window of at most $\epsilon$ seconds are considered as valid encounters, thereby limiting the window of opportunity for relay attacks. Other decentralized schemes like \cite{GAEN:crypto, DP3T:WhitePaper, PACT-MIT} cannot impose such limitations on the timestamps of ephemeral IDs, because the involved tracing apps can not mutually verify the actual time point of when contacts take place due to the fact that only one-way communication is used.}

Due to this, the \gap API~\cite{GAEN:crypto} allows a two-hour time window for synchronizing $\mathit{RPI}$, i.e., \advWormhole can have up to two hours to conduct relay attacks (cf. Sect. \ref{sec:gaen-security}). 
In DH-based schemes, this $\epsilon$ could be limited to seconds when assuming that smartphones used for contact tracing apps can sync their clocks via an Internet connection or during the exchange of the public keys. Note that all contact tracing apps need a frequent Internet connection for uploading and downloading encounter information.

Thus, the combination of these two advantages, requirement of two-way communication and small time window help DH-based schemes such as \TraceCORONA  to significantly reduce the impact of relay attacks on the system.

\subsubsection{Ethics}
\label{sec:dh-ethic}
Like BlueTrace, DH-based systems like \ourname can be implemented
with complete access to the source code, guaranteeing transparency. It is a standalone app that does not depend on any built-in contact tracing APIs running deep inside the mobile operating systems such as Android or iOS, thus satisfying requirements with regard to transparency and (\rTransparancyIdx) and independence (\rIndependenceIdx).
 \fixedAlC{This is in stark contrast to proprietary and closed \gap systems strictly enforced by Google and Apple. Especially in Apple's iOS systems independent contact tracing applications that continuously need to use BLE in the background are blocked by the operating system so that effective BLE sensing as required by contact tracing apps is in practice not possible. Instead, Apple forces all contact tracing approaches to rely on their closed and proprietary \gap API whose functionality can not be independently examined nor verified. It is therefore highly debatable, whether this approach is ethical, as Apple in fact forces users into using their \gap solution, having to involuntarily accept all the deficiencies discussed in this paper, or, refrain from using contact tracing solutions at all.} 

\subsection{Summary of Benefits of DH-based Approaches and Comparison to Other Approaches} 
\begin{figure}[ht]
    \centering
    \includegraphics[width=\columnwidth]{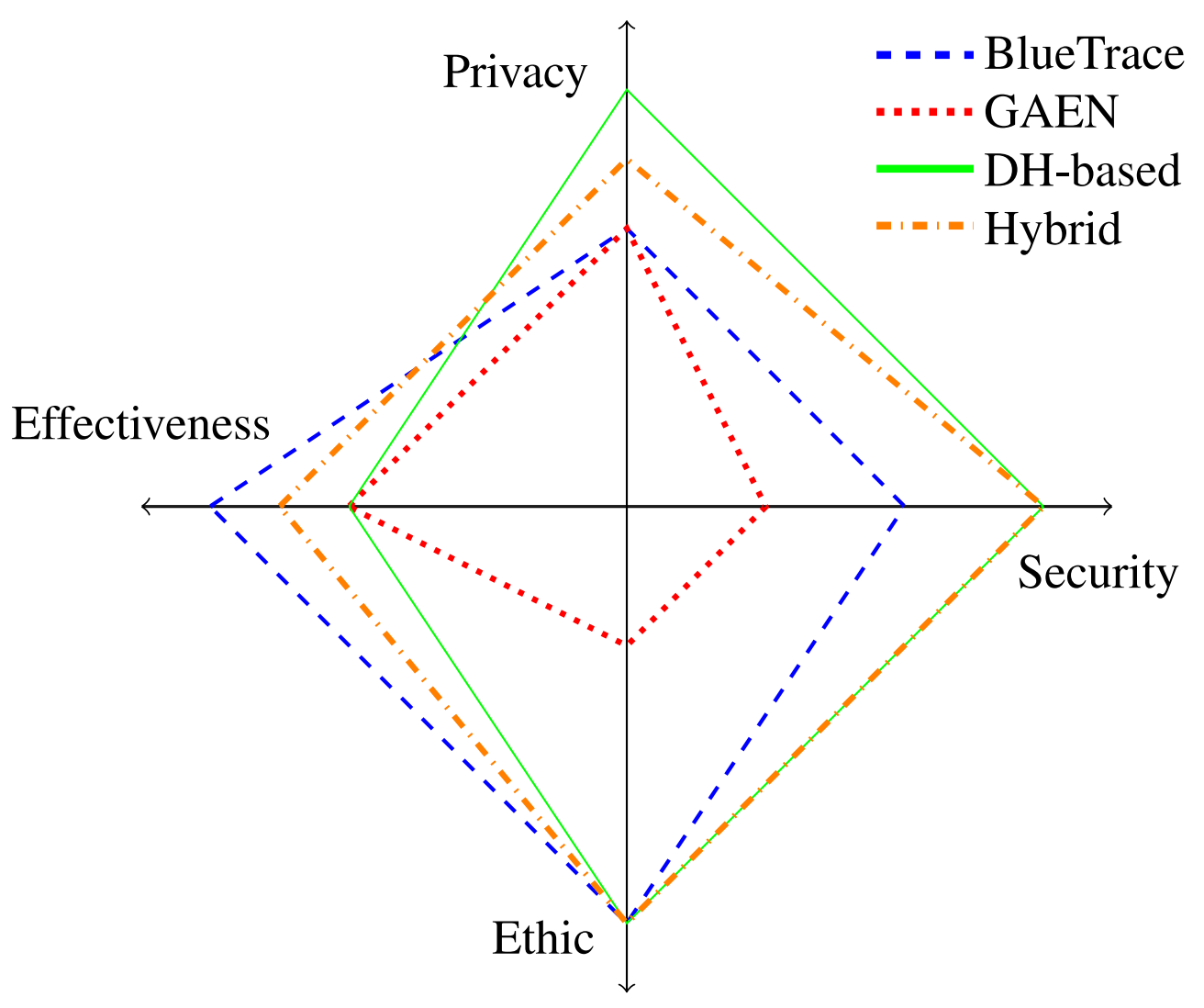}
    \caption{Quadrilemma comparison of prominent contact tracing scheme.}
    \label{fig:quadrilemma-comparison}
\end{figure}

We summarize key differences and security and privacy advantages of DH-based systems in comparison to existing approaches in Tab. \ref{tab:comparison-esp}. 
Further, for a better overview, we illustrate the comparisons in Fig. \ref{fig:quadrilemma-comparison}. It shows that \gap collects lower scores among all the alternatives. The DH-based systems provide better security and privacy protection than all other discussed solutions. For example, DH-based approaches are resilient to fake exposure claim attacks and wormhole adversary (i.e., narrowing the attack window time and requiring more communication effort as the adversary would have to operate real-time two-way communication relays). Moreover, comparing to the most widely spread contact tracing framework by Apple and Google, which is vulnerable to profiling attacks as the adversary can track the movements of infected users, DH-based systems guarantee a better protection. Interesting but not surprisingly, BlueTrace is the best w.r.t to fulfilling effectiveness requirements since it can potentially detect \rSuperspreader and \rCAII and provide useful data to epidemiologists while this could be challenging to other approaches. In terms of ethics, \gap again is on the lower end because it received many criticisms due to their coercion and the lack of transparency. More importantly, our hybrid approach inherits the advantages of DH-based approaches in terms of security and ethical aspects, while being on par with centralized approaches with regard to effectiveness.

\section{Implementation and Beta Test}
\label{sec:implementation}
We prototyped \ourname for the Android smartphone platform and tested it in a public beta test. 
The implementation uses Kotlin, an emerging programming language for Android. We have not implemented \ourname on iOS because it does not allow apps to use Bluetooth communication in the background. We use the native Android Bluetooth Low Energy APIs to implement the Encounter Token Establishment protocol. Further, our cryptographic functions, e.g., ECDH are based on the Bouncy Castle library. For the server acting as $\service$, the code is written in Java and run on Ubuntu Server operating system. 
In principle, our app can run on any Android smartphone that supports Bluetooth LE, i.e., Android 5.0 and later. 

\noindent\textbf{Alpha testing.} We internally tested the app with 25 devices covering various models and manufacturers\footnote{https://tracecorona.net/list-of-device-models-on-which-tracecorona-has-been-testedhttps-tracecorona-net-download-tracecorona/}. The results show that our app works without any problems and consumes 5-8 \% battery for a whole day (24 hours) of contact tracing without further optimizations. It is worth noting that, for the apps based on Bluetooth proximity detection, the main power consumption emerges from advertising and scanning for other devices. Therefore, our approach would consume more or less the same amount of battery as state-of-the-art approaches like BlueTrace \cite{BlueTrace} or \gap \cite{GAEN:crypto} if the time windows and frequencies of advertising and scanning processes are set the similar.   

\noindent\textbf{Beta test.} We published the \ourname app on our website and interestingly the app has drawn a lot of attention\footnote{https://tracecorona.net/download-tracecorona/}. Indeed, more than 2000 users have downloaded and tested the app. We have received many positive feedbacks on the app features and performance, except received criticism that the app does not work on very old devices that do not support Bluetooth LE. However, this is a technical limitation that is out of our control.

\noindent\textbf{Implementation on wearable devices.} To demonstrate the possibility of deploying \ourname even on on wearable devices like wristbands (For a full description please refer to Sect. III in the supplementary materials), we have implemented our design on Adafruit HUZZAH32 (ESP32), a MCU developer board that costs about US \$20. Our program size (including libraries) is 1.12 MB. Our evaluation shows that the device needs only 666 milliseconds to synchronize all the keys with the phone for a whole day. It can exchange Ephemeral Identification $EI$ and Ephemeral ECDH Public Key $Q$ with multiple other ESP32 devices and smartphones simultaneously without any delay compared to the communication between smartphones.

\section{Related Work}
\label{sec:relatted-work}
\subsection{DH-based approaches}
\label{sec:existing-DH}
\textbf{\ProntoCC \cite{avitabile2020ProntoC2}}
The main problem of DH-based approach is that the size of the public key might exceed the space limit of BLE advertising messages (i.e., identifier beacons). The minimum requirement for a standardized ECDH key is 256 bits (or even 384 bits tp provide security against a powerful adversary) while in a typical BLE advertising message there is space for 128 bits only. \ProntoCC{'s authors} propose to store the public keys on a bulletin board that can be maintained by the $\service$ or can be decentralized, and even implemented with a blockchain. Hence, instead of broadcasting the public keys via BLE, the devices only beacon the references (i.e., addresses) of the keys in the bulletin.
When a user is infected, a cryptographic hash
of encounter tokens is uploaded to the bulletin board. \fixedAlC{As discussed in Sect. \ref{sec:design-consideration}, \TraceCORONA solves this problem by utilizing BLE connections to transfer public keys without any data restrictions.} 

\textbf{CleverParrot \cite{canetti2020cleverParrot}}
To deal with the issue of fitting a DH public key in a BLE advertising message, CleverParrot proposes using a minimum key size of 224 bits (28 bytes) based on the elliptic curve P-224. They choose this key size since it is the same as the one use in Apple's Find My protocol\footnote{“Find My overview”, Apple, \url{https://support.apple.com/guide/security/locating-missing-devices-sece994d0126/1/web/1}.}. However, it is worth noting that Apple's Find My protocol is a special function in iOS. In fact, both Android and iOS support only 128-bit BLE advertising messages. Therefore, CleverParrot cannot be implemented in practice unless Google and Apple change their BLE platform or they have to adopt and treat CleverParrot as a special function like Apple's Find My. 

\textbf{DH with Private Set Intersection Cardinality (PSI-CA)}
 Epione \cite{trieu2020epione} leverages Function Secret Sharing (FSS) techniques \cite{boyle2015function} to prevent other users from learning information about the encounter tokens uploaded by infected users. In particular, this approach enables clients (user Apps) in collaboration with the servers $\service$ to learn matching encounter tokens, i.e., $\userj$ can know how many encounters with infected users it has without downloading these encounters as shown in \textbf{Step 3} Fig. \ref{fig:generic-dh}. 
 
\subsection{Survey on existing \dct schemes, apps and challenges}
There are a number of works that survey existing \dct schemes, apps and challenges. Those works can be categorized into two groups: (i) discussing technical specifications, operations and issues of the rolled out apps \cite{sun2021empirical, Wen2020study} and (ii) studying certain aspects of some \dct schemes \cite{ahmed2020survey, vaudenay2020dilemma}. 

Sun et al. \cite{sun2021empirical} focus on investigating the security and privacy issues of \dct apps on Android. Wen et al. \cite{Wen2020study} vet privacy issues of 41 country apps that have rolled our worldwide, in which they focus on analysis of documentation but also binary code to figure out what data an app collects and discuss the potential privacy risks. 

Unlike those works that focus on the apps, Vaudenay et al. \cite{vaudenay2020dilemma} focus on investigating the security and privacy issues of several schemes along with their architectures. The most relevent to our work is the study provided by Ahmed et al. \cite{ahmed2020survey}. They discuss 8 different potential attacks on 12 country apps divided in three groups: centralized, decentralized and hybrid architectures. However, those works do not provide an abstraction that groups evaluation requirements of similar schemes as we do in our work. 

\fixedAC{While existing works point out a number of privacy problems of \gap (cf. Sect. \ref{sec:gaen-privacy}), Ahmed et al. claim that \gap protects privacy of users and criticize that existing attacks are unrealistic \cite{ahmed2021privacy}. However, they do not provide arguments and evidence for their claim, i.e., it is not clear how \gap can defend against such attacks. In fact, their main experiments only confirm the principal design requirements of \gap like Randomness of Bluetooth addresses or RPI intervals that are also included in existing attack models \cite{baumgartner2020mind, DP3T:WhitePaper, avitabile2020ProntoC2, vaudenay2020analysis}. Unfortunately, the paper also gives some misleading information. For example, it states that: \say{in normal operation, the TEK downloaded are not readily available to the user and the exposure assessment is done away from the user.} However, the uploaded TEK keys of infected users are in fact by design public information that is accessible to any moderately sophisticated adversary\footnote{An archive collecting $\tek$s of the German Contact Tracing App: \url{https://ctt.pfstr.de/}}.}
\ifarXiv 
 We summarize existing works on analyzing \dct in Tab. \ref{tab:existing-attacks-analyses} (Appendix \ref{sec:existing-analysis}). 
\fi

\section{Conclusion}
In our work we have considered existing digital contact tracing \dct architectures, schemes, technologies and apps. We provide a systematic analysis and comparison of existing approaches based on a natural quadrilemma (i.e., obtaining simultaneously effectiveness, security, privacy while taking ethical aspects into account) in digital contact tracing. Our study shows that \gap which is adopted by many EU countries and states in the USA, unfortunately
is less effective and worse in protecting sensitive data
compared to other approaches like BlueTrace and DH-based schemes like ProntoC2 and CleverParrot. We propose \ourname that addresses security and privacy challenges of existing contact tracing approaches while providing comparable effectiveness. In contrast to state-of-the-art approaches that are based on exchanging ephemeral IDs, \ourname allows users to anonymously establish encounter-specific tokens using short-range wireless communication like Bluetooth. The encounter tokens can be later on used to warn users of potential exposure risks with infected persons. We systematically and extensively analyze the security and privacy of \ourname in comparison to existing approaches in Sect. \ref{sec:proposal} to show that \ourname is resilient to various attacks and thus provides better security and privacy guarantees than other approaches. We have implemented \ourname and tested it with 25 different devices of various brands and models. Further, we have published a beta test version of \ourname that has been downloaded and used by more than 2000 users without any major functional problems demonstrating that \ourname is practical.

\section*{Acknowledgments}
This research was funded by the Deutsche Forschungsgemeinschaft (DFG) SFB-1119 CROSSING/236615297. The material resources used in this research were partly co-financed by the European Union through the European Regional Development Fund.

\bibliographystyle{IEEEtran}
\bibliography{tc.bib}

\begin{thebibliography}{10}
\providecommand{\url}[1]{#1}
\csname url@samestyle\endcsname
\providecommand{\newblock}{\relax}
\providecommand{\bibinfo}[2]{#2}
\providecommand{\BIBentrySTDinterwordspacing}{\spaceskip=0pt\relax}
\providecommand{\BIBentryALTinterwordstretchfactor}{4}
\providecommand{\BIBentryALTinterwordspacing}{\spaceskip=\fontdimen2\font plus
\BIBentryALTinterwordstretchfactor\fontdimen3\font minus
  \fontdimen4\font\relax}
\providecommand{\BIBforeignlanguage}[2]{{%
\expandafter\ifx\csname l@#1\endcsname\relax
\typeout{** WARNING: IEEEtran.bst: No hyphenation pattern has been}%
\typeout{** loaded for the language `#1'. Using the pattern for}%
\typeout{** the default language instead.}%
\else
\language=\csname l@#1\endcsname
\fi
#2}}
\providecommand{\BIBdecl}{\relax}
\BIBdecl

\bibitem{ferretti2020quantifying}
L.~Ferretti, C.~Wymant, M.~Kendall, L.~Zhao, A.~Nurtay, L.~Abeler-Doerner,
  M.~Parker, D.~Bonsall, and C.~Fraser, ``Quantifying sars-cov-2 transmission
  suggests epidemic control with digital contact tracing,'' \emph{Science},
  2020.

\bibitem{wymant2021NHSapp}
\BIBentryALTinterwordspacing
C.~Wymant, L.~Ferretti, D.~Tsallis, M.~Charalambides, L.~Abeler-Dörner,
  D.~B.~R. Hinch, M.~Kendall, L.~Milsom, M.~Ayres, C.~Holmes, M.~Briers, and
  C.~Fraser, ``The epidemiological impact of the nhs covid-19 app.''
  \emph{Nature}, vol. 594, no.~4, 2021. [Online]. Available:
  \url{https://doi.org/10.1038/s41586-021-03606-z}
\BIBentrySTDinterwordspacing

\bibitem{GAEN:crypto}
Apple and Google, ``{Exposure Notification: Cryptography Specification,
  v1.2},'' April 2020, \url{https://www.apple.com/covid19/contacttracing}.

\bibitem{gvili2020security}
Y.~Gvili, ``{Security Analysis of the COVID-19 Contact Tracing Specifications
  by Apple Inc. and Google Inc.}'' Cryptology ePrint Archive, Report 2020/428,
  April 2020, \url{https://eprint.iacr.org/2020/428}.

\bibitem{baumgartner2020mind}
L.~Baumg{\"a}rtner, A.~Dmitrienko, B.~Freisleben, A.~Gruler, J.~H{\"o}chst,
  J.~K{\"u}hlberg, M.~Mezini, M.~Miettinen, A.~Muhamedagic, T.~D. Nguyen
  \emph{et~al.}, ``Mind the gap: Security \& privacy risks of contact tracing
  apps,'' in \emph{19th IEEE International Conference on Trust, Security and
  Privacy in Computing and Communications (TrustCom)}, 2020.

\bibitem{Iovino2020timetravel}
V.~Iovino, S.~Vaudenay, and M.~Vuagnoux, ``On the effectiveness of time travel
  to inject covid-19 alerts,'' \emph{The Cryptographer's Track at the RSA
  Conference, CT-RSA2021}, 2021, \url{https://eprint.iacr.org/2020/1393}.

\bibitem{avitabile2020tenku}
G.~Avitabile, D.~Friolo, and I.~Visconti, ``Tenk-u: Terrorist attacks for fake
  exposure notifications in contact tracing systems,'' \emph{19th International
  Conference on Applied Cryptography and Network Security, ACNS2021}, 2021,
  \url{https://eprint.iacr.org/2020/1150}.

\bibitem{White2021why}
\BIBentryALTinterwordspacing
L.~White and P.~van Basshuysen, ``Without a trace: Why did corona apps fail?''
  \emph{Journal of Medical Ethics}, 2021. [Online]. Available:
  \url{https://jme.bmj.com/content/early/2021/01/08/medethics-2020-107061}
\BIBentrySTDinterwordspacing

\bibitem{lanzing2020ethical}
M.~Lanzing, ``{Contact tracing apps: an ethical roadmap},'' 2020,
  \url{https://doi.org/10.1007/s10676-020-09548-w}.

\bibitem{IsraelHashomer}
B.~Pinkas and E.~Roneny, ``Hashomer: A proposal for a privacy-preserving
  bluetooth based contact tracing scheme for hamagen,'' 2020,
  \url{https://github.com/eyalr0/HashomerCryptoRef}.

\bibitem{leith2020gaendata}
D.~Leith and S.~Farrell, ``Contact tracing app privacy: What data is shared by
  europe’s gaen contact tracing apps,''
  \url{https://www.scss.tcd.ie/Doug.Leith/pubs/contact_tracing_app_traffic.pdf}.

\bibitem{lucaApp2021}
``Luca app,'' 2021, \url{https://www.luca-app.de/}.

\bibitem{stadler2021preliminary}
T.~Stadler, W.~Lueks, K.~Kohls, and C.~Troncoso, ``Preliminary analysis of
  potential harms in the luca tracing system,'' 2021.

\bibitem{BlueTrace}
\BIBentryALTinterwordspacing
J.~Bay, J.~Kek, A.~Tan, C.~S. Hau, L.~Yongquan, J.~Tan, and T.~A. Quy,
  ``Bluetrace: A privacy-preserving protocol for community-driven contact
  tracing across borders,'' Apr. 2020. [Online]. Available:
  \url{https://bluetrace.io/static/bluetrace\_whitepaper-938063656596c104632def383eb33b3c.pdf}
\BIBentrySTDinterwordspacing

\bibitem{leith2020measurement}
\BIBentryALTinterwordspacing
D.~J. Leith and S.~Farrell, ``Measurement-based evaluation of google/apple
  exposure notification api for proximity detection in a light-rail tram,''
  \emph{PLOS ONE}, vol.~15, no.~9, pp. 1--16, 09 2020. [Online]. Available:
  \url{https://doi.org/10.1371/journal.pone.0239943}
\BIBentrySTDinterwordspacing

\bibitem{istomin2021janus}
T.~Istomin, E.~Leoni, D.~Molteni, A.~L. Murphy, G.~P. Picco, and M.~Griva,
  ``Janus: Efficient and accurate dual-radio social contact detection,'' 2021.

\bibitem{MIT-SonicPACT}
J.~Meklenburg, M.~Specter, M.~Wentz, H.~Balakrishnan, A.~Chandrakasan, J.~Cohn,
  G.~Hatke, L.~Ivers, R.~Rivest, G.~J. Sussman, and D.~Weitzner, ``Sonicpact:
  An ultrasonic ranging method for the private automated contact tracing (pact)
  protocol,''
  \url{https://pact.mit.edu/wp-content/uploads/2020/11/SonicPACT\_Final\_v2-with-logos-revA.pdf}.

\bibitem{Trivedi2020digital}
\BIBentryALTinterwordspacing
A.~Trivedi and D.~Vasisht, ``Digital contact tracing: Technologies,
  shortcomings, and the path forward,'' \emph{SIGCOMM Comput. Commun. Rev.},
  vol.~50, no.~4, p. 75–81, Oct. 2020. [Online]. Available:
  \url{https://doi.org/10.1145/3431832.3431841}
\BIBentrySTDinterwordspacing

\bibitem{DP3T:WhitePaper}
\BIBentryALTinterwordspacing
C.~Troncoso, M.~Payer, J.~Hubaux, M.~Salath{\'{e}}, J.~R. Larus, E.~Bugnion,
  W.~Lueks, T.~Stadler, A.~Pyrgelis, D.~Antonioli, L.~Barman, S.~Chatel, K.~G.
  Paterson, S.~Capkun, D.~A. Basin, J.~Beutel, D.~Jackson, M.~Roeschlin,
  P.~Leu, B.~Preneel, N.~P. Smart, A.~Abidin, S.~F. G{\"{u}}rses, M.~Veale,
  C.~Cremers, M.~Backes, N.~O. Tippenhauer, R.~Binns, C.~Cattuto, A.~Barrat,
  D.~Fiore, M.~Barbosa, R.~Oliveira, and J.~Pereira, ``Decentralized
  privacy-preserving proximity tracing,'' \emph{CoRR}, vol. abs/2005.12273,
  2020. [Online]. Available: \url{https://arxiv.org/abs/2005.12273}
\BIBentrySTDinterwordspacing

\bibitem{reichert2020lighthouses}
L.~Reichert, S.~Brack, and B.~Scheuermann, ``Lighthouses: A warning system for
  super-spreader events,'' Cryptology ePrint Archive, Report 2020/1473, 2020,
  \url{https://eprint.iacr.org/2020/1473}.

\bibitem{vaudenay2020analysis}
\BIBentryALTinterwordspacing
S.~Vaudenay, ``{Analysis of DP-3T},'' Cryptology ePrint Archive, Report
  2020/399, April 2020. [Online]. Available:
  \url{https://eprint.iacr.org/2020/399}
\BIBentrySTDinterwordspacing

\bibitem{pepppt}
\BIBentryALTinterwordspacing
PEPP-PT, ``pepp-pt,'' 2020. [Online]. Available:
  \url{https://www.pepp-pt.org/content}
\BIBentrySTDinterwordspacing

\bibitem{castelluccia2020desire}
A.~Boutet, C.~Castelluccia, M.~Cunche, V.~Roca, A.~Baud, P.-G. Raverdy, and
  C.~Lauradoux., ``Desire: Leveraging the best of centralized and decentralized
  contact tracing systems,'' ACM Digital Threats: Research and Practice,
  Special Issue on Security and Privacy for Covid-19, 2021., 2021.

\bibitem{PACT-MIT}
R.~L. Rivest \emph{et~al.}, ``The pact protocol specification,'' 2020,
  \url{https://pact.mit.edu/wp-content/uploads/2020/11/The-PACT-protocol-specification-2020.pdf}.

\bibitem{avitabile2020ProntoC2}
G.~Avitabile, V.~Botta, V.~Iovino, and I.~Visconti, ``Towards defeating mass
  surveillance and sars-cov-2: The pronto-c2 fully decentralized automatic
  contact tracing system,'' CoronaDef Workshop at NDSS 2021, 2021,
  \url{https://www.ndss-symposium.org/ndss-paper/auto-draft-164/}.

\bibitem{canetti2020cleverParrot}
R.~Canetti, Y.~T. Kalai, A.~Lysyanskaya, R.~L. Rivest, A.~Shamir, E.~Shen,
  A.~Trachtenberg, M.~Varia, and D.~J. Weitzner, ``Privacy-preserving automated
  exposure notification,'' Cryptology ePrint Archive, Report 2020/863, 2020,
  \url{https://eprint.iacr.org/2020/863}.

\bibitem{trieu2020epione}
N.~Trieu, K.~Shehata, P.~Saxena, R.~Shokri, and D.~Song, ``Epione: Lightweight
  contact tracing with strong privacy,'' 2020.

\bibitem{reichert2021ovid}
L.~Reichert, S.~Brack, and B.~Scheuermann, ``Ovid: Message-based automatic
  contact tracing,'' CoronaDef at NDSS 2021, 2021,
  \url{https://www.ndss-symposium.org/ndss-paper/auto-draft-165/}.

\bibitem{vaudenay2020dilemma}
S.~Vaudenay, ``Centralized or decentralized? the contact tracing dilemma,''
  Cryptology ePrint Archive, Report 2020/531, 05 2020,
  \url{https://eprint.iacr.org/2020/531}.

\bibitem{dp3tPeppptSecurity}
\BIBentryALTinterwordspacing
``Security and privacy analysis of the document 'pepp-pt: Data protection and
  information security architecture,'' DP-3T project, Apr. 2020. [Online].
  Available:
  \url{https://github.com/DP-3T/documents/blob/master/Security\%20analysis/PEPP-PT\_\%20Data\%20Protection\%20Architechture\%20-\%20Security\%20and\%20privacy\%20analysis.pdf}
\BIBentrySTDinterwordspacing

\bibitem{knight2021linkability}
Z.~Brighton-Knight, J.~Mussared, and A.~Tiu, ``Linkability of rolling proximity
  identifiers in google’s implementation of the exposure notification
  system,'' Technical report,
  \url{https://github.com/alwentiu/contact-tracing-research/blob/main/GAEN.pdf}.

\bibitem{avitabile2020ProntoB2}
G.~Avitabile, V.~Botta, V.~Iovino, and I.~Visconti, ``Towards defeating mass
  surveillance and sars-cov-2: The pronto-c2 fully decentralized automatic
  contact tracing system,'' Cryptology ePrint Archive, Report 2020/493, 2020,
  \url{https://eprint.iacr.org/2020/493}.

\bibitem{white2021privacy}
L.~White and P.~van Basshuysen, ``Privacy versus public health? a reassessment
  of centralised and decentralised digital contact tracing,'' in \emph{Science
  and Engineering Ethics}, 2021,
  \url{https://doi.org/10.1007/s11948-021-00301-0}.

\bibitem{vaudenay2020dark}
S.~Vaudenay, ``The dark side of swisscovid,'' 2020,
  \url{https://lasec.epfl.ch/people/vaudenay/swisscovid.html}.

\bibitem{hoepman2021critique}
J.-H. Hoepman, ``A critique of the google apple exposure notification (gaen)
  framework,'' 2021.

\bibitem{reardon2021why}
J.~Reardon, ``Why google should stop logging contact-tracing data,''
  blog.appcensus.io, 2021,
  \url{https://blog.appcensus.io/2021/04/27/why-google-should-stop-logging-contact-tracing-data/}.

\bibitem{github2020immunirelay}
``Replay attack ``in the past'','' 2020,
  \url{https://github.com/immuni-app/immuni-app-android/issues/278}.

\bibitem{Gennaro2020}
R.~Gennaro, A.~Krellenstein, and J.~Krellenstein, ``Exposure notification
  system may allow for large-scale voter suppression,'' 2020,
  \url{https://static1.squarespace.com/static/5e937afbfd7a75746167b39c/t/5f47a87e58d3de0db3da91b2/1598531714869/Exposure_Notification.pdf}.

\bibitem{dehaye2020swisscovid}
P.-O. Dehaye and J.~Reardon, ``Swisscovid: a critical analysis of risk
  assessment by swiss authorities,'' 2020,
  \url{https://arxiv.org/abs/2006.10719}.

\bibitem{CoronaWarnAppGermany}
D.~Telekom and {SAP}, ``{Corona-Warn-App - The Official COVID-19 Exposure
  Notification App for Germany},'' \url{https://github.com/corona-warn-app}.

\bibitem{boyle2015function}
E.~Boyle, N.~Gilboa, and Y.~Ishai, ``Function secret sharing,'' in
  \emph{Advances in Cryptology - EUROCRYPT 2015}, E.~Oswald and M.~Fischlin,
  Eds.\hskip 1em plus 0.5em minus 0.4em\relax Berlin, Heidelberg: Springer
  Berlin Heidelberg, 2015, pp. 337--367.

\bibitem{sun2021empirical}
R.~Sun, W.~Wang, M.~Xue, G.~Tyson, S.~Camtepe, and D.~C. Ranasinghe, ``An
  empirical assessment of global covid-19 contact tracing applications,'' 2021.

\bibitem{Wen2020study}
H.~Wen, Q.~Zhao, Z.~Lin, D.~Xuan, and N.~Shroff, ``A study of the privacy of
  covid-19 contact tracing apps,'' in \emph{Security and Privacy in
  Communication Networks}, N.~Park, K.~Sun, S.~Foresti, K.~Butler, and
  N.~Saxena, Eds.\hskip 1em plus 0.5em minus 0.4em\relax Cham: Springer
  International Publishing, 2020, pp. 297--317.

\bibitem{ahmed2020survey}
N.~{Ahmed}, R.~A. {Michelin}, W.~{Xue}, S.~{Ruj}, R.~{Malaney}, S.~S.
  {Kanhere}, A.~{Seneviratne}, W.~{Hu}, H.~{Janicke}, and S.~K. {Jha}, ``A
  survey of covid-19 contact tracing apps,'' \emph{IEEE Access}, vol.~8, pp.
  134\,577--134\,601, 2020.

\bibitem{ahmed2021privacy}
S.~Ahmed, Y.~Xiao, C.~Fung, M.~Yung \emph{et~al.}, ``Privacy guarantees of ble
  contact tracing: A case study on covidwise,'' \emph{arXiv preprint
  arXiv:2111.08842}, 2021.

\bibitem{TousAntiCovid}
``Tousanticovid,'' Government of France, 2020,
  \url{https://solidarites-sante.gouv.fr/soins-et-maladies/maladies/maladies-infectieuses/coronavirus/tousanticovid}.

\bibitem{covidradar}
``Covidradar,'' covidradar.mx, 2020, \url{https://covidradar.mx/}.

\bibitem{virusradar}
``Virusradar,'' virusradar.hu, 2020, \url{https://virusradar.hu/}.

\bibitem{BlueZone}
``Bluezone,'' bluezone.gov.vn, 2020, \url{https://bluezone.gov.vn/}.

\bibitem{RakningC19}
``Rakning c-19,'' www.covid.is, 2020, \url{https://www.covid.is/app/en}.

\bibitem{safepaths-mit}
``Safe paths,'' safepaths.mit.edu, 2020, \url{https://safepaths.mit.edu/}.

\bibitem{Tawakkalna}
``Tawakkalna,'' ta.sdaia.gov.sa, 2020, \url{https://ta.sdaia.gov.sa/}.

\bibitem{ViruSafe}
``Virusafe,'' coronavirus.bg, 2020, \url{https://app.coronavirus.bg/}.

\bibitem{Shlonik}
``Shlonik,'' Kuwait Central Agency for Information TechnologyHealth \& Fitness,
  2020,
  \url{https://play.google.com/store/apps/details?id=com.healthcarekw.app&hl=en_US&gl=US}.

\bibitem{AarogyaSetuIndia}
``{Aarogya Setu Mobile App},'' Government of India,
  \url{https://www.mygov.in/aarogya-setu-app/}.

\bibitem{BeAwareBahrain}
``Beaware bahrain,'' iga.gov.bh, 2020, \url{https://bahrain.bh/}.

\bibitem{Ehteraz}
``Ehteraz,'' acta.gov.qa, 2020, \url{https://www.acta.gov.qa/en/ehteraz/}.

\bibitem{MorChana}
``Morchana,'' Digital Government Development Agency, Thailand, 2020,
  \url{https://www.dga.or.th/}.

\bibitem{PeduliLindungi}
``Pedulilindungi,'' Ministry of communication and informatics, Indonesia, 2020,
  \url{https://www.pedulilindungi.id/}.

\bibitem{ChinaCTApps}
``In coronavirus fight, china gives citizens a color code, with red flags,''
  nytimes.com, 2020,
  \url{https://www.nytimes.com/2020/03/01/business/china-coronavirus-surveillance.html}.

\bibitem{hoepman2021hansel}
J.-H. Hoepman, ``Hansel and gretel and the virus: Privacy conscious contact
  tracing,'' 2021.

\bibitem{buccafurri2020}
F.~Buccafurri, V.~De~Angelis, and C.~Labrini, ``A privacy-preserving solution
  for proximity tracing avoiding identifier exchanging,'' in \emph{2020
  International Conference on Cyberworlds (CW)}, 2020, pp. 235--242.

\bibitem{PACT-UW}
J.~Chan, D.~Foster, S.~Gollakota, E.~Horvitz, J.~Jaeger, S.~Kakade, T.~Kohno,
  J.~Langford, J.~Larson, P.~Sharma, S.~Singanamalla, J.~Sunshine, and
  S.~Tessaro, ``Pact: Privacy sensitive protocols and mechanisms for mobile
  contact tracing,'' 2020.

\bibitem{co100}
``South korea to step-up online coronavirus tracking,'' smartcitiesworld.net,
  2020,
  \url{https://www.smartcitiesworld.net/news/news/south-korea-to-step-up-online-coronavirus-tracking-5109
  }.

\bibitem{IsraelHamagen}
M.~o.~H. Israel~govement, ``Hamagen,'' 2020,
  \url{https://govextra.gov.il/ministry-of-health/hamagen-app/download-en/}.

\bibitem{tracecorona-website}
``Tracecorona: Anonymous decentralized contact tracing for pandemic response,''
  TraceCORONA, May 2020, \url{tracecorona.net}.

\bibitem{dittmer2020}
S.~Dittmer, Y.~Ishai, S.~Lu, R.~Ostrovsky, M.~Elsabagh, N.~Kiourtis,
  B.~Schulte, and A.~Stavrou, ``Function secret sharing for psi-ca: With
  applications to private contact tracing,'' Cryptology ePrint Archive, Report
  2020/1599, 2020, \url{https://eprint.iacr.org/2020/1599}.

\bibitem{Hasan2021blockchain}
H.~R. Hasan, K.~Salah, R.~Jayaraman, I.~Yaqoob, M.~Omar, and S.~Ellahham,
  ``Covid-19 contact tracing using blockchain,'' \emph{IEEE Access}, vol.~9,
  pp. 62\,956--62\,971, 2021.

\bibitem{berke2020assessing}
\BIBentryALTinterwordspacing
A.~Berke, M.~A. Bakker, P.~Vepakomma, R.~Raskar, K.~Larson, and A.~S. Pentland,
  ``Assessing disease exposure risk with location histories and protecting
  privacy: {A} cryptographic approach in response to {A} global pandemic,''
  \emph{CoRR}, vol. abs/2003.14412, 2020. [Online]. Available:
  \url{https://arxiv.org/abs/2003.14412}
\BIBentrySTDinterwordspacing

\bibitem{OpenMined2020maximizing}
``Maximizing privacy and effectiveness in covid-19 apps,'' OpenMined, 2020,
  \url{https://blog.openmined.org/covid-app-privacy-advice/}.

\bibitem{cheng2020KHOVID}
\BIBentryALTinterwordspacing
X.~Cheng, H.~Yang, A.~S. Krishnan, P.~Schaumont, and Y.~Yang, ``{KHOVID:}
  interoperable privacy preserving digital contact tracing,'' \emph{CoRR}, vol.
  abs/2012.09375, 2020. [Online]. Available:
  \url{https://arxiv.org/abs/2012.09375}
\BIBentrySTDinterwordspacing

\bibitem{Salesforce2020Private}
Salesforce, ``Track employee health-related interactions safely and securely,''
  2020, \url{https://www.salesforce.com/products/contact-tracing/overview/}.

\bibitem{IBM2020Private}
IBM, ``Watson works: A safer way to return to the workplace,'' 2020,
  \url{https://www.ibm.com/watson/watson-works?lnk=hpmcov_bs&lnk2=learn}.

\bibitem{ourPamphlet}
A.~Boutet, C.~Castelluccia, M.~Cunche, A.~Dmitrienko, V.~Iovino, M.~Miettinen,
  T.~D. Nguyen, V.~Roca, A.-R. Sadeghi, S.~Vaudenay, I.~Visconti, , and
  M.~Vuagnoux, ``Contact tracing by giant data collectors: Opening pandora’s
  box of threats to privacy, sovereignty and national security,'' 2020,
  \url{https://tracecorona.net/wp-content/uploads/2020/12/Digital_Contact_Tracing.pdf}.

\bibitem{crocker2020challenge}
\BIBentryALTinterwordspacing
A.~Crocker, K.~Opsahl, and B.~Cyphers, ``{The Challenge of Proximity Apps For
  COVID-19 Contact Tracing},'' 2020. [Online]. Available:
  \url{https://www.eff.org/de/deeplinks/2020/04/challenge-proximity-apps-covid-19-contact-tracing}
\BIBentrySTDinterwordspacing

\bibitem{danz2020security}
N.~Danz, O.~Derwisch, A.~Lehmann, W.~Puenter, M.~Stolle, and J.~Ziemann,
  ``Security and privacy of decentralized cryptographic contact tracing,''
  Cryptology ePrint Archive, Report 2020/1309, 2020,
  \url{https://eprint.iacr.org/2020/1309}.

\bibitem{kojaku2020effectiveness}
S.~Kojaku, L.~Hébert-Dufresne, E.~Mones, S.~Lehmann, and Y.-Y. Ahn, ``The
  effectiveness of backward contact tracing in networks,'' 2020.

\bibitem{leith2020coronavirus}
D.~J. Leith and S.~Farrell, ``Coronavirus contact tracing: Evaluating the
  potential of using bluetooth received signal strength for proximity
  detection,'' 2020.

\bibitem{DelValle2007}
\BIBentryALTinterwordspacing
S.~{Del Valle}, J.~Hyman, H.~Hethcote, and S.~Eubank, ``Mixing patterns between
  age groups in social networks,'' \emph{Social Networks}, vol.~29, no.~4, pp.
  539--554, 2007. [Online]. Available:
  \url{https://www.sciencedirect.com/science/article/pii/S0378873307000330}
\BIBentrySTDinterwordspacing

\end{thebibliography}

\section*{Biography}
\noindent\textbf{Thien Duc Nguyen} is a research assistant and a PhD candidate at the System Security Lab of
Technical University of Darmstadt (TU Darmstadt), Germany. He is currently working in various topics related to machine learning-based security mechanisms for IoT, security for federated machine learning, contextual security and digital contact tracing.\\

\noindent\textbf{Markus Miettinen} is a postdoctoral researcher at
the System Security Lab of TU Darmstadt, Germany. He graduated as Dr.-Ing. from TU Darmstadt in 2018. Before joining academia, he acquired more than a decade of professional experience in industrial research at the Nokia Research Center in Helsinki,
Finland and Lausanne, Switzerland. His research interests lie in utilizing machine learning methods for realizing autonomous security solutions for IoT and mobile computing environments.\\

\noindent\textbf{Alexandra Dmitrienko} is an associate professor and the head of the Secure Software Systems group at the University of Wuerzburg in Germany. She holds a PhD degree in Security and Information Technology from TU Darmstadt (2015). She received numerous awards for her research, including Intel Doctoral Student Honor Award (2013) and ERCIM STM WG  Award (2016). Her research interests focus on secure software engineering, systems security and privacy, and security and privacy of mobile, cyber-physical, and distributed systems.\\

\noindent\textbf{Ahmad-Reza Sadeghi} is a full professor for Computer Science at TU Darmstadt,
where he directs the System Security Lab, the Intel Private AI Center, and Open S3 Lab. He received his PhD from the University of Saarland. He has served on the editorial board of ACM TISSEC and as Editor-in-Chief for IEEE Security and Privacy Magazine. \\

\noindent\textbf{Ivan Visconti} is a full professor of Computer Science in the Computer and Electrical Engineering and Applied Mathematics Department of the University of Salerno.
His research interests focus mainly on designing provably secure and privacy-preserving cryptographic protocols and securing blockchains and their applications.
He served for two years as Senior Area Editor for the journal IEEE Transactions on Information Forensics and Security. Several of his results have been published in the most competitive conferences in cryptography (i.e., CRYPTO, EUROCRYPT, TCC).

\ifarXiv
    \appendix
\subsection{List of Country Apps}
\label{sec:list-country-app}

\begin{table*}[ht]
\centering
\caption{List of centralized contact tracing approaches and their adoption. *) Contacts are anonymous in France and UK systems; **) Not in France and UK systems; ***) Location can be any source used to locate a user, e.g., GPS coordinates captured by the user's phones or the address of a shop that the user has made a credit card transaction. Personal data can be name, address, gender, or age, etc.
}
\label{tab:ct-centralized}
\begin{tabular}{llllll}
\multicolumn{1}{c}{Approach}                                                                                               & \multicolumn{1}{c}{Tech}                                                                        & \multicolumn{1}{c}{Notification}                                                                       & \multicolumn{1}{c}{\begin{tabular}[c]{@{}c@{}}Data Collected\\ by Server\end{tabular}}                                                                                          & \multicolumn{1}{c}{\begin{tabular}[c]{@{}c@{}}Data Collected\\ by Client\end{tabular}}                                           & \multicolumn{1}{c}{Adoption}                                                                                                                                                                                          \\ \hline
\begin{tabular}[c]{@{}l@{}}BlueTrace \cite{BlueTrace},\\ PEPP-PT \cite{pepppt},\\ TousAntiCovid \cite{TousAntiCovid},\\ CovidRada~\cite{covidradar},\\ E7mi,\\ VirusRadar \cite{virusradar},\\ BlueZone~\cite{BlueZone}\end{tabular} & Bluetooth                                                                                        & \begin{tabular}[c]{@{}l@{}}Server via App/SMS, \\ or the Health Authority \\ via telephone\end{tabular} & \begin{tabular}[c]{@{}l@{}}Users' Temporary \\ identifiers (TempIDs),\\ Who encountered infected users*,\\ Phone number**,\\ Personal data (e.g., \\ name or age)**\end{tabular} & \begin{tabular}[c]{@{}l@{}}TempIDs of the\\ encountered users,\\ Phone number,\\ Personal data (e.g.,\\ name or age)\end{tabular} & \begin{tabular}[c]{@{}l@{}}Australia, France, Singapore, \\ UK (1st version), Czech \\ Republic, Fiji, Gibraltar, \\ Hungary, Malaysia, Mexico, \\ North Macedonia, Philippines, \\ Tunisia, UAE, Vietnam\end{tabular} \\ \hline
\begin{tabular}[c]{@{}l@{}}Rakning C-19~\cite{RakningC19},\\ SafePaths(MIT)~\cite{safepaths-mit}\\ Tawakkalna ~\cite{Tawakkalna},\\ ViruSafe~\cite{ViruSafe},\\ Shlonik~\cite{Shlonik}\end{tabular}                   & Location***                                                                                      & \begin{tabular}[c]{@{}l@{}}Server via App/SMS, \\ or the Health Authority \\ via telephone\end{tabular} & \begin{tabular}[c]{@{}l@{}}Location of all users,\\ Who encountered whom,\\ Phone number,\\ Personal data (e.g., \\ name, age)\end{tabular}                                      & \begin{tabular}[c]{@{}l@{}}Location,\\ Phone number,\\ Personal data (e.g.,\\ name, age)\end{tabular}                             & \begin{tabular}[c]{@{}l@{}}Bulgaria, Cyprus, \\ Kuwait, Saudi Arabia; \\ North Dakota, South \\ Dakota, Wyoming (US)\end{tabular}                                                                                      \\ \hline
\begin{tabular}[c]{@{}l@{}}Aarogya Setu~\cite{AarogyaSetuIndia},\\ BeAware~\cite{BeAwareBahrain},\\ Ehteraz~\cite{Ehteraz},\\ MorChana~\cite{MorChana},\\ PeduliLindungi~\cite{PeduliLindungi} \end{tabular}                    & \begin{tabular}[c]{@{}l@{}}Bluetooth \\ and\\ location\end{tabular}                              & \begin{tabular}[c]{@{}l@{}}Server via App/SMS, \\ or the Health Authority \\ via telephone\end{tabular} & \begin{tabular}[c]{@{}l@{}}Location of all users,\\ Who encountered whom,\\ Phone number,\\ Personal data (e.g., \\ name, age, gender, \\ occupation)\end{tabular}               & \begin{tabular}[c]{@{}l@{}}Location,\\ Phone number,\\ Personal data (e.g.,\\ name, age, gender, \\ occupation\end{tabular}       & \begin{tabular}[c]{@{}l@{}}Bahrain, Bangladesh, India, \\ Indonesia, Qatar, Thailand, \\ Turkey; Rhode Island (US)\end{tabular}                                                                                        \\ \hline
\begin{tabular}[c]{@{}l@{}}Chinese health \\ code system\\ e.g., Hangzhou~\cite{ChinaCTApps}\end{tabular}                                       & \begin{tabular}[c]{@{}l@{}}Combination\\ (Location, \\ credit card \\ transactions)\end{tabular} & \begin{tabular}[c]{@{}l@{}}Server via App/SMS, \\ or the Health Authority \\ via telephone\end{tabular} & \begin{tabular}[c]{@{}l@{}}Location of all users,\\ Who encountered whom\end{tabular}                                                                                            & \begin{tabular}[c]{@{}l@{}}Location,\\ Phone number\end{tabular}                                                                  & China                                                                                                                                                                                                                  \\ \hline
Janus~\cite{istomin2021janus}                                                                                                                        & \begin{tabular}[c]{@{}l@{}}Bluetooth \\ and UWB\end{tabular}                                     & NA                                                                                                      & NA                                                                                                                                                                               & NA                                                                                                                                & NA                                                                                                                                                                                                                     \\ \hline
Hoepman et at.~\cite{hoepman2021hansel}                                                                                                               & Bluetooth                                                                                        & Server via App                                                                                          & TempIDs                                                                                                                                                                          & Encrypted TempIDs                                                                                                                 & NA                                                                                                                                                                                                                     \\ \hline

Buccafurri et at.~\cite{buccafurri2020}                                                                                                                        & \begin{tabular}[c]{@{}l@{}}Location and\\ Bluetooth\end{tabular}                                     & NA                                                                                                      & NA                                                                                                                                                                               & NA                                                                                                                                & NA                                                               \\ \hline                                                                                                                                                

\end{tabular}
\end{table*}

\begin{table*}[htb]
\centering
\caption{List of decentralized contact tracing approaches and adoption.
}
\label{tab:ct-decentralized}

\begin{tabular}{llllll}
\hline
\multicolumn{1}{c}{Approach}                                                 & \multicolumn{1}{c}{Tech} & \multicolumn{1}{c}{Notification}                                    & \multicolumn{1}{c}{\begin{tabular}[c]{@{}c@{}}Data Collected\\ by Server\end{tabular}}          & \multicolumn{1}{c}{\begin{tabular}[c]{@{}c@{}}Data Collected\\ by Client\end{tabular}}                       & \multicolumn{1}{c}{Adoption}                                                                                                                                                                                                                                 \\ \hline
\begin{tabular}[c]{@{}l@{}}GAEN~\cite{GAEN:crypto},\\ DP3T-1~\cite{DP3T:WhitePaper}\end{tabular}                         & Bluetooth                 & \begin{tabular}[c]{@{}l@{}}The App notifies \\ the User\end{tabular} & \begin{tabular}[c]{@{}l@{}}Temporary exposure\\ keys (TEKs) of \\ infected users\end{tabular}    & \begin{tabular}[c]{@{}l@{}}TEKs of \\ infected users,\\ TempIDs of the\\ encounter users\end{tabular}         & \begin{tabular}[c]{@{}l@{}}Belgium, Canada, Denmark, \\ Estonia, Finland, Germany, \\ Ireland, Italy, Japan, \\ New Zealand, Northern \\ Ireland, Norway, Poland, \\ Saudi Arabia, South Africa, \\ Switzerland, UK, \\ and 23 states in the USA\end{tabular} \\ \hline
\begin{tabular}[c]{@{}l@{}}DP3T-2~\cite{DP3T:WhitePaper}, \\ MIT-PACT~\cite{PACT-MIT},\\ UW-PACT~\cite{PACT-UW} \end{tabular}                  & Bluetooth                 & \begin{tabular}[c]{@{}l@{}}The App notifies \\ the User\end{tabular} & \begin{tabular}[c]{@{}l@{}}Temporary identifiers\\ (TempIDs) of\\ infected users\end{tabular}    & \begin{tabular}[c]{@{}l@{}}TempIDs of \\ infected users,\\ TempIDs of the \\ encountered users\end{tabular}   & NA                                                                                                                                                                                                                                     \\ \hline
Pronto-B2~\cite{avitabile2020ProntoB2}                                                                      & Bluetooth                 & \begin{tabular}[c]{@{}l@{}} the App notifies \\ the User\end{tabular} & \begin{tabular}[c]{@{}l@{}} Hashed pairs of \\ TempIDs in\\ encounters with \\ infected users\end{tabular}    & \begin{tabular}[c]{@{}l@{}}TempIDs of \\ infected users,\\ TempIDs of the \\ encountered users\end{tabular}   & NA                                                                                                                                                                                                                                                            \\ \hline
\begin{tabular}[c]{@{}l@{}}Co100~\cite{co100},\\ HaMagen~\cite{IsraelHamagen}\end{tabular}                       & Location                  & \begin{tabular}[c]{@{}l@{}}The App notifies \\ the User\end{tabular} & \begin{tabular}[c]{@{}l@{}}Location of \\ infected users\end{tabular}                            & Location                                                                                                      & South Korea, Israel                                                                                                                                                                                                                                           \\ \hline
\begin{tabular}[c]{@{}l@{}}TraceCORONA ~\cite{tracecorona-website},\\ Pronto-C2~\cite{avitabile2020ProntoC2},\\ CleverParrot~\cite{canetti2020cleverParrot}\end{tabular} & Bluetooth                 & \begin{tabular}[c]{@{}l@{}}The App notifies \\ the User\end{tabular} & \begin{tabular}[c]{@{}l@{}}Encrypted encounter \\ tokens (ETs) of \\ infected users\end{tabular} & \begin{tabular}[c]{@{}l@{}}Encrypted ETs of \\ infected users,\\ ETs of the \\ encountered users\end{tabular} & NA                                                                                                                                                                                                                                                            \\ \hline
Epione~\cite{trieu2020epione}, \\ Dittmer et al. \cite{dittmer2020}                                                                         & Bluetooth                 & \begin{tabular}[c]{@{}l@{}}The App notifies \\ the User\end{tabular} & TempIDs of infected users                                                                        & \begin{tabular}[c]{@{}l@{}}TempIDs of the\\ encountered users\end{tabular}                                    & NA                                                                                                                                                                                                                                                            \\ \hline
\end{tabular}
\end{table*}

\noindent\textbf{List of centralized approaches.}
\label{sec:list-centralized-app}
Table~\ref{tab:ct-centralized} shows an overview of prominent centralized contact tracing solutions. \changeT{In the table, the apps are categorized by employed technology, e.g., notification method and type of data collected. Further, adoption information is also provided, from which one can see that centralized-based approaches are more preferred in Asian countries than on other continents. In terms of technologies, besides Bluetooth, location is also widely used in many countries. Noticeably, many if not most deployed apps, e.g., \cite{BlueTrace, covidradar, AarogyaSetuIndia}  collect sensitive data, e.g., phone numbers, names, ages of the users.}  

It is worth noting that some schemes have been proposed to improve the security and privacy of centralized approaches \cite{hoepman2021hansel, istomin2021janus}. For example, Hoepman et al. \cite{hoepman2021hansel} propose two centralized protocols that aim to reduce the risk of tracking users’ locations and replay attacks. The first handshake (peer-to-peer) protocol establishes logs of encounters. In particular,  $\useri$ broadcasts a public key, and $\userj$ in the vicinity will respond with $\userj$’s ID encrypted by $\useri$’s public key and vice-versa.  However, this peer-to-peer protocol fails if either of the two-way messages is lost. Therefore, Hoepman et al. introduce a second approach that utilizes a central server to tackle this issue so that devices only need to broadcast a random public key. The responses with encrypted ID information will then be sent via the server, i.e., do not need to be sent directly via the BLE channel.

\noindent\textbf{List of decentralized approaches.}
\label{sec:list-decentralized-app}
Table~\ref{tab:ct-decentralized} shows an overview of prominent solutions. \changeT{It shows that most approaches use Bluetooth, some use location while a very few approaches using other technologies, e.g., \cite{istomin2021janus} using UWB and \cite{MIT-SonicPACT} using ultrasound.  
Further, the figure shows that \gap is widely used in Western Europe and North America while South Korea and Israel use their own location-based approaches. Unfortunately, other decentralized approaches that claim to have better security and privacy guarantees in comparison to \gap, e.g., Pronto-C2\cite{avitabile2020ProntoC2}, ClerverParrot\cite{canetti2020cleverParrot}, and Epione \cite{trieu2020epione} have not been adopted yet. In terms of data collection, decentralized apps do not collect personal data of users like phone numbers, email addresses or names. However,  the system (consisting of $\service$ and Apps) collects encounter information of infected users and additionally, the apps also record information about encounters with other users in the vicinity. Such encounter information can be generally divided into two categories based on which cryptographic approach is used. Apps based on the use of symmetric cryptography collect temporary IDs ($\tempId$s) of other users and match these against IDs that they derive from the temporary exposure keys ($\tek$s) of all infected users they download from $\service$. In contrast, apps based on asymmetric cryptography establish encounter-specific cryptographic tokens for each encounter. To enable identification of at-risk encounters, infected users upload hashed values of their encounter tokens to $\service$, from where all other users download them and compare them to the hashed versions of their collected encounter tokens. In case there is a match, this is an indication that an at-risk encounter with an infected person has taken place.}

\subsection{Advanced Privacy Techniques}
\label{sec:privacy-tech}

There are several (mainly cryptography-based) privacy techniques that have been considered to enhance privacy of \dct. Table \ref{tab:advanced-pri-tech} shows an overview of such approaches. Unfortunately, none of these contact tracing approaches have been used in practice yet. Next, we will elaborate on these approaches. 

\noindent\textbf{Blind Signature.} Some approaches, e.g., \cite{reichert2021ovid, avitabile2020ProntoC2} propose using a blind signature scheme to verify the authenticity of encounter information, e.g., the temporary keys that infected users upload to the tracing service provider $\service$. In contrast to common approaches in which $\healthauthority$ gives a unique transaction number (TAN) to each of the infected users for uploading their data, allowing $\healthauthority$ to link infected users in a \dct to $\healthauthority$'s patient data, blind signature-based approaches prevent this by enabling $\healthauthority$ to "blindly" sign the temporary keys of infected users without knowing the actual content. Therefore, neither $\healthauthority$ nor any other party can know based on verification information to which user of the system published temporary keys belong.

\noindent\textbf{Blockchain.} Instead of using a centralized server to collect and forward encounter information of infected users to other users, some approaches by, e.g., Avitabile et al. \cite{avitabile2020ProntoC2} or Hasan et al. \cite{ Hasan2021blockchain} leverage a blockchain to decentralize the process of publishing, verifying, and matching encounter information. This makes the system transparent to users in that it does not require a trusted server.

\noindent\textbf{Private Set Intersection (PSI) and binary filters.} The core function of a \dct app is to find a match between encounter information of infected users and other users. However, this process raises privacy concerns.
In centralized approaches, $\service$ has access to encounter information of all infected users, whereas in decentralized approaches, any user can receive information about the encounters of all infected users, providing the potential for possibly inferring information about the user's location and social graph. To solve this problem, Trieu et al.~\cite{trieu2020epione} and Dittmer et al.~\cite{dittmer2020} leverage a private set intersection (PSI) approach utilizing Function Secret Sharing (FSS) techniques \cite{boyle2015function}.
The idea is that $\service$ (which has a set of $\tempId$s of all infected users) and a potentially affected user $\userj$ (who has a set of $\tempId$s of the users whom $\userj$ has encountered) collaborate to perform encounter matching in a way that only the matching results (e.g., the number of matching $\tempId$s) become known to $\userj$. 
This means that $\service$ does not get to know encounter information of $\userj$ (e.g., $\tempId$s) and vice versa.
However, a PSI-based approach often incurs high computation and communication overhead which makes it less practical. A more simple approach is to use a  cuckoo filter~\cite{DP3T:WhitePaper}, where the $\service$ builds a filter from the $\tempId$s of the infected users and sends it to user Apps. Apps can check for matches by inputting each of their observed $\tempId$s into the filter and receive an output whether the $\tempId$ is in the filter or not. Thus, Apps can do matching without having access to the complete list of $\tempId$s of infected users.

\noindent\textbf{Secret Sharing.} In order to prevent $\tempId$s to be captured or relayed easily, Troncoso et al.~\cite{DP3T:WhitePaper} propose to use secret sharing by dividing a $\tempId$ into $n$ parts, so-called secret shares that are broadcast into the user's vicinity over time. This means that any other user needs to remain in vicinity of the user to capture at least $k$ out of $n$ shares to be able to reconstruct the $\tempId$, making it more difficult for an adversary to stage successful attacks by just relaying $\tempId$s of people passing by them. \fixedMC{However, this approach has several limitations and drawbacks. Firstly, this approach is ineffective in main attack scenarios like shopping, public events or restaurants where adversary has enough time to capture at least $k$ shares. Secondly, this approach would introduce a significant overhead, latency, and error rate because it requires from $k$ to $n$ times more communication overhead (compared to BlueTrace or \gap) and significant computation to reconstruct the $\tempId$ from the shares. Further, the authors argue that the communication overhead is insignificant if the number of shares is equal to the number of broadcasts that the App makes within an epoch (the lifetime of a $\tempId$). However, this is a flaw as many encounters would not be recorded if the encounters do start at the beginning of the epoch.}

\noindent\textbf{Use of Anonymization Networks.} A malicious $\service$ may track IP addresses of users uploading and downloading encounter information, which potentially could (1) leak personal information of the user, e.g., home addresses and (2) be used to link users with encounter information they upload. Therefore, a number of existing approaches (e.g., \cite{castelluccia2020desire}) propose to use an anonymization network like Tor or a Mixnet to prevent such potential leakage.

\begin{table}[ht]
    \centering
    \caption{Advanced privacy techniques for DCT systems}
    \begin{tabular}{l|l}
    Techniques & References \\ \hline
    Blind signature & \cite{reichert2021ovid, avitabile2020ProntoC2} \\
    Blockchain & \cite{avitabile2020ProntoC2, Hasan2021blockchain} \\
    Private set intersection/ a filter         &  \cite{trieu2020epione, dittmer2020, berke2020assessing, OpenMined2020maximizing, DP3T:WhitePaper}\\
    Secret sharing &  \cite{DP3T:WhitePaper} \\ 
    Tor/Mixnet & \cite{castelluccia2020desire} \\
    Others & \cite{cheng2020KHOVID}\\ \hline
    \end{tabular}
    
    \label{tab:advanced-pri-tech}
\end{table}

\subsection{Other Application Scenarios of \ourname}
\label{sec:discussion}

\noindent\textbf{Smartphone and Wearable Devices.}
\label{sec:wearable}
Smartphones are prohibited or inconvenient to use in many scenarios like in schools, hospitals, corporate offices, sports and other events, beaches, waterparks, funfairs, etc. Therefore, we propose using wearable devices as a complementary approach for contact tracing. Figure \ref{fig:wristband} shows our smartband-based \ourname. The Bluetooth tracing function is integrated in the smartbands to emit and record the Ephemeral Public Keys $Q$s and ephemeral IDs which are generated by the \app{s} on the smartphones. The smartphones are also responsible for calculating ETs $k$s from private key $d$s and public key $Q$s collected by the smartbands. The protocol is summarized as follows:

\begin{enumerate}
	\item The smartphones generate ECDH private key $d$s and public $Q$s, and ephemeral IDs $EI$s for the next day or the next few days.
	\item The smartphones send $Q$s and $EI$s to the smartbands every day or every few days or when users interact with the apps. 
	\item The smartbands exchangse $Q$s and $EI$s with other devices (smartbands or smartphone tracing apps) in vicinity.
	\item The smartbands send $Q'$s and $EI'$s that they have received from other devices along with associated metadata (timestamp and duration) to the smartphones.
	\item The smartphones calculate the ETs from their own private keys $d$s and other public keys $Q'$s.
	\item The smartphones perform all other phases: infection verification, token information upload and token information download.	
\end{enumerate}  

In order to avoid double encounter recording as both the phone and its pair smartband exchange public keys to other devices in vicinity, the system can detect the co-presence of two devices using a separate secure Bluetooth channel (since both the phone and the smartband are paired) and let only one of the devices perform tracing. In case the phone and its pair smartband are our of range, the smartband will run the tracing functionality but the phone can pop-up a message to ask user whether they prefer do tracing on the phone or not, e.g., if the smartband is running out of battery.

\begin{figure}[ht]
	\centering
	\includegraphics[width=\columnwidth]{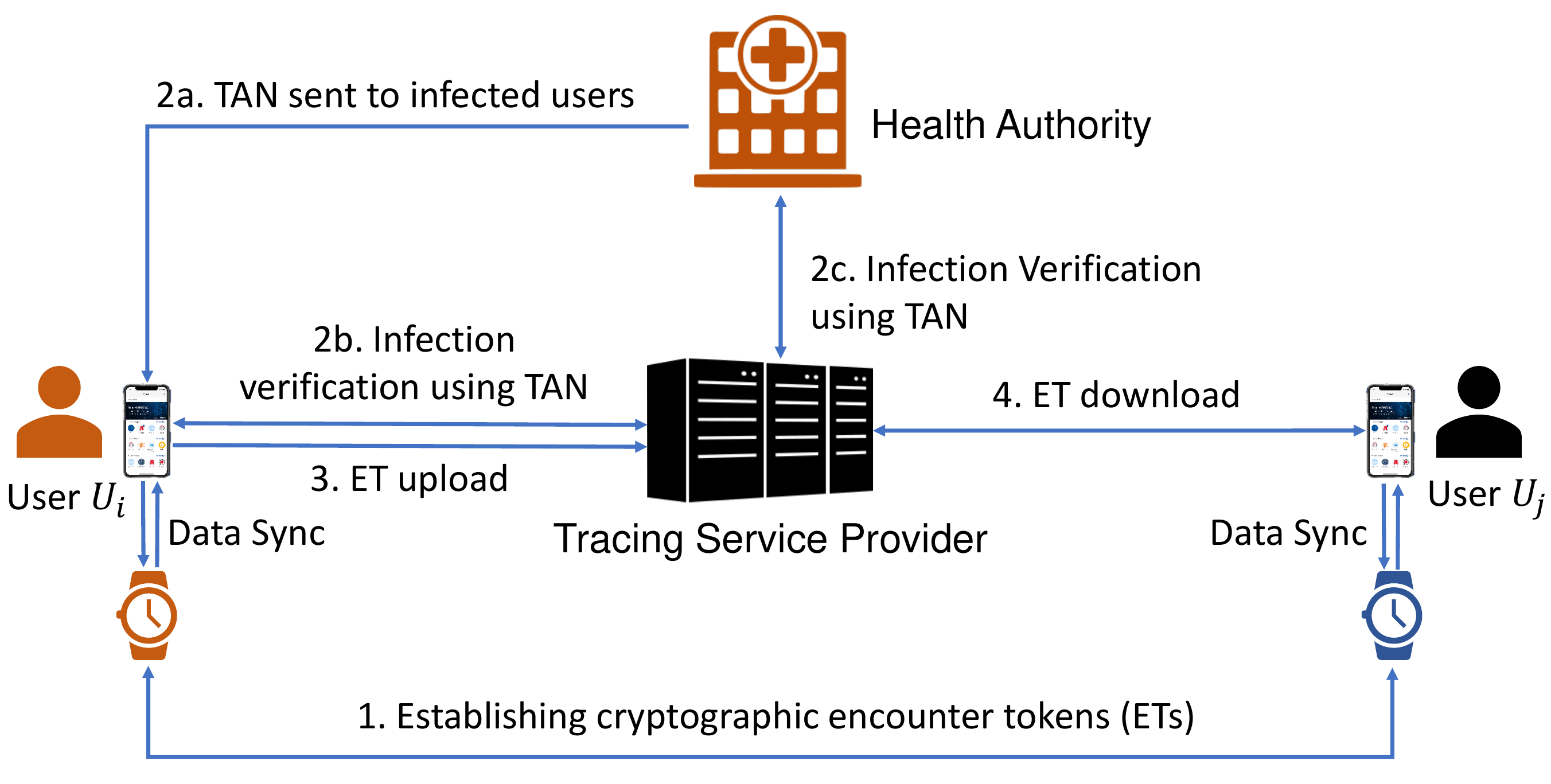}
	\caption{TraceCORONA system using wristband.}
	\label{fig:wristband}
\end{figure}

The wearable device and the smartphone frequently synchronize their data, so that the smartphone sends the $Q$s that it generates to the wearable device and the wearable device sends the $Q'$s it has received from other devices in proximity along with metadata associated with them to the smartphone. This synchronization function can be executed periodically, e.g., daily or when the phone is being charged or when the user interacts with the \ourname app, i.e., when the app is running in the foreground. For one day, the smartphone sends 96 $EI$s and 96 $EPK$s ( 24 hours x 4 EPKs/hour if $T = 15$ minutes). In total, the phone sends 96x(65+520) = 56,064 (bits) or 58 kbits. The amount of data that the wearable device sends to the phone depends on how much encounters (contacts) a user has during the day. Recent research shows that on average, a user has about 20 contacts (with a duration over 15 minutes) with other users excluding household members. Therefore, we estimate that the wearable device sends ca. 31 kbits to the phone daily. In theory, Bluetooth LE can transfer from 125 kbit/s to 2 Mbit/s meaning that the daily synchronization process can be done in seconds.

\noindent\textbf{Proximity Communication.} In this paper, we focus on Bluetooth Low Energy (Bluetooth LE) as a prominent example. However, \ourname can work with any type of short-range wireless communication protocols like ZigBee or Z-Wave. Further, other communication protocols or data transfer channels like NFC (Near Field Communication), ultrasound or QR code can also be used.  
   
\noindent\textbf{Public and Private Contact Tracing Systems.}
Many organizations like schools or corporations have needs for performing contact tracing that do not necessarily align well with the needs and requirements of a national or country-level tracing solution \cite{Salesforce2020Private, IBM2020Private}. Many businesses and organizations would like to implement their own (private) tracing solutions to have the flexibility of better managing quarantining of employees. To tackle this problem, some manual private contact tracing solutions, e.g., \cite{Salesforce2020Private} have been introduced. \ourname is flexible in the sense that it can be applied to private contact tracing directly. For example, a \ourname service provider can provide infrastructure (e.g., the server, the apps and dashboard control) to organizations or corporations for a private deployment. Since administrators like personnel departments know who in their organizations or companies are tested positive for COVID-19, they can issue $TAN$s and verify infection state as explained in Sect. \ref{sec:verification-upload}. Further, our smartband tracing solution can also help to cover many use cases in which using smartphones is forbidden e.g., in many schools or companies as mentioned above.

\begin{table*}[ht]
\centering
\caption{Attacks and analyses on existing contact tracing approaches. IDU: Identifying Users, PIU: Profiling Infected Users, ISC: Inferring Social Graph, FEC: Fake exposure Claim, RA: Relay Attack, DCT: Digital Contact Tracing.}
\label{tab:existing-attacks-analyses}

\begin{tabular}{|l|c|c|c|c|c|c|l|l|}
\hline
\multicolumn{1}{|c|}{}                         & IDU & PIU & ISC & FEC & RA & Effectiveness & \multicolumn{1}{c|}{Other}                                    & \multicolumn{1}{c|}{Target}                                              \\ \hline
Avitabile et al., \cite{avitabile2020ProntoC2}   & x   & x   & x   &     & X  &               &                                                               & \begin{tabular}[c]{@{}l@{}}GAEN\\ DP3T-1\\ PEPP-PT\\ ROBERT\end{tabular} \\ \hline
Avitabile et al., \cite{avitabile2020tenku}    &     &     &     &     & X  &               &                                                               & GAEN                                                                     \\ \hline
Baumgaertne et al., \cite{baumgartner2020mind} &     & x   & x   &     & x  &               &                                                               & GAEN                                                                     \\ \hline
Boutet et al., \cite{ourPamphlet}              & x   & x   & x   & x   & x  &               &                                                               & GAEN                                                                     \\ \hline
Crocker et al., \cite{crocker2020challenge}    & x   & x   & x   &     & x  & x             &                                                               & GAEN                                                                     \\ \hline
Danz et al., \cite{danz2020security}           & x   & x   & x   &     & x  &               &                                                               & \begin{tabular}[c]{@{}l@{}}GAEN\\ DP3T-1\end{tabular}                    \\ \hline
Dehaye et al., \cite{dehaye2020swisscovid}     &     &     &     &     & x  &               &                                                               & GAEN                                                                     \\ \hline
Gennaro et at., \cite{Gennaro2020}             &     &     &     &     & x  &               &                                                               & GAEN                                                                     \\ \hline
Gvili et al., \cite{gvili2020security}         & x   & x   & x   & x   & x  &               & DoS*                                                          & GAEN                                                                     \\ \hline
Iovino et al., \cite{Iovino2020timetravel}     &     &     &     &     & x  &               &                                                               & GAEN                                                                     \\ \hline
Kojaku et al., \cite{kojaku2020effectiveness}  &     &     &     &     &    & x             &                                                               & GAEN                                                                     \\ \hline
Lanzing et al., \cite{lanzing2020ethical}      &     & x   & x   &     &    &               & \begin{tabular}[c]{@{}l@{}}Monopolist\\ Coercion\end{tabular} & GAEN                                                                     \\ \hline
Leith et al., \cite{leith2020coronavirus}      &     &     &     &     &    & x             &                                                               & Bluetooth                                                                \\ \hline
Leith et al., \cite{leith2020gaendata}         & x   & x   & x   &     &    &               & Data collection                                               & GAEN                                                                     \\ \hline
Vaudenay et al., \cite{vaudenay2020analysis}   &     & x   &     &     & x  &               &                                                               & \begin{tabular}[c]{@{}l@{}}GAEN\\ DP3T\end{tabular}                      \\ \hline
Vaudenay et al., \cite{vaudenay2020dilemma}    & x   & x   & x   & x   & x  &               &                                                               & \begin{tabular}[c]{@{}l@{}}GAEN\\ DP3T\\ PEPP-PT\end{tabular}            \\ \hline
Wen et al., \cite{Wen2020study}                &     &     &     &     &    & x             & Data collection                                               & 41 apps                                                                  \\ \hline
White et al., \cite{White2021why}              & x   & x   & x   &     &    & x             & \multicolumn{1}{c|}{}                                         & DCT                                                                      \\ \hline

White et al., \cite{white2021privacy}                & x     & x     &  x   &     &    & x             & Ethics                                               &  \dct                                                                 \\ \hline
\end{tabular}
\end{table*}

\subsection{Parameter Settings and Estimations}
\label{sec:parameters}
We now analyze the data use of the \ourname protocols and define following parameters for our analysis.
\begin{itemize}
	\item $EI$ - Ephemeral Identification (128 bits). $EI$ is used to temporarily identify devices in proximity.
	item $Q$ - Ephemeral ECDH Public Key (384 bits). $Q$ is used together with associated secret keys to establish encounter tokens. 
	\item $UUID$ - Universally Unique Identifier (128 bits). $UUID$ is used to identify the \ourname app.
	\item $ET$ - Encounter Token (256 bits). $ET$ is used to uniquely identify the encounter between two users.		
	\item $AM$ - Advertising Message (256 bits). $AM$ is used to advertise (broadcast) $UUID$s and $EI$s of devices. It changes every $T$ minutes.
	\item $RSSI$ - Received Signal Strength Indicator.	$RSSI$ indicates the strength of a received Bluetooth signal that can be roughly used to estimate the distances between the sender and the receiver.
\end{itemize}

In the following, we present estimation of data exchange (network bandwidth) and key parameters. 

\noindent\textbf{Proximity Detection.} Each \ourname App constantly advertises and scans for $AM$ messages. We aim to make sure that a device records other devices in proximity every 2 minutes. The app periodically advertises $AM$s  every minute, during which devices run Bluetooth LE advertising for 40 seconds and are in idle mode for 20 seconds. The app periodically scans $AM$s every 50 seconds, during which devices run Bluetooth LE scanning for 30 seconds and are in idle mode for 20 seconds. These advertising and scanning patterns are empirically selected to ensure that \ourname keeps track of other devices every 2 minutes.   

\noindent\textbf{Public Key Exchange.} When a device finds a new device in proximity, the two devices start exchanging public key $Q$. First, one device starts the Bluetooth client and another device starts the Bluetooth server and both devices exchange their public keys $Q$s (384-bit length). Both devices store the $Q$s which later on are used to calculate $ET$s. $Q$ is changed every $T$ minutes along with $EI$. During a $T$-minute lifetime, to save energy, $Q$ is only sent when the device finds a new device.

\noindent\textbf{Encounter distance - 2 meters.}
In theory, the Bluetooth Low Energy (BLE) signal range is up to 50 meters. In the context of contact tracing, we consider the distance of two meters or smaller.

\noindent\textbf{Encounter time – 15 minutes.}
The encounter time that is equal to or more than 15 minutes should be considered as a high exposure risk. 

\noindent\textbf{Expected number of Encounter Tokens per day - 20.}
In average, a user has approximately 16 encounters (15 minutes or longer) with other users excluding known contacts e.g., household members or colleagues in the same office \cite{DelValle2007}. 

\noindent\textbf{Amount of data uploaded by the app to the server – 4.3 kB.}
The app of the infected user uploads the hashes (128 bits) of $ET$s for 14 days (20 $ET$s per day as mentioned above). Hence, the total amount of data should be, 14*20*128 = 35,840 (bits) or 4.3 (kB).

\noindent\textbf{Amount of data downloaded by the app from the server – 8.6 MB per day.}
If we assume there are 10,000 new COVID-19 cases per day, the app will download 10,000* 4.3 = 43,000 (kB) or 43 MB per day.

\noindent\textbf{Amount of data received by the server (daily) – 43 MB per day.}
The same to the data that an app downloaded.

\noindent\textbf{Amount of data sent by the server (daily) – 2150 TB per day.}
If we assume that there are 50 million app users, the server will send 50,000,000* 43 = 2,150,000,000 (MB) or 2150 TB per day.

\noindent\textbf{Maximum numbers of devices in proximity that the App can establish the Encounter Tokens per second - 100.}
In theory, we can estimate the amount of encounter established based on the bandwidth of BLE communication. To establish an $ET$, the device need to send and receive an AM and $Q$ that are 192 + 520 = 712 (bits) or 712 *2 = 1,424 (bits) for both sending and receiving. In the worst case, the BLE bandwidth is 125 (kbit/s) that means the app can establish 125,000/ 1,424 = 175 $ET$s per second. In the ideal case, the BLE bandwidth is 2 Mbit/s that means the app can establish 2,000,000/1,424  = 1,404 $ET$s per second. However, since the latency to establish a BLE connection is ca. 6 ms, the app can only establish 1000/6 = 160 $ET$s per second if we consider the data transmission time is much smaller than the connection time. Thus, it is fair to estimate that the app can establish 100 encounters per second.

\subsection{List of Existing Works on  Digital Contact Tracing}
\label{sec:existing-analysis}
We summarize attacks on \dct in Tab. \ref{tab:existing-attacks-analyses}.
\else

\fi

\end{document}